\documentclass[fleqn,usenatbib]{mnras}

\usepackage{newtxtext,newtxmath}

\usepackage[T1]{fontenc}

\DeclareRobustCommand{\VAN}[3]{#2}
\let\VANthebibliography\thebibliography
\def\thebibliography{\DeclareRobustCommand{\VAN}[3]{##3}\VANthebibliography}

\usepackage{graphicx}	
\usepackage{amsmath}	

\usepackage{subcaption}
\usepackage{caption}
\usepackage{multirow} 
\usepackage{float} 
\usepackage{natbib}
\usepackage{hyperref}
\defcitealias{chandro-gomez_inprep}{Paper~II}

\title[MQGs at $z\ge2$ in COLIBRE -- I]{Unveiling the population of massive quenched galaxies at $z\ge2$ in the COLIBRE simulations -- I. Galaxy demographics}

\author[\'Angel Chandro-G\'omez et al.]{
\parbox[t]{\textwidth}{
\vspace{-0.5cm}
\'Angel Chandro-G\'omez$^{1,2}$\thanks{E-mail: angel.chandrogomez@research.uwa.edu.au}, 
Claudia del P. Lagos$^{1,2}$,
Chris Power$^{1,2}$,
\textcolor{black}{William} M. Baker$^{3}$,
Alejandro Benítez-Llambay$^{4}$,
Evgenii Chaikin$^{5,6}$,
Harry G. Chittenden$^{7,8}$,
Camila Correa$^{6}$,
Carlos S. Frenk$^{5}$,
Filip Huško$^{6}$,
Robert J. McGibbon$^{6}$,
Themiya Nanayakkara$^{7,8}$,
Sylvia Ploeckinger$^{9}$,
Alexander J. Richings$^{10,11}$,
Matthieu Schaller$^{6,12}$,
Joop Schaye$^{6}$
and James W. Trayford$^{13}$}
\vspace*{8pt} \\
$^{1}$ International Centre for Radio Astronomy Research (ICRAR), The University of Western Australia, 35 Stirling Highway, Crawley, WA 6009, Australia\\
$^{2}$ ARC Centre for All-Sky Astrophysics in 3 Dimensions (ASTRO 3D)\\
$^{3}$ DARK, Niels Bohr Institute, University of Copenhagen, Jagtvej 155A, DK-2200 Copenhagen, Denmark\\
$^{4}$ Dipartimento di Fisica G. Occhialini, Universit\`a degli Studi di Milano Bicocca, Piazza della Scienza, 3 I-20126 Milano MI, Italy\\
$^{5}$ Institute for Computational Cosmology, Department of Physics, University of Durham, South Road, Durham, DH1 3LE, UK\\$^{6}$ Leiden Observatory, Leiden University, PO Box 9513, NL-2300 RA Leiden, The Netherlands\\
$^{7}$ Centre for Astrophysics and Supercomputing, Swinburne University of Technology, P.O. Box 218, Hawthorn VIC 3122, Melbourne, Australia\\
$^{8}$ \textit{JWST} Australia Data Centre, Swinburne Advanced Manufacturing and Design Centre, John Street, Hawthorn VIC 3122, Australia\\
$^{9}$ Department of Astrophysics, University of Vienna, Türkenschanzstrasse 17, A-1180 Vienna, Austria\\
$^{10}$ Centre for Data Science, Artificial Intelligence and Modelling, University of Hull, Cottingham Road, Hull, HU6 7RX, UK\\
$^{11}$ E. A. Milne Centre for Astrophysics, University of Hull, Cottingham Road, Hull, HU6 7RX, UK\\
$^{12}$ Lorentz Institute for Theoretical Physics, Leiden University, PO Box 9506, NL-2300 RA Leiden, The Netherlands\\
$^{13}$ Institute of Cosmology and Gravitation, University of Portsmouth, Dennis Sciama Building, Burnaby Road, Portsmouth PO1 3FX, UK\\
\vspace*{-0.5cm}}

\date{Accepted XXX. Received YYY; in original form ZZZ}

\pubyear{2025}

\begin{document}
\label{firstpage}
\pagerange{\pageref{firstpage}--\pageref{lastpage}}
\maketitle

\begin{abstract}
The \textit{James Webb Space Telescope} has uncovered a substantial population of Massive ($M_{\star} > 10^{10}\,\mathrm{M_{\odot}}$), Quenched Galaxies (MQGs) in the early Universe ($z \ge 2$), whose properties challenge current galaxy formation models. In this \textcolor{black}{series}, we examine this population of MQGs within the new {\sc COLIBRE} cosmological hydrodynamical simulations\textcolor{black}{, which introduce key innovations in their sub-grid physics. \textcolor{black}{In this first paper,} we find a dependence of MQG number densities on both mass resolution and the Active Galactic Nucleus feedback implementation, as well as a significant impact from potential observational uncertainties. Using the fiducial $(200\,\rm cMpc)^3$ volume L200m6 simulation, which provides adequate volume, mass and spatial resolution to study these systems,} we report number densities and stellar mass functions in broad agreement with the latest observations. \textcolor{black}{The predicted quenching and formation timescales are qualitatively consistent with observational inferences, indicating extended formation (medians $t_{50}\approx0.5-1.5\,\mathrm{Gyr}$) followed by rapid quenching (medians $t_{\mathrm{q}}\lesssim0.6\,\mathrm{Gyr}$) with strong starburst episodes.} Leveraging the state-of-the-art physics in {\sc COLIBRE}, the model predicts that MQGs have dust and $\rm H_{2}$ fractions more than $1$~dex lower than their massive star-forming counterparts; \textcolor{black}{generally consistent with the (scarce) observational estimates}. \textcolor{black}{MQGs and massive star-forming systems show broadly similar stellar sizes and kinematics}, \textcolor{black}{suggesting that size or morphological transformations occur after quenching in {\sc COLIBRE}}. Our results provide robust predictions for MQGs \textcolor{black}{and} show that tensions with observations are reduced when an effective observational uncertainty is forward-modelled. 
\end{abstract}




\section{Introduction}
\label{sec:intro} 

The \textit{James Webb Space Telescope} (\textit{JWST}) has ushered in a new era in observational astronomy, thanks to its high sensitivity and resolution, providing an unprecedented view of the early Universe \citep[see, for example,][for a review]{adamo25}. The first \textit{JWST} results revealed a surprisingly large population of Massive, Quenched Galaxies (hereafter MQGs) at high redshift ($z > 3$), with stellar masses $M_{\star} > 10^{10}\,\mathrm{M}_{\odot}$ and negligible star formation \citep{nanayakkara22, carnall23}. Such galaxies had already been detected \citep{dunlop96, cimatti04, glazebrook17, schreiber18, valentino20, weaver22}, but \textit{JWST} showed that their number densities are higher than previously thought, that older MQGs exist and that they are found at even higher redshifts \citep{weibel25}, implying that this population is not as rare as once believed. 

The high number densities inferred from the first \textit{JWST} observations were initially difficult to reconcile with the hierarchical $\Lambda$ Cold Dark Matter (hereafter $\Lambda$CDM) model; these objects appeared too massive for the early epochs at which they are detected and would require implausibly high star formation efficiencies \citep{boylan-kolchin23}. However, as observational samples have grown, through large photometric surveys \citep[e.g.][]{baker25, stevenson25} and spectroscopic confirmations \citep[e.g.][]{nanayakkara25, zhang25} extending to $z \gtrsim 2$, the inferred number densities have consistently decreased, easing much of the tension. In fact, the initially reported high number densities by \citet{carnall23} have since been revised downward in light of more recent analyses. However, they remain more than a factor of two higher than earlier ground-based estimates. New surveys are beginning to unveil the diverse properties of MQGs, extending stellar mass and star formation rate (SFR) measurements to include dust content \citep{lee24}, sizes \citep{kawinwanichakij25} and kinematics \citep{pascalau25}, and they are enabling comparisons with lower-$z$ counterparts.

The short timescales on which MQGs form and quench are also puzzling. Spectroscopic observations indicate that some quenched systems must have formed rapidly, as early as when the Universe was about or even less than one Gyr old \citep[with examples at $z \approx 11$;][]{glazebrook24}. These studies suggest that strong starburst episodes drive this early mass assembly \citep{forrest20}. In the local Universe, feedback processes explain the quiescent state of galaxies \citep[e.g.][]{man18}; these include internal mechanisms such as supernova (SN) and active galactic nucleus (AGN) feedback and external influences such as interactions with, and regulation by, the environment. SN feedback is insufficient to quench massive systems and predominantly affects lower-mass galaxies \citep{benson03}, even at high redshifts \citep[e.g.][]{gelli23}; while it is believed that environment is already important at high redshifts \citep[e.g.][]{jin24}, its role in quenching massive galaxies seems to be minor \citep[e.g.][]{lee15, shuntov25}. Given the short quenching timescales of MQGs, AGN feedback is therefore considered the most plausible mechanism. This process involves the injection of energy and momentum by central supermassive black holes (BHs) into their surroundings and is supported by growing observational evidence of AGN activity in these systems \citep{martinez-marin24, d'eugenio24, park24, stevenson25, baker25b}. 

The last decade has witnessed a remarkable growth in galaxy formation simulations that model statistical samples of galaxies within cosmological volumes, capable of producing population properties that are in broad agreement with observations of galaxies in the local Universe \citep[e.g.][]{schaye15, pilepich18}. They accomplish this by calibrating parameterised sub-grid models for, for example, BH growth and feedback, to recover particular sets of observational data \citep[e.g.][]{crain15,chaikin25}, typically from low-redshift galaxy surveys, while still enabling genuine predictions for other galaxy properties. Such simulations provide an ideal framework to test what role AGN feedback plays in quenching MQGs at high redshift, and to connect them to their low-$z$ descendants, 
a link that is difficult to establish observationally \citep[see e.g. the reviews by][]{somerville15, crain23}. 

\textcolor{black}{Because of this, a plethora of simulation studies has increasingly produced predictions for these passive systems in recent years to help interpret the \textit{JWST} observations. These efforts span a wide range of galaxy formation and evolution models with different sub-grid physics implementations: large-volume hydrodynamical simulations such as {\sc IllustrisTNG} \citep{hartley23, kurinchi-vendhan24}, {\sc Magneticum} \citep{kimmig25, remus25} or {\sc Simba-C} \citep{szpila25}; zoom-in hydrodynamical simulations such as {\sc FLARES} \citep{Lovell23, turner25} or {\sc Thesan} \citep{chittenden25}; semi-analytic models such as {\sc Shark v2.0} \citep{lagos24} or {\sc GAEA} \citep{de-lucia24, xie24}; as well as comparisons between multiple modelling approaches \citep{lustig23, lagos25, weller25}. These studies reveal substantial differences between models, primarily arising from variations in AGN feedback implementations \citep{lagos25}, and in some cases from numerical limitations such as limited statistics due to small simulation volumes and finite spatial resolution. Moreover, these models continue to struggle to reproduce the observational properties of these systems, \textcolor{black}{most notably their abundance, in general underpredicting the observed number densities at $z\gtrsim4$. Discrepancies also remain in reproducing their stellar mass distribution and other physical properties \citep[e.g.][]{lagos25, baker25}}. Together, these factors have so far limited the ability of current simulations to provide a fully consistent and comprehensive description of MQGs.}

\textcolor{black}{We present here a series of papers, including this work (Paper~I), and \citet[][hereafter \citetalias{chandro-gomez_inprep}]{chandro-gomez_inprep}, intended to be read together, where we investigate the properties and origins of MQGs at high redshift in the new {\sc COLIBRE} cosmological hydrodynamical simulations \citep{schaye25, chaikin25}. {\sc COLIBRE} introduces several advances over previous cosmological simulations, including the self-consistent modelling of a cold molecular phase in the interstellar medium (ISM) coupled to live dust grains, together with updated \textcolor{black}{BH growth and} AGN feedback prescriptions.} \citet{chaikin25b} show that the evolution of the galaxy stellar mass function (SMF) in {\sc COLIBRE} is reproduced across the full redshift range ($z=0-12$) for which observational data are available. That work presents the first results on the abundance of MQGs in these simulations, demonstrating good agreement with \textit{JWST} data. \textcolor{black}{In this first paper of the series}, we extend \textcolor{black}{this initial investigation} with a more comprehensive analysis of the {\sc COLIBRE} MQGs, providing further details on their abundance and assessing whether the {\sc COLIBRE} model can reproduce the many other key characteristics of MQGs as revealed by recent \textit{JWST} surveys. We focus on a comprehensive set of galaxy properties, including stellar masses, SFRs, dust and molecular gas content, stellar sizes and kinematics; and compare them with the latest observational constraints. \textcolor{black}{In \citetalias{chandro-gomez_inprep}, we identify} the physical mechanisms responsible for their rapid formation and quenching. By tracing the evolutionary histories of individual MQGs, we explore the role of BH growth and their associated AGN feedback, as well as environmental factors in driving quenching of these systems. The tracking is performed down to $z=0$, enabling us to examine the fate of MQGs after quenching.


The structure of this \textcolor{black}{first} paper is as follows. In \S~\ref{sec:colibre} we describe the {\sc COLIBRE} simulations \textcolor{black}{and their sub-grid physics models}.
\S~\ref{sec:prop} outlines our sample selection, as well as the properties analysed in this work. Predictions across a range of observables, including number densities, SMFs, star formation histories (SFHs), dust and molecular gas fractions, sizes and kinematics; are presented \textcolor{black}{and compared with observations} in \S~\ref{sec:res1}. 
Finally, \S~\ref{sec:conclusions} summarises our results.

\section{The COLIBRE simulation suite}
\label{sec:colibre}

\begin{table*}
	\centering
	\caption{{\sc COLIBRE} hydrodynamical simulations used in this work. $L$: the periodic box size in comoving Mpc (cMpc); $N_{\rm b}$: the initial number of baryonic particles in the volume; $N_{\rm DM}$: the number of DM particles in the volume; $m_{\rm g}$: the mean initial gas particle mass in M$_{\odot}$; $m_{\rm DM}$: the mean DM particle mass in M$_{\odot}$; $\epsilon_{\rm com}$: the gravitational softening length in ckpc; $\epsilon_{\rm prop}$: the maximum gravitational softening length in proper kpc (pkpc); and the type of AGN feedback implemented (thermal or hybrid including thermal+jet); $\mathrm{N_{MQG}}(z=3)$: the number of MQGs selected at $z=3$ using the criteria defined in \S~\ref{ssec:prop-mq-def}. We highlight in bold the simulation adopted as fiducial in this work.}
	\label{tab:runs}
	\begin{tabular}{cccccccccc}
		\hline
		{\sc COLIBRE} sim. & $L$/cMpc & $N_{\rm b}$ & $N_{\rm DM}$ & $m_{\rm g}$/M$_{\odot}$ & $m_{\rm DM}$/M$_{\odot}$ & $\epsilon_{\rm com}$/ckpc & $\epsilon_{\rm prop}$/pkpc & AGN feedback model & $\mathrm{N_{MQG}}(z=3)$\\
		\hline
        L050m5 & 50 & 1504$^3$ & $4\times1504^3$ & 2.30$\times$10$^5$ & 3.03$\times$10$^5$ & 0.9 & 0.35 & thermal & 2\\
        \textcolor{black}{L100m5} & \textcolor{black}{100} & \textcolor{black}{3008}$^3$ & $4\times\textcolor{black}{3008}^3$ & 2.30$\times$10$^5$ & 3.03$\times$10$^5$ & 0.9 & 0.35 & thermal & \textcolor{black}{12}\\
        L100m6 & 100 & 1504$^3$ & $4\times1504^3$ & 1.84$\times$10$^6$ & 2.42$\times$10$^6$ & 1.8 & 0.7 & thermal & 79\\
        \textbf{L200m6} & $\boldsymbol{200}$ & $\boldsymbol{3008^3}$ & $\boldsymbol{4\times3008^3}$ & $\boldsymbol{1.84\times10^6}$ & $\boldsymbol{2.42\times10^6}$ & $\boldsymbol{1.8}$ & $\boldsymbol{0.7}$ & \textbf{thermal} & $\boldsymbol{470}$\\
        L200m7 & 200 & 1504$^3$ & $4\times1504^3$ & 1.47$\times$10$^7$ & 1.94$\times$10$^7$ & 3.6 & 1.4 & thermal & 1527\\
        L400m7 & 400 & 3008$^3$ & $4\times3008^3$ & 1.47$\times$10$^7$ & 1.94$\times$10$^7$ & 3.6 & 1.4 & thermal & 12359\\
        L100m6h & 100 & 1504$^3$ & $4\times1504^3$ & 1.84$\times$10$^6$ & 2.42$\times$10$^6$ & 1.8 & 0.7 & hybrid (thermal+jet) & 5\\
        L200m7h & 200 & 1504$^3$ & $4\times1504^3$ & 1.47$\times$10$^7$ & 1.94$\times$10$^7$ & 3.6 & 1.4 & hybrid (thermal+jet) & 310\\
		\hline
	\end{tabular}
\end{table*}

For this analysis, we use the new {\sc COLIBRE}\footnote{\url{https://colibre-simulations.org/}} large-volume cosmological hydrodynamical simulations \citep{schaye25, chaikin25}. The {\sc COLIBRE} suite comprises several simulations that vary in volume (periodic box size of $L=25-400\rm\,cMpc$ per side), resolution (mean initial gas particle mass, $m_{\rm g}$, and mean DM particle mass, $m_{\rm DM}$, $\sim10^5-10^7\,\mathrm{M}_{\odot}$), and AGN feedback model (either fiducial thermal or hybrid thermal+jet). Table~\ref{tab:runs} summarises the simulation set used in this analysis, which allows us to assess how resolution, volume and feedback variations influence the properties of MQGs. Among these, we adopt the L200m6 simulation as our fiducial choice, as it offers the best compromise between volume ($L=\rm 200\,cMpc$ per side), sufficient to sample rare MQGs, and resolution ($m_{\rm g}$~and~$m_{\rm DM}\rm \sim10^6\,M_{\odot}$), adequate to capture their internal structure more accurately. The relatively small gravitational softening lengths ($\epsilon_{\rm com}$ and $\epsilon_{\rm prop}$) of this run are important for more reliably analysing galaxy sizes and kinematics, as discussed in \S~\ref{ssec:res1-size} and Appendix~\ref{appendix:l400m7-size-kin}.

\subsection{Simulation methods}
\label{ssec:colibre-method} 

The initial conditions (ICs) are generated with {\sc MONOFONIC} \citep{hahn20, michaux21} and evolved using the {\sc SWIFT} code \citep{schaller24}. The density field is sampled with cold DM and baryonic particles, with the latter initially representing gas but able to convert into collisionless stellar or BH particles. To suppress spurious energy transfer between DM and stellar particles, the ICs of the simulations contain four times more DM particles than baryonic ones \citep{ludlow19}.
Stellar and gas particles can increase or decrease their mass due to stellar mass loss.
Gas particles exceeding $4\,m_{\rm g}$ are split into two nearly co-spatial equal-mass particles to avoid baryonic particles with very different masses. Gravity is solved with a 4th-order fast multipole method \citep{dehnen14} and softened using a kernel with a functional form given by \citet{wendland95}, while hydrodynamics is handled with the {\sc SPHENIX} implementation of smoothed particle hydrodynamics (SPH) \citep{borrow22} using a quartic spline kernel. The simulations are run from $z=63$ to $z=0$, adopting the DES Y3 ‘3x2pt + All Ext.’\footnotemark $\Lambda$CDM cosmology \citep{abbott22}: $h=0.681$, $\Omega_{\Lambda}=0.693922$, $\Omega_{\rm m}=0.306$, $\Omega_{\rm CDM}=0.256011$, and $\Omega_{\rm b}=0.0486$.

\footnotetext{Maximum-posterior likelihood values from the Dark Energy Survey year three (DES Y3) assume a spatially flat universe and combine constraints from three two-point correlation functions: cosmic shear, galaxy clustering, and galaxy-galaxy lensing, with external data from baryon acoustic oscillations (BAO), redshift-space distortions, type Ia SNe, and Planck observations of the Cosmic Microwave Background (CMB) (including CMB lensing), Big Bang nucleosynthesis, and local measurements of the Hubble constant.}

Between $z=30$ and $z=0$, the simulations output 36 full snapshots and 92 reduced ``snipshots'' (which record only BH properties and a limited subset of gas and stellar properties). Haloes are first identified with a friends-of-friends (FoF) algorithm that links DM particles separated by less than 0.2 times the mean inter-particle distance, with baryonic particles attached to the nearest DM neighbour. FoF groups with fewer than 32 particles are discarded. Subhaloes are then identified using the {\sc HBT-HERONS} code \citep{forouhar25}, an extension of {\sc HBT+} \citep{han18}, which tracks the most bound particles of each structure from the time it was first identified. This explicit temporal tracking improves detection of substructures near halo centres \citep{forouhar25} and prevents unphysical mass fluctuations in merger trees \citep{chandro-gomez25}. Finally, the {\sc SOAP} code \citep{mcgibbon25} computes halo, subhalo and galaxy properties across multiple 3D and projected apertures, defined either by fixed sizes \textcolor{black}{(e.g. 1, 10, 50 pkpc) or overdensity thresholds (e.g. 200 times the mean density of the Universe, 200 or 500 times the critical density of the Universe), using all or only the bound particles.}

\subsection{Sub-grid prescriptions}
\label{ssec:colibre-sub-grid} 

Compared to other large-volume hydrodynamical simulations, {\sc COLIBRE} introduces key innovations in its sub-grid prescriptions, which capture processes occurring below the simulation’s spatial resolution. These improvements are briefly outlined below, although we refer the reader to \citet{schaye25} and the individual papers describing each sub-grid model for a detailed discussion. The parameters of these models are calibrated \citep{chaikin25, husko25} to reproduce the $z=0$ galaxy SMF \citep{driver22}, galaxy sizes \citep{hardwick22}, and BH masses \citep{graham23}. \textcolor{black}{The hybrid AGN feedback model additionally incorporates the $z=0.2$ bolometric AGN luminosity function \citep{shen20} as a calibration constraint.} Each {\sc COLIBRE} resolution and AGN feedback model uses a separate calibration \citep[see table~1 in][]{schaye25}.

\begin{enumerate}
    \item \textit{Radiative cooling}: unlike most other hydrodynamical simulations that impose an effective equation of state to model the ISM, {\sc COLIBRE} explicitly tracks the chemical processes in the cold ISM, allowing gas to cool down to $\approx 10$~K using the {\sc HYBRID-CHIMES} model \citep{ploeckinger25}. {\sc HYBRID-CHIMES} uses the {\sc CHIMES} reaction network \citep{richings14a, richings14b} to compute radiative cooling and heating rates, treating hydrogen and helium in full non-equilibrium and metals in equilibrium with corrections for non-equilibrium free electrons.
    \item \textit{Dust}: the formation and evolution of dust grains is modelled directly on-the-fly, explicitly tracking three grain species and two grain sizes \citep{trayford25}. Grains are nucleated in the ISM by ejecta from asymptotic giant branch (AGB) stars \citep{dell'agli17}, as well as by \textcolor{black}{core collapse SNe (CCSNe)} \citep{zhukovska08}. In the ISM, grains grow by accretion \citep{hirashita14}, primarily in molecular clouds, and are destroyed through sputtering \citep{tsai95}, astration, and \textcolor{black}{feedback shocks}. Grain sizes evolve via \textcolor{black}{coagulation \citep{aoyama17} and shattering \citep{granato21}}. The dust model is coupled to the radiative cooling module, so that reactions and cooling rates account for dust presence and properties.
    \item \textit{Star formation}: star formation follows the local gravitational instability criterion from \citet{nobels24}. SFRs of gas particles that satisfy this criterion are calculated using a \citet{schmidt59} law with a fixed efficiency of gas-to-stars conversion per free-fall time of 1\%. This model reproduces the Kennicutt–Schmidt relation \citep{kennicutt98} self-consistently at kpc scales, without explicit tuning to that relation; including the observed scatter and its dependence on local properties such as stellar surface density and gas metallicity \citep{lagos25b}.
    \item \textit{Chemical enrichment}: yields have been updated with data from the literature to account for stellar mass from AGB stars, massive stars, Type Ia SNe and \textcolor{black}{CCSNe}. Abundances of key elements for radiative cooling, along with s-process elements (Ba, Sr) important in AGB stars and the r-process element (Eu) important in neutron star mergers, common-envelope jets SNe and collapsars, are tracked to enable more realistic ISM enrichment. Turbulent diffusion is included to capture small-scale mixing of metals and dust. These updates are described in \textcolor{black}{\citet{correa26}}.
    \item \textit{Stellar feedback}: early stellar feedback is implemented through stellar winds, radiation pressure, and H~\textsc{ii} regions \citep{benitez-llambay25}. The model for CCSN feedback is based on \citet{chaikin23}, with 90\% of the energy injected thermally \citep[by heating gas particles stochastically, driving galactic winds;][]{dalla-vecchia&schaye12} and the remaining 10\% injected kinetically (through low-velocity kicks, driving turbulence in the ISM). SN energy injections are distributed statistically isotropically around young stellar particles, following \citet{chaikin22}.
    \item \textit{BH growth}: supermassive BHs are modelled, including seeding, growth via mergers and gas accretion, and repositioning to account for dynamical friction \citep{springel05, bahe22}. Gas accretion can exceed the Eddington rate by a factor of 100, but is regulated by radiative feedback in the thermal model, and by a combination of radiation, accretion disc winds, and jets in the hybrid model. In the hybrid case, mass loss between the outer accretion disc and the BH horizon is also included; while the accretion disc state (set by the BH accretion rate) determines distinct regimes with different accretion and feedback efficiencies. Further details are provided in \citet{husko25} and \textcolor{black}{\citetalias{chandro-gomez_inprep}}.
    \item \textit{AGN feedback}: the fiducial purely thermal model is based on \citet{booth&schaye09} (in which the BH injects energy thermally by increasing the temperature of gas particles) while several simulations adopt a hybrid model that combines a thermal component, modelling winds and radiation, with kinetic jets (in which the BH injects energy kinetically giving a velocity kick to gas particles). The jets are regulated by the BH spin, which also determines their directions and feedback efficiencies \citep{husko25}. A more detailed description is provided in \textcolor{black}{\citetalias{chandro-gomez_inprep}, which focuses on analysing the effect of AGN feedback in MQGs.}
\end{enumerate}

\section{Galaxy properties and sample selection}
\label{sec:prop} 

Galaxies in {\sc COLIBRE} are embedded within subhaloes, and so determining galaxy properties is intimately linked to identifying subhaloes and quantitatively characterising their contents. Subhaloes in {\sc COLIBRE} are identified with {\sc HBT-HERONS}, which tracks particles across timesteps. Galaxy properties are then measured using {\sc SOAP} in different apertures centred on the most bound particle. Throughout this paper, we adopt a fiducial 50 pkpc spherical aperture to analyse high-$z$ MQGs, considering only particles bound to the subhalo, as in \citet{schaye25}. Unless otherwise stated, this aperture is used. 

We also test other aperture radii \textcolor{black}{(e.g. 1 pkpc and 10 pkpc)}. For reference, \textit{JWST} observations typically employ apertures of $\approx10\,\rm ckpc$ to measure colours for MQGs \citep[see section 2.3 in][]{lagos25}\textcolor{black}{, corresponding to $\approx 1-3$ pkpc over the redshift range $z=2-8$. We find that the measured properties converge for the 10 pkpc aperture and, to a lesser extent, for the 1 pkpc aperture. In the latter case, we obtain slightly higher number densities, likely because this selection excludes star-forming outskirts in more extended star-forming systems (Appendix~\ref{appendix:mqg-selection})}. This reflects the compact nature of MQGs at high redshift (see \S~\ref{ssec:res1-size}) and confirms that the galaxy properties we use are robust to the \textcolor{black}{fiducial} choice of aperture. 

Below, we summarise the main galaxy properties and definitions used throughout the paper.

\begin{itemize}
    \item \textit{Stellar mass} ($M_\star$): computed as the sum of all stellar particles. 
    \item \textit{SFR}: calculated as the sum of the SFRs of gas particles flagged as star-forming (i.e., locally gravitationally unstable; \citealt{nobels24}). We adopt this measure of instantaneous SFR. \textcolor{black}{We verify that using a $\rm10\,Myr$ averaged SFR does not significantly affect the results, whereas a $\rm100\,Myr$ average has a noticeable impact on the number densities at $z>3$ (see Appendix~\ref{appendix:mqg-selection}). These averaged estimates are more directly comparable to observational tracers \citep{kennicutt12}. In particular, photometric studies are typically sensitive to timescales of order $\sim100\,\rm Myr$ \citep[e.g.][]{alberts24, baker25}, while spectroscopic analyses can probe shorter timescales of around $\sim10\,\rm Myr$ \citep[e.g.][]{zhang25}.}
    \item \textit{Dust mass} ($M_{\rm dust}$): the total dust mass in gas particles (the sum over the $6$ species described in \S~\ref{sec:colibre}).
    \item \textit{Molecular hydrogen mass} ($M_{\rm H_2}$): the sum of $\rm H_2$ mass in gas particles.
    \item \textit{Stellar half-mass radius} ($r_\star$): the radius enclosing half of the stellar mass, measured in projected 50 pkpc apertures along the $x$, $y$, and $z$-axes, considering bound particles only. The final value is the average of the three projections to mimic the random viewing orientation of observed galaxies.
\end{itemize}

\subsection{Selecting the high-$z$ MQG sample}
\label{ssec:prop-mq-def} 

Multiple definitions for MQGs at high-$z$ exist in the literature. We adopt the following criteria, focusing on systems at $z\geq2$ where there appears to be various degrees of tension with galaxy formation models and simulations \citep{lagos25}.

\begin{enumerate}
    \item Massive criterion: $\boldsymbol{M_\star>10^{10}\,\rm M_\odot}$.
    \item Quenched criterion: $\boldsymbol{\mathrm{sSFR}<0.2/t_{\rm age}}$, where the specific star formation rate is defined as $\mathrm{sSFR=SFR}/M_{\star}$ and $t_{\rm age}$ is the age of the Universe at the redshift of selection. Galaxies above this threshold are classified as star-forming.
\end{enumerate}

The ``massive'' threshold is defined following common practice. For the definition of quiescence, several approaches exist, including colour–colour cuts, fixed specific sSFR thresholds, or redshift-dependent sSFR thresholds. We opt for a redshift- and mass-dependent cut following \citet{franx08}, since relevant timescales (e.g. gravitational or cooling times) scale with the Hubble time and gas density; while colour-based methods can be contaminated by obscured star formation or emission lines \citep{schreiber18} and may even miss quiescent galaxies that fall outside the traditional strict criteria \citep{merlin18}. Table~\ref{tab:runs} shows the number of MQGs identified at $z=3$ using this definition. A comparison of alternative criteria is presented in Appendix~\ref{appendix:mqg-selection}.

\subsection{Computing star formation histories}
\label{ssec:prop-sfh} 

\begin{figure}
\centering
\includegraphics[width=0.48\textwidth]{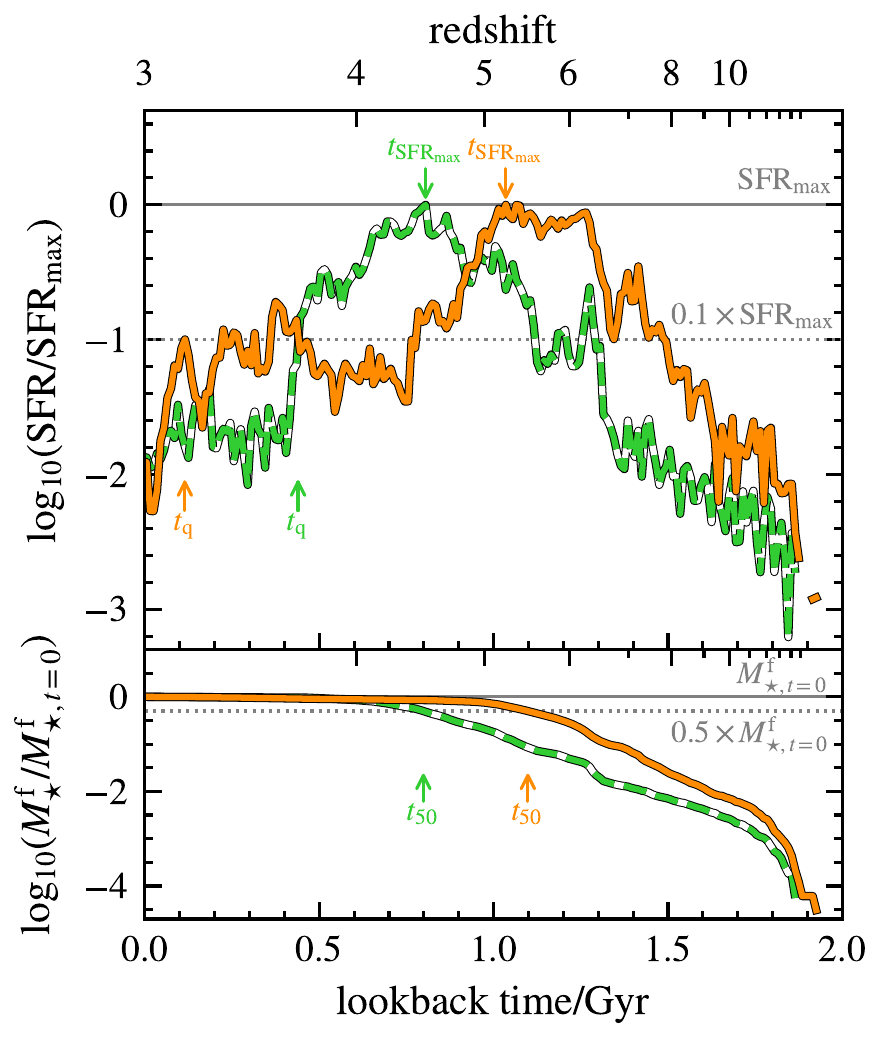}
\caption{\textit{Top panel}: normalised SFHs ($\rm SFR/SFR_{max}$ versus lookback time, with redshift indicated by the top $x$-axis) of two example galaxies identified to be massive and quenched at $z=3$ in L200m6. The horizontal solid line corresponds to the peak SFR value, $\rm SFR_{max}$, while the dotted line shows the $\rm 0.1\times SFR_{max}$ threshold used to define the quenching timescale, $t_{\rm q}$. Arrows indicate $t_{\rm q}$ for each galaxy, defined as the latest time the galaxy's SFR drops below this threshold; as well as the SFR peak time, $t_{\rm SFR_{max}}$. The green curve represents a galaxy that, once it has crossed the threshold indicated by the dotted line, never rises above it again. The orange curve represents a galaxy that experiences several star formation episodes above the threshold after its SFR peak. \textit{Bottom panel}: normalised stellar mass assembly histories, \textcolor{black}{$M_{\star}^{\mathrm{f}}/M_{\star,t=0}^{\mathrm{f}}$,} versus lookback time, with redshift indicated by the top $x$-axis) for the same galaxies. The horizontal solid line shows the maximum \textcolor{black}{``formed''} stellar mass value, \textcolor{black}{$M_{\star,t=0}^{\mathrm{f}}$}; while the dotted line shows the \textcolor{black}{$0.5\times M_{\star,t=0}^{\mathrm{f}}$} threshold used to define the formation timescale, $t_{\rm 50}$. Arrows indicate $t_{\rm 50}$.} 
\label{fig:sfh-example} 
\end{figure}

SFHs are computed from stellar particles gravitationally bound to each galaxy within the fiducial $\rm 50\,pkpc$ spherical aperture, using their stellar ages and stellar masses at birth to construct a mass-weighted histogram of formation times. It is important to note that this method does not distinguish between in-situ (stars formed within the galaxy) and ex-situ (stars formed outside the galaxy’s main progenitor) star formation. Consequently, star formation bursts could reflect the peak formation of an accreted ex-satellite rather than intrinsic activity in the galaxy itself.

We adopt a bin size of $\rm 10\,Myr$ to sample the SFHs, which gives reasonable results, being much longer than the simulation time step and comparable to observational studies capable of probing SFR variations on similar short timescales via spectral features \citep{kennicutt12}. We also test a $\rm 20\,Myr$ bin size, which yields similar results but smooths out variability, making it less effective at capturing short-timescale features in the SFHs.

Observationally, different SFH features are extracted to characterise key evolutionary events in a galaxy, rather than comparing their overall shapes, which have less quantitative meaning. To enable meaningful comparisons with observations, we focus on a subset of these features, measured in lookback time from the epoch of selection, following common practices in the literature:

\begin{itemize}
    \item The quenching timescale, $t_{\rm q}$, indicates when a galaxy effectively ceases star formation. We define $t_{\rm q}$ as the lookback time from the selection redshift (not from $z=0$) when the SFR last dropped below 10\% of its peak value: $\mathrm{SFR}(t_{\rm q}) = 0.1 \times \mathrm{SFR_{max}}$. Our approach is inspired by \citet{nanayakkara25}, although they define quenching as the longest continuous period below this threshold after the peak.
    
    The top panel of Fig.~\ref{fig:sfh-example} shows a single-starburst SFH of a $z=3$ MQG (green line), where the SFR remains below the threshold once crossed. For MQGs with multiple starburst episodes after the peak (orange line), we adopt the last time the galaxy crosses it, reflecting when it finally transitions to quiescence before the selection snapshot. The fraction of galaxies with multiple starbursts after the peak is dominant in general with $\approx0.76$ at $z=2$, $\approx0.64$ at $z=3$, $\approx0.46$ at $z=4$, $\approx0.09$ at $z=5$, and $\approx0.50$ at $z=6$ for L200m6. Multiple episodes of SF, before or after the peak could be connected to `mini-quenching' discussed in the literature \citep{strait23, looser24, baker25d}.
    
    \item The formation timescale describes how rapidly a galaxy assembles its stellar mass. It can be defined in several ways: for example, the time at which 50\% ($t_{50}$) or 90\% ($t_{90}$) of the stellar mass has formed, or the interval $t_{90}-t_{50}$. We adopt $t_{50}$ as the primary measure of formation time, which we compute in lookback time from the selection redshift as follows. Using the SFH, we calculate the cumulative stellar mass by integrating over time bins and estimate when its \textcolor{black}{``formed''} mass first exceeds 50\% of its integrated mass at \textcolor{black}{selection, $0.5 \times M_{\star,t=0}^{\mathrm{f}}$}. This is shown in the bottom panel of Fig.~\ref{fig:sfh-example}. Note that \textcolor{black}{$M_{\star}^{\mathrm{f}}$} ignores stellar mass loss and may therefore differ from the actual \textcolor{black}{``observed''} galaxy stellar mass \textcolor{black}{$M_{\star}^{\mathrm{obs}}=M_\star$}.
\end{itemize}

\subsection{Kinematic tracer $v_{\star}/\sigma_\star$ definition}
\label{ssec:prop-vsigma} 

To characterise the kinematics of a galaxy, we focus on the stellar component and measure $v_{\star}/\sigma_\star$ as a tracer. This approach has recently been applied to massive quenched systems \citep{pascalau25, slob25, forrest25}. Here, $v_{\star}$ represents the stellar rotational velocity, indicating the galaxy's rotational support, while $\sigma_\star$ quantifies the stellar velocity dispersion, indicating the random motion contribution. This measure differs from the classical spin parameter $\lambda_r$, although the two can be converted \citep{cappellari16}.

To compute these quantities, we must first define a reference frame: 1. measure the galactic centre, and 2. rotate the galaxy to align it with the stellar angular momentum vector. We define our centre as the stellar centre of mass and its corresponding velocity in a $\rm 50\,pkpc$ spherical aperture, using only bound particles. Other options are: (i) the stellar centre of mass and its velocity within other apertures such as $10$~or~$\rm 30\,pkpc$, (ii) the position and velocity of the most massive BH particle within a $\rm 50\,pkpc$ spherical aperture, considering only bound particles, or (iii) the most bound particle of the subhalo (subhalo centre) and stellar centre of mass velocity within $\rm 10\,pkpc$. We find that our approach gives more stable results, minimising the mean position and velocity within our fiducial aperture. Secondly, we rotate the galaxy so that the $z$-axis aligns with the stellar angular momentum vector of the $\rm 10\,pkpc$ core.

Once aligned, we compute the stellar kinematics using cylindrical coordinates for each bound particle ($v_r, v_\phi, v_z$) in the fiducial 50 pkpc aperture centred on the most bound particle. The rotational velocity is given by the azimuthal component, $v_\phi$, and the total galaxy’s stellar rotational velocity is computed as the mass-weighted root mean square (RMS):
\begin{equation}
    v_{\star}=\sqrt{\frac{\sum_{i\in R}{m_{i}v_{\phi,i}^{2}}}{\sum_{i\in R}{m_{i}}}}.
\end{equation}

\noindent The perpendicular component to the plane traces the vertical velocity dispersion in the disc, $v_z$, and its final value is similarly computed as the mass-weighted RMS:
\begin{equation}
    \sigma_{\star}=\sqrt{\frac{\sum_{i\in R}{m_{i}v_{z,i}^{2}}}{\sum_{i\in R}{m_{i}}}}.
\end{equation}

We caution the reader that our $v_{\star}/\sigma_\star$ measurements (\S~\ref{ssec:res1-size-results}) are not quantitatively comparable to observations because of limitations such as beam smearing and projection effects \citep{harborne19, harborne20}, which make our values effectively upper limits \textcolor{black}{for this ratio\footnotemark.} Observations measure line-of-sight velocity and velocity dispersion (which differ from our definition) and are sensitive to galaxy orientation, although inclination corrections are applied. Observations also typically measure $v_{\star}/\sigma_{\star}$ within the half-light radius, whereas we use all stellar particles within the aperture.

\footnotetext{\textcolor{black}{The rotational velocity we measure corresponds to the intrinsic rotational component and is therefore typically underestimated in observations, while the velocity dispersion inferred observationally is often overestimated due to observational limitations such as beam smearing or projection effects \citep[see a detailed discussion and demonstration of these in][]{harborne20}.}}

\section{The demographics of MQGs in COLIBRE}
\label{sec:res1} 

We present predictions from the {\sc COLIBRE} model for MQGs (as defined in \S~\ref{ssec:prop-mq-def}) at high-redshift ($z\ge 2$). Our analysis considers a broad set of properties, including those that can be constrained observationally and others that remain largely theoretical due to current observational limitations. 

\subsection{Number densities of MQGs}
\label{ssec:res1-ndens} 

We begin by analysing the evolution of the number density (i.e. number of objects per unit comoving volume) of MQGs. \citet{lagos25} have shown that galaxy formation and evolution models face challenges in consistently reproducing the number density of MQGs inferred from \textit{JWST} observations. \citet{chaikin25b} perform a {\sc COLIBRE} analysis using only two simulations with the thermal AGN feedback model; here, we extend this study.


\subsubsection{Results}
\label{sssec:res1-ndens-results} 

\begin{figure}
\centering
\includegraphics[width=0.48\textwidth]{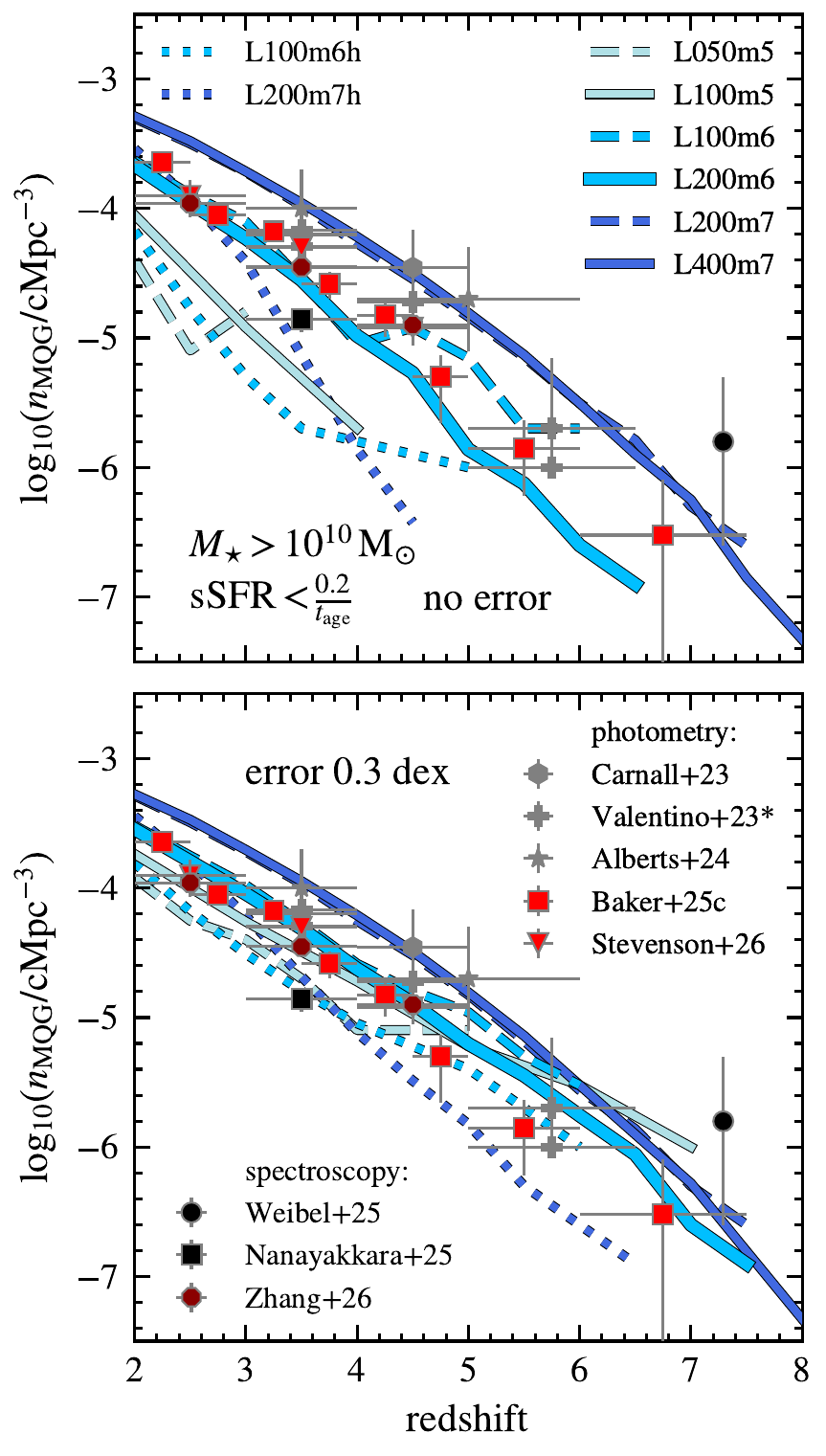}
\caption{Comoving number density of MQGs, defined by $M_{\star}>10^{10}\,\rm \textcolor{black}{M_\odot}$ and $\mathrm{sSFR}<0.2/t_{\rm age}$, as a function of redshift. Lines represent predictions from various {\sc COLIBRE} boxes (spanning different volumes, resolutions, and AGN feedback models) in Table~\ref{tab:runs}, with the selection at $2 \le z \le 8$. \textcolor{black}{Lines that terminate at $z<8$ indicate that no galaxies are identified as MQGs beyond those redshifts}. These are compared with the latest \textit{JWST} observational estimates: black (smaller area) and dark red (larger area) points show spectroscopic measurements, while grey (smaller area) and red (larger area) points correspond to photometric measurements. Observational data are taken from \citet{carnall23, valentino23, alberts24, baker25, weibel25, nanayakkara25, stevenson25, zhang25}. \textit{Top panel}: predicted simulation values. \textit{Bottom panel}: $M_{\star}$ and SFR values in the simulations are convolved independently with a Gaussian distribution (mean 0, standard deviation $0.3$~dex) representing a reasonable error budget for these quantities.}
\label{fig:ndens-boxes} 
\end{figure}

The top panel of Fig.~\ref{fig:ndens-boxes} presents results from the various {\sc COLIBRE} simulations listed in Table~\ref{tab:runs}; these differ in volume, resolution, and AGN feedback implementation. Number densities are computed at intervals of 0.5 in redshift over $z=2-8$ \textcolor{black}{(except for L100m5, which is only analysed at integer redshifts)}. When no galaxies are identified as MQGs at a given redshift, the corresponding redshift bin is omitted; thus, some lines end at $z<8$. These predictions are compared with the latest \textit{JWST} observational constraints. 


Fig.~\ref{fig:ndens-boxes} reveals that there is significant variation among the different simulations. At fixed colour, the dashed and solid lines correspond to different simulated volumes at fixed mass resolution. Larger volumes capture rarer galaxies at the highest redshifts, reducing cosmic variance \citep[particularly important for these systems, e.g.][]{lim25} and improving number statistics (notably \textcolor{black}{at $z\gtrsim2$ for the m5 runs}, at $z\gtrsim4$ for m6 runs and at $z\gtrsim6$ for m7 runs), as well as including longer-wavelength fluctuations that give rise to the largest overdensities capable of forming these galaxies. However, the bump seen in L100m6 around $z\approx5$ is not solely due to Poisson noise but reflects a pronounced overdensity in \textcolor{black}{the ICs\footnotemark of the $100\,\mathrm{cMpc}$ volume} traced at high redshift \citep[final part of appendix A1 in][]{chaikin25b}. 

\footnotetext{\textcolor{black}{The ICs are identical for simulations with the same volume, independently of the mass resolution or AGN feedback implementation.}}

\textcolor{black}{However,} resolution effects are also apparent. Darker (lighter) lines represent lower (higher) mass resolution simulations. There is a systematic \textcolor{black}{clear} offset between the different resolution simulations, indicating that galaxies in higher-resolution simulations are less quenched\textcolor{black}{, e.g. when comparing L200m6 (solid cyan) and L200m7 (dashed dark blue) \textcolor{black}{or L100m5 (solid light blue) and L100m6 (dashed cyan)}}. Two factors contribute to this trend: (1) at higher resolution, gas particles trace smaller effective volumes, producing denser gas around BHs, and so stronger feedback is required to eject or heat it; (2) AGN feedback becomes less bursty at higher resolution, due to the resolution-dependent nature of the feedback prescriptions \textcolor{black}{(equations~5~and~8 of \citetalias{chandro-gomez_inprep})}. These effects make AGN feedback less effective at higher resolution. Differences in BH seed masses may also play a role, since higher-resolution simulations use lower seed masses, delaying BH growth and the onset of AGN feedback.

The dotted lines represent simulations that include hybrid AGN feedback, thermal plus jet feedback, instead of purely thermal AGN feedback. This hybrid model produces significantly less quenching. \textcolor{black}{In \textcolor{black}{\citetalias{chandro-gomez_inprep}}},
we show that the difference arises because the hybrid model features BHs that grow more slowly in the early Universe compared to the fiducial, purely thermal AGN feedback model. We can see this by considering the ratio, $M_{\rm BH}/M_{\star}$. The medians are comparable in the thermal L200m7 and hybrid L200m7h simulations at $z=2$ ($\sim0.67$); however, at $z=5$, the medians are 0.70 (thermal) versus 0.64 (hybrid), and 0.70 versus 0.60 at $z=8$. Moreover, jet feedback appears to take longer to quench galaxies than thermal feedback. At fixed stellar mass and for $z \lesssim 3$, the same cumulative injected BH energy leads to a higher fraction of quenched systems in the thermal feedback model (see \textcolor{black}{fig.~2 of \citetalias{chandro-gomez_inprep}}). 
These effects do not appear to depend on mass resolution; the same trend is evident in both the m7 and m6 runs, with the pronounced overdensity visible around $z\approx5$ in L100m6h as well. By $z<2$, however, hybrid feedback becomes more efficient at quenching because the weaker early feedback allows galaxies to build more stellar mass, which must then converge by $z=0$ \citep[see fig.~13 in][]{chaikin25b}. 

Ultimately, runs at different resolutions, and within the same resolution, runs with different AGN feedback models (thermal or hybrid), were independently tuned to reproduce $z=0$ observations. Each run should therefore be seen as a distinct model. 
We remind the reader that all of them have been calibrated to consistently reproduce the $z=0$ SMF, the galaxy size–mass relation, and the galaxy BH–mass relation \citep[figs.~12,~14~and~15 in][]{schaye25}; and additionally the $z=0.2$ bolometric AGN luminosity function for the hybrid AGN feedback model \citep[fig. 10 in][]{husko25}.



Compared with observations, 
the top panel of Fig.~\ref{fig:ndens-boxes} 
shows that the highest-resolution m5 and m6 {\sc COLIBRE} thermal AGN feedback runs, as well as the hybrid AGN feedback runs, underpredict the high number densities inferred from \textit{JWST}. However, the tension is significantly alleviated if we consider potential errors associated with the derivation of stellar mass and SFR of galaxies. \citet{bellstedt25} showed that the many decisions behind the derivation of stellar masses and SFRs \citep[i.e. adopted functional form of the star formation and metallicity histories; the choice of stellar population models; the IMF; to mention a few; see][for a review]{conroy13} can lead to differences in the derived stellar masses and SFRs of $0.1-0.2$~dex and $\approx 0.2$~dex, respectively, in the $z\approx 0$ spectroscopic GAMA survey. We refer the reader to other works for further discussion of these uncertainties (e.g. \citealt{conroy09, robotham20, cochrane25, jones25}). The uncertainties reported by \citet{bellstedt25} could be seen as lower limits for the uncertainties one would expect for high-z galaxies with potentially more bursty SFHs, poorer wavelength coverage and potentially with photometric redshifts only. It is thus natural to expect uncertainties to be even larger at high-$z$. 

To exemplify the impact such an error budget could have, we show in the lower panel of Fig.~\ref{fig:ndens-boxes} the number densities after convolving the stellar masses and SFRs of {\sc COLIBRE} galaxies with a Gaussian centred at $0$ with a width of $\sigma=0.3$~dex. \textcolor{black}{The lack of a systematic shift in the recovered quantities (in a population sense) can be seen as a conservative case, in which the individual uncertainties coming from SED fitting, photometric redshifts, etc., contribute to broadening the scatter of the recovered quantities, consistent with the findings of \citet{bellstedt25} at $z=0$}. \textcolor{black}{A Gaussian error convolution is, however, only an approximate representation, since e.g. observational selections based on colours do not strictly follow this behaviour.} We carry out the error convolution for stellar masses and SFRs in an independent manner, which is \textcolor{black}{another} simplification, as errors could potentially be correlated. In Appendix~\ref{appendix:error}, we explore the impact of adopting different values for the standard deviation (Fig.~\ref{fig:ndens-selection-error}). This procedure produces perturbed values ($M_\star^{\rm conv}$ and $\mathrm{SFR}^{\rm conv}$), to which we then apply the MQG selection (\S~\ref{ssec:prop-mq-def}). To reduce stochastic noise introduced by the convolution, the procedure is repeated $100$ times, and the median of the resulting measurements is taken. This error convolution is just for reference, and the aim is to make the visual argument that quantifying uncertainties in the observational estimates is crucial to isolate the level of tension between models and the observations. For a more detailed analysis of the impact of errors on the quenched fraction, SMF and number densities of galaxies, we refer the reader to appendix~D in \citet{chaikin25b}. A more realistic calibration of observational systematics, based on generating mock photometric and spectroscopic observations of simulated {\sc COLIBRE} MQGs followed by analysis with modern SED-fitting pipelines under controlled assumptions, is left for future work (Nanayakkara et al., in prep.).




\subsubsection{Comparison with observations}
\label{ssec:res1-ndens-obs} 

\begin{figure*}
\centering
\includegraphics[width=\textwidth]{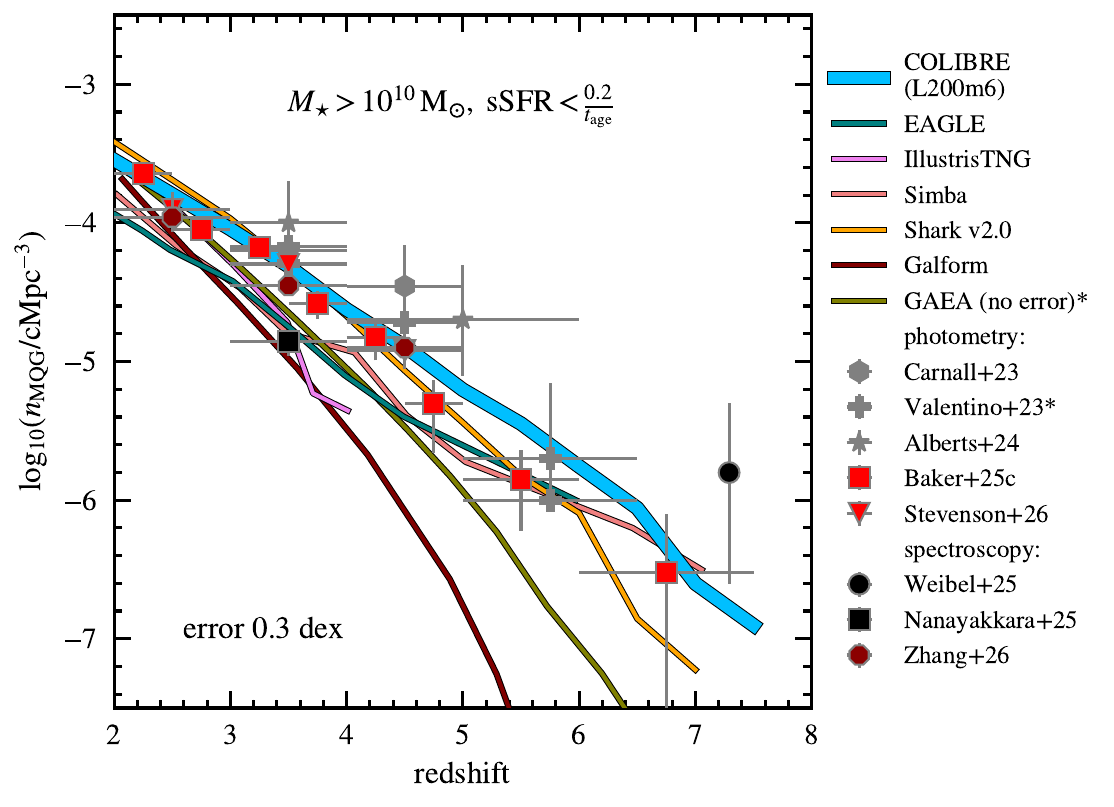}
\caption{Comoving number density of MQGs, defined by $M_{\star}>10^{10}\,\rm \textcolor{black}{M_\odot}$ and $\mathrm{sSFR}<0.2/t_{\rm age}$, as a function of redshift. Light-blue line shows predictions from our fiducial L200m6 {\sc COLIBRE} simulation, with the selection at $2 \le z \le 8$, compared to results from other galaxy formation and evolution models in the literature ({\sc GAEA} defines quenching using a fixed cut of $\rm sSFR = 10^{-10}yr^{-1}$). \textcolor{black}{Lines that terminate at $z<8$ indicate that no galaxies are identified as MQGs beyond those redshifts}. For all simulation results (except {\sc GAEA}), $M_{\star}$ and SFR values in the simulations are convolved independently with a Gaussian distribution (mean 0, standard deviation $0.3$~dex) representing a reasonable error budget for these quantities. These are compared with the latest \textit{JWST} observational estimates: black (smaller area) and dark red (larger area) points show spectroscopic measurements, while grey (smaller area) and red (larger area) points correspond to photometric measurements. Observational data are taken from \citet{carnall23, valentino23, alberts24, baker25, weibel25, nanayakkara25, stevenson25, zhang25}.} 
\label{fig:ndens-models} 
\end{figure*}

For the remainder of \S~\ref{sec:res1}, we primarily focus on the fiducial L200m6 simulation. This run provides a sufficiently large volume to contain dozens of MQGs while also resolving their internal structure more reliably (see \S~\ref{ssec:res1-size} and Appendix~\ref{appendix:l400m7-size-kin}). This run also performs better in reproducing the observed number density evolution after we consider errors. 
We additionally include results from the L400m7 simulation where relevant. We do not show results from the L200m7h hybrid run henceforth, but confirmed that the results look broadly similar, highlighting the places where this may not be the case.

In Fig.~\ref{fig:ndens-models}, we show the L200m6 predictions (including Gaussian-distributed uncertainties in the stellar masses and SFRs of width $0.3$~dex)
%
alongside the same observational data shown in Fig.~\ref{fig:ndens-boxes}. We have already discussed observational uncertainties arising from some of the choices behind SED modelling. However, there are additional, important caveats and biases that affect observationally derived number densities that are worth discussing in more detail. 

\begin{itemize}
\item \textit{Selection criteria:} The observations by \citet{carnall23}, \citet{baker25} and \citet{stevenson25} are most directly comparable to our data; they adopt the same criteria described in \S~\ref{ssec:prop-mq-def} ($M_{\star}>10^{10}\,\mathrm{M_{\odot}}$ and $\mathrm{sSFR}<0.2/t_{\mathrm{age}}$). \citet{baker25} also apply a colour-colour selection to clean their sample. However, other studies use different definitions. For example, \citet{alberts24} apply a lower stellar mass cut ($M_{\star}\ge10^{9.5}\mathrm{M_{\odot}}$), while \citet{valentino23}\footnotemark use a UVJ or NUVUVJ colour–colour selection in addition to $M_{\star}\ge10^{9.5}\mathrm{M_{\odot}}$. Similarly, \citet{nanayakkara25} select galaxies via UVJ colours but with $M_{\star}\gtrsim10^{10.5}\mathrm{M_{\odot}}$. Colour-colour selections have been reported to suffer from contamination of obscured star formation or strong emission lines \citep{schreiber18, zhang25} and can also miss significant numbers of quiescent galaxies that fall outside the traditional criteria \citep{merlin18}, motivating the development of alternative methods \citep[e.g.][]{long24}. Finally, \citet{weibel25} and \citet{zhang25} report galaxies with $\mathrm{sSFR}<10^{-10}\mathrm{yr^{-1}}$. The former work for a single $M_{\star}=10^{10.23}\mathrm{M_{\odot}}$ galaxy (hence the large error) and at higher redshift, where our threshold $\mathrm{sSFR}<0.2/t_{\mathrm{age}}\approx3\times10^{-10}$ diverges more, biasing simulation results higher); the latter for $M_{\star}>10^{10.3}\mathrm{M_{\odot}}$ ($\rm m_{F444W}<24$ in their table~1, corresponding to a complete sample above that stellar mass). 

A more detailed comparison, matching the exact observational selection criteria used to compute number densities, is left for future work. For reference, \citet{chaikin25b} present {\sc COLIBRE} results with different $M_{\star}$ cuts ($>10^{10}\rm\,M_{\odot}$, $>10^{10.5}\rm\,M_{\odot}$ and $>10^{11}\rm\,M_{\odot}$), although they adopt an $\mathrm{sSFR}<10^{-10}\,\mathrm{yr^{-1}}$ criterion for quiescence.

\footnotetext{We homogenise the \citet{valentino23} results by combining $10^{9.5} \le M_{\star}/\mathrm{M_{\odot}} < 10^{10.6}$ and $M_{\star} > 10^{10.6}\,\mathrm{M_{\odot}}$.}


\item \textit{Photometric versus spectroscopic measurements:} Only two studies provide a fully robust spectroscopic sample \citep{nanayakkara25, zhang25}, plus the \citet{weibel25} galaxy, which is spectroscopically confirmed as well. Most other works rely on photometric redshifts, which can misidentify MQGs due to dust obscuration, old stellar ages, AGN activity, or reduced star formation \citep{forrest24, zhang25}. Some studies account for errors in the photometric \citep{valentino23, alberts24} or spectroscopic redshifts \citep{nanayakkara25, zhang25}, or apply contamination corrections \citep{stevenson25}, whereas others consider only Poisson errors.

\item \textit{Area coverage:} Sample size and coverage statistics also play a role; some datasets cover relatively small areas of the sky, potentially targeting over- or under-dense regions. This is particularly true for the photometric samples from \citet{carnall23,valentino23,alberts24} (all in overdense regions, with at most dozens of objects), as well as the spectroscopic samples from \citet{weibel25} (a single galaxy in a pronounced overdensity) and \citet{nanayakkara25} (covering a smaller region). Some works account for cosmic variance in their error bars \citep{baker25, zhang25}, which explains the larger uncertainties.

\item \textit{Choice of SFR timescale:} The SFR timescale deduced from modelling of the SED can depend on the photometric bands or spectral features used to infer the galaxy’s SFR \citep{kennicutt12}. Some photometric studies adopt a $\rm 100\,Myr$ timescale \citep[e.g.][]{alberts24,baker25}, while \textcolor{black}{spectroscopic analyses} use shorter timescales of around $\rm 10\,Myr$ \citep[e.g.][]{zhang25}. We tested the impact of varying these timescales and found that the results \textcolor{black}{are significantly affected at $z>3$ only when using the $\rm 100\,Myr$ averaged SFRs (see Appendix~\ref{appendix:mqg-selection}). This indicates that MQG SFHs in {\sc COLIBRE} can remain bursty over timescales of $100\,\mathrm{Myr}$. The dominant MQG population exhibiting multiple starburst episodes (\S~\ref{ssec:prop-sfh}) may therefore experience a final burst shortly before quenching, strong enough to affect the time-averaged SFR. Observationally, MQGs at $3<z<5$ may be dominated by recently quenched systems within the last $\lesssim150\,\mathrm{Myr}$ \citep{merlin25}.}

\item \textit{Wavelength coverage:} The wavelength range used in SED fitting is critical for mitigating dust contamination. E.g. UV–\textcolor{black}{near-infrared(NIR)} photometry alone is insufficient to constrain AGN contributions \citep{chang25}. The incorporation of the mid-\textcolor{black}{infrared} (MIR) data, such as \textit{JWST}/MIRI observations \citep[e.g.][]{wangt25}, has proven to provide stronger constraints. \citet{valentino23, weibel25} and \citet{baker25} include them when available.

\end{itemize}

Taking these factors into account, larger, more robust samples provide more consistent and reliable results. In Fig.~\ref{fig:ndens-models}, these are some photometric data \citep[e.g.][]{baker25, stevenson25} (light red) and spectroscopic data with incompleteness corrections \citep[e.g.][]{zhang25} (dark red). These more robust datasets show better agreement with each other, and the simulations are no longer as far from the observed values, reducing the tension highlighted in previous works \citep[e.g.][]{carnall23, valentino23}. The observational landscape is evolving rapidly, especially with increasing sample sizes and a better understanding of the systematic effects associated with different selection criteria, as demonstrated by the recent works of \citet{yang25} and \citet{shuntov25}.

We conclude that {\sc COLIBRE} predictions are broadly consistent with the most recent and robust \textit{JWST} measurements (red and dark red points), particularly after invoking an error budget of $0.3$~dex in stellar masses and SFRs.


\subsubsection{Comparison with other galaxy formation models}
\label{ssec:res1-ndens-models} 

We also place our {\sc COLIBRE} results in the context of other recent simulation predictions in Fig.~\ref{fig:ndens-models}. Here we compare our fiducial run with hydrodynamical galaxy formation simulations: {\sc EAGLE} \citep{schaye15}, {\sc IllustrisTNG} \citep{pilepich18} and {\sc Simba} \citep{dave19}; and semi-analytic models: {\sc Shark v2.0} \citep{lagos24}, {\sc Galform} \citep{lacey16}, and {\sc GAEA} \citep{de-lucia24}. Except for {\sc GAEA}, all models adopt the same quenching definition (\textcolor{black}{with $t_{\rm age}$ in \S~\ref{ssec:prop-mq-def} computed using the corresponding cosmology of each simulation}) and include Gaussian-distributed errors in their measurements, although the convolution is not repeated 100 times as for the {\sc COLIBRE} results. \textcolor{black}{As a result, the reported number densities may be affected by the specific random realisation of the Gaussian errors, leading to values that are either higher or lower than the ensemble median}. {\sc GAEA} defines quenching using a fixed threshold of $\mathrm{sSFR}=10^{-10}\,\mathrm{yr}^{-1}$.

For {\sc EAGLE}, we use the ``{\sc L100N1504}'' simulation with $L=100\rm\,cMpc$, $m_{\rm g}\sim10^{6}\rm\, M_{\odot}$ and $m_{\rm DM}\sim10^{7}\rm\, M_{\odot}$; for {\sc IllustrisTNG}, the ``{\sc TNG100}'' run with $L\approx111\,\mathrm{cMpc}$, $m_{\rm g}\sim10^{6}\rm\, M_{\odot}$ for the initial gas cell mass and $m_{\rm DM}\sim10^{7}\rm\, M_{\odot}$; while for {\sc Simba}, the ``{m1000n1024}'' run with $L\approx147\,\mathrm{cMpc}$, $m_{g}\sim10^{7}\rm\, M_{\odot}$ per gas element mass and $m_{\rm DM}\sim10^{8}\rm\, M_{\odot}$;. In the case of the SAMs, {\sc Shark v2.0} has been run on a $L=800\rm\,cMpc$ box with $m_{\rm DM}\sim10^{8}\rm\, M_{\odot}$; {\sc Galform} on a $L\approx710\rm\,cMpc$ box with $m_{\rm DM}\sim10^{9}\rm\, M_{\odot}$; while {\sc Galform} on a $L\approx685\rm\,cMpc$ per side with $m_{\rm DM}\sim10^{9}\rm\, M_{\odot}$. The larger volumes of the SAMs provide better number statistics, reducing cosmic variance, probing lower number densities (as seen in Fig.~\ref{fig:ndens-models}) and capturing more extreme overdensities.

{\sc COLIBRE} and {\sc Shark v2.0} tend to predict higher number densities compared to the other models over the redshift range, with  {\sc COLIBRE} about $0.5$~dex higher for $z>5$. In {\sc COLIBRE}, \textcolor{black}{\citet{chaikin26} show clearly that this arises from the allowance for BH accretion above the Eddington limit. The explicit modelling of the cold gas phase enables the resolution of higher gas densities, which in turn lead to higher BH accretion rates, as the accretion rate scales directly with the local gas density (see equation~2 of \citetalias{chandro-gomez_inprep}). This allows BHs to enter the super-Eddington regime, common at these high redshifts \citep{husko25b}, thereby accelerating BH growth and triggering stronger AGN feedback and earlier quenching.}

The other models underpredict the observations: {\sc EAGLE} and {\sc Simba} at $z\lesssim5$, while {\sc Galform} and {\sc GAEA} at $z\gtrsim3$. \textcolor{black}{Although the {\sc GAEA} number densities are not convolved with observational errors, we verified that its raw simulated number densities are still lower than those of {\sc COLIBRE} in the $z\gtrsim4$ regime.} These discrepancies reflect differences in sub-grid physics, especially AGN feedback, which plays a central role in regulating the quenching of massive galaxies \citep{lagos25}. Other simulations, not included in Fig.~\ref{fig:ndens-models}, predict higher number densities without accounting for observational uncertainties: for example, {\sc Magneticum} \citep{kimmig25} and {\sc Thesan} \citep{chittenden25}. However, we note that {\sc Magneticum} has only been run down to $z\approx2$, where it overpredicts the number densities \citep{lagos25}; while {\sc Thesan}, \textcolor{black}{which explores a regime of suppressed AGN feedback efficiency,} focuses on $z\geq5.5$ and has not been tested in the local Universe. \textcolor{black}{{\sc Astrid}, also not included in the comparison, predicts significantly lower number densities than {\sc IllustrisTNG} \citep{weller25}.}

\subsection{Stellar mass functions of MQGs}
\label{ssec:res1-smf} 

\begin{figure*}
\centering
\includegraphics[width=\textwidth]{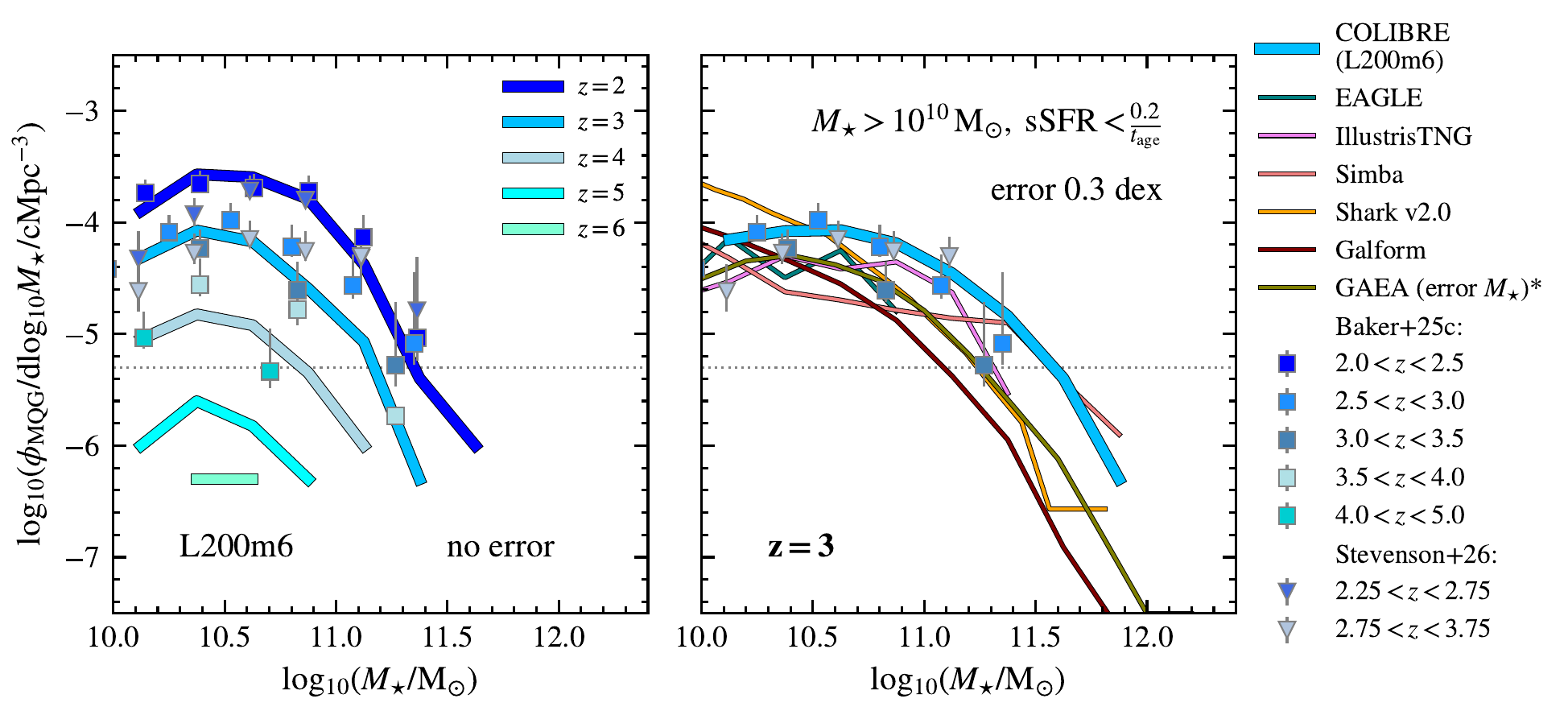}
\caption{SMF of MQGs, defined by $M_{\star}>10^{10}\,\rm \textcolor{black}{M_\odot}$ and $\mathrm{sSFR}<0.2/t_{\rm age}$, from the fiducial L200m6 simulation. \textit{Left panel}: predicted simulation values at different redshifts $2 \le z \le 6$ (colour-coded as labelled in the left panel), compared with observational estimates from \citet{baker25} and \citet{stevenson25} within the same redshift range, as labelled.  \textit{Right panel}: the SMF at $z=3$ from {\sc COLIBRE} in light blue (fiducial L200m6) compared with results from other galaxy formation and evolution models, where $M_{\star}$ and SFR values in the simulations are convolved independently with a Gaussian distribution (mean 0, standard deviation $0.3$~dex) representing a reasonable error budget for these quantities ({\sc GAEA} defines quenching using a fixed cut of $\rm sSFR = 10^{-10}\,yr^{-1}$ and its results are convolved \textcolor{black}{only in $M_{\star}$} with potential observational errors). The horizontal dotted line denotes the threshold of 10 galaxies \textcolor{black}{for {\sc COLIBRE}}, below which the statistics become unreliable.} 
\label{fig:smf} 
\end{figure*}

The works of \citet{lagos25} and \citet{baker25} show that the SMF is an important diagnostic of predictions of MQGs. While the number densities can compare reasonably well with observations, the SMF can reveal deficiencies and, in some cases, sharp tensions with the observations. 

\subsubsection{Results} 

We analyse the SMF of MQGs in Fig.~\ref{fig:smf}. The left panel presents the SMF predicted by {\sc COLIBRE} at different redshifts $2 \le z \le 7$ for the fiducial L200m6 simulation. The predictions show a clear increase in the number of MQGs with decreasing redshift, as expected: galaxies need time to assemble enough stellar mass to be classified as massive, and additional time for quenching mechanisms to suppress their star formation. Consequently, the population shifts towards higher stellar masses at lower redshifts. Interestingly, these galaxies exhibit a peak around $M_{\star} \approx 10^{10.5}\,\mathrm{M_{\odot}}$, where quenching is most efficient: galaxies are massive enough for the quenching mechanism to act strongly, yet extremely massive systems remain rare. Once quenched, they undergo little further stellar mass growth.


For comparison, in Fig.~\ref{fig:smf} we also show observational results from \citet{baker25} and \citet{stevenson25}. The {\sc COLIBRE} results are broadly consistent with current observational constraints, which also show a similar peak \citep[at around $M_{\star}\approx10^{10.6}\,\mathrm{M_{\odot}}$ in][]{baker25}. A comparable peak is also reported in \citet{shuntov25} using the larger COSMOS-Web survey, although that study has reduced filter coverage and depth.
We further test the impact of excluding satellite galaxies from the MQG selection and find that the SMF still peaks at a similar stellar mass to the case where both centrals and satellites are included. Differences between the two selections emerge only at the low-mass end across $2 \le z \le 6$, with a small fraction of quenched satellites already present by $z=5$, potentially consistent with recent observational discoveries of such systems in overdense regions \citep{baker25c}.

While the observational data used here are robust in terms of large sample sizes and reliable selection of massive quenched systems, they are photometrically derived and thus may suffer from systematic uncertainties \citep{forrest24}. The lack of similarly large spectroscopic samples to measure SMFs highlights the need for future follow-up campaigns. Upcoming wide-area photometric surveys such as MINERVA \citep{muzzin25}, with extensive NIR and MIR coverage, will help reduce systematics and provide valuable targets for spectroscopic confirmation.

For completeness, we note that the SMF for the L400m7 simulation shows an excess of galaxies around $M_{\star} \approx 10^{10.6}\,\mathrm{M_{\odot}}$ at all redshifts, consistent with the overestimated number densities discussed in \S~\ref{sssec:res1-ndens-results}. In contrast, the L200m7h SMF at $z=2$ predicts fewer galaxies at $M_{\star} \lesssim 10^{10.4}\,\mathrm{M_{\odot}}$ and more at the high–mass end ($M_{\star} \gtrsim 10^{11.2}\,\mathrm{M_{\odot}}$), while at $z>2$ it yields lower SMF values across the full mass range.

\subsubsection{Comparison with galaxy formation models}
\label{ssec:res1-smf-models} 

The right panel of Fig.~\ref{fig:smf} presents the SMF after convolving stellar masses and SFRs with a Gaussian-distributed error of width $0.3$~dex (as was computed for the number densities) at $z=3$ in {\sc COLIBRE} (L200m6). These results are compared with similarly error-convolved predictions from other galaxy formation and evolution models for the $z=3$ SMF with the same definition of ``massive quenched'' (except for {\sc GAEA}, \textcolor{black}{for which the quenching definition at this redshift is effectively similar, since $0.2/t_{\rm age} \approx 10^{-10}\,\rm yr^{-1}$, and where Gaussian-distributed errors are applied only to $M_{\star}$}). A general trend emerges: models either accurately reproduce the low-mass end of the SMF ({\sc EAGLE}, {\sc Galform}), the high-mass end ({\sc Shark v2.0}) or both ends but not the intermediate mass regime ({\sc Simba}), but none match the full mass range simultaneously. This has been noted in \citet{lagos25} and \citet{baker25} and linked to the different AGN feedback prescriptions. In contrast, {\sc COLIBRE} performs well across the full stellar mass range. While {\sc GAEA} and {\sc IllustrisTNG} may appear to match the full mass range, \textcolor{black}{{\sc GAEA} produces fewer galaxies above $M_{\star}\sim10^{10.5}\,\mathrm{M_{\odot}}$. We verified that the different error convolution does not primarily drive this, since the raw SMF also remains below that of {\sc COLIBRE} at the high-mass end.} {\sc IllustrisTNG} suffers from small-number statistics at the high-mass end, with the number density of MQGs that declines rapidly beyond $z>3$ (Fig.~\ref{fig:ndens-models}). Overall, the consistency of both SMFs and number densities in {\sc COLIBRE}, once reasonable uncertainties in stellar masses and SFRs are considered, highlights its robust performance in reproducing MQG properties compared to other models. The SMF results derived with the error convolution across the full redshift range are shown in Appendix~\ref{appendix:error-smf}.




\subsection{Star formation histories of MQGs}
\label{ssec:res1-ages} 

SFHs provide key insights into how galaxies assemble their stellar mass (through a single intense starburst, multiple bursts, or an extended period of star formation) and how quenching occurs, either abruptly or gradually. Here, we reconstruct SFHs (\S~\ref{ssec:prop-sfh}) for the selected MQGs in {\sc COLIBRE} and compare with observational data. Before we present our main results, it is important to note some caveats for the following subsections in \S~\ref{sec:res1} compared to observations:
\begin{itemize}
\item To select massive quenched systems, we use the predicted $M_{\star}$ and SFR of galaxies in COLIBRE, ignoring any potential uncertainties in these quantities.
\item The observational samples considered here consist of spectroscopically confirmed galaxies. This is important, as photometrically selected galaxies are less robust as explained earlier, particularly at high redshift \citep{forrest24}.
\item The definition of ``massive quenched'' varies across observational studies; this should be kept in mind when interpreting results.
\end{itemize}

\begin{figure}
\centering
\includegraphics[trim={0 0 0 0},clip, width=0.49\textwidth]{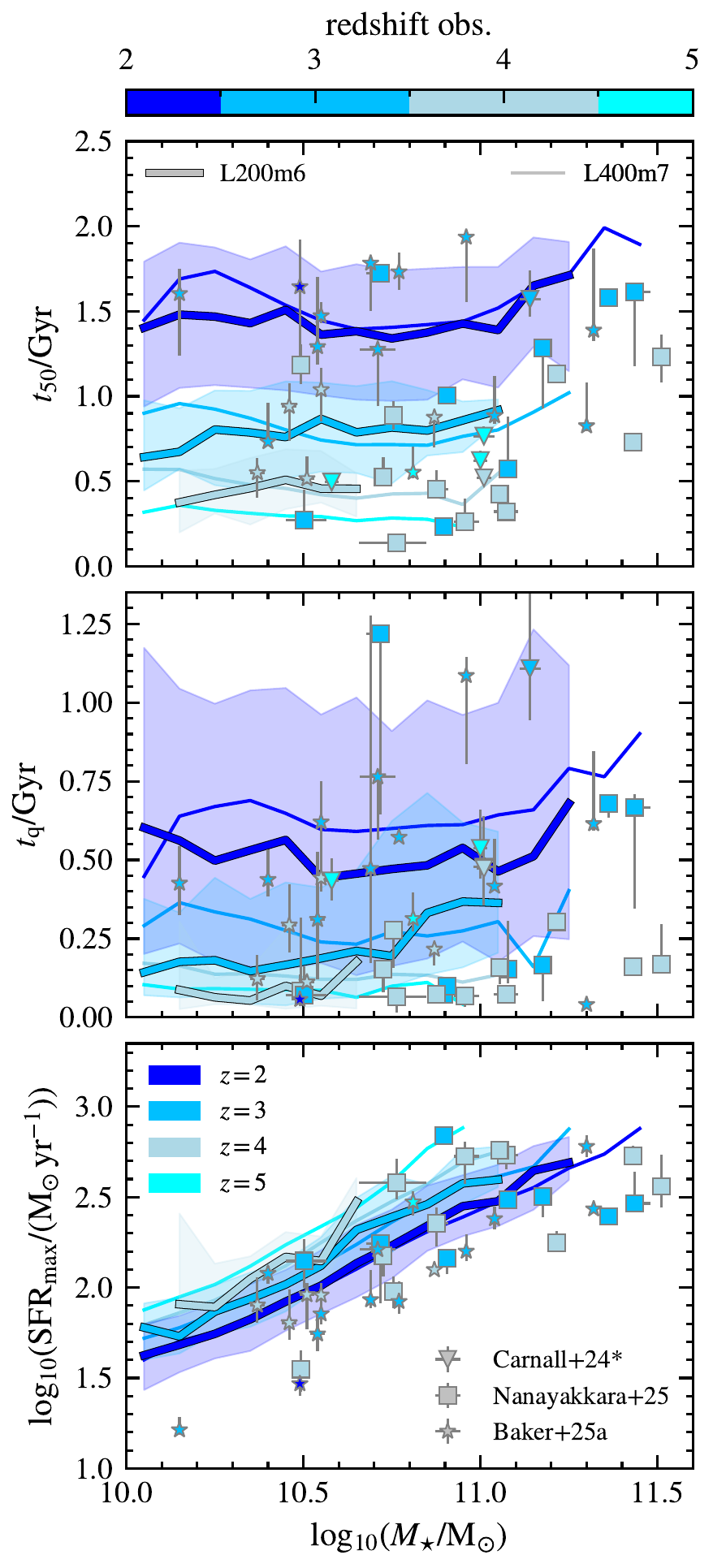}
\caption{SFH properties of MQGs as a function of $M_{\star}$. Thick solid lines show the median predictions from L200m6 at different redshifts $2 \le z \le 5$ (as labelled in the bottom panel), with shaded regions indicating the 16th and 84th percentile range. At $z=5$, MQGs are too few to split in stellar mass bins (9 galaxies). Thin solid lines show the corresponding medians from L400m7. These are compared to recent \textit{JWST} spectroscopic measurements \citep{carnall24, nanayakkara25, baker25b}, shown as symbols colour-coded by observed galaxy redshift, with the corresponding colorbar at the top. \textit{Top panel}: formation times ($t_{50}$), defined in \S~\ref{ssec:prop-sfh}. \textit{Middle panel}: quenching times ($t_{\rm q}$), defined in \S~\ref{ssec:prop-sfh} \textcolor{black}{\citep[for][defined as the earliest lookback time at which the galaxy first crosses $\mathrm{sSFR}<0.2/t_{\rm age}$]{carnall24}}. \textit{Bottom panel}: SFR peak ($\mathrm{SFR}_{\rm max}$) \citep[no data from][]{carnall24}.} 
\label{fig:ages} 
\end{figure}

In Fig.~\ref{fig:ages}, we show the fiducial {COLIBRE} simulation predictions for SFH properties as a function of $M_{\star}$, compared to recent \textit{JWST} spectroscopic observations, displayed as individual data points colour-coded by spectroscopic redshift. Observationally, SFHs are derived from SED modelling.
For the data from \citet{nanayakkara25, baker25b}, we use the full SFH Markov chain Monte Carlo (MCMC) output provided by the SED-fitting code {\sc Prospector} \citep{johnson21}, computing each galaxy’s properties as the median and the 16th and 84th percentile range of the MCMC. For \citet{nanayakkara25}, we adopt an exponentially increasing continuity prior between SFH bins rather than the manuscript’s fiducial choice, which is broadly consistent with expectations from $\Lambda$CDM cosmology \citep{dekel13}\textcolor{black}{, as it favours a gradual stellar mass build-up at high redshift and reduces biases toward older timescales \citep{turner25}}. For \citet{carnall24}, we use the values reported directly in the paper.

\subsubsection{Formation timescale}
\label{ssec:res1-ages-results-tf} 

In the top panel of Fig.~\ref{fig:ages}, we show the formation timescale for MQGs (the lookback time from the time of selection at which 50\% of the $z=0$ stellar mass without accounting for mass losses was formed; see \S\ref{ssec:prop-sfh}) as a function of $M_{\star}$. Each colour represents the median relation at a different selection redshift, with shaded areas indicating the corresponding 16th and 84th percentile range. Here, we focus on MQGs selected at $2 \le z \le 5$, although there are too few galaxies at $z>4$ for L200m6 to compute reliable medians per stellar mass bin. In L200m6, the simulated galaxies undergo a rapid mass assembly, forming the bulk of their mass early in cosmic time, with no significant dependence on stellar mass. \textcolor{black}{This differs from a recent observational result \citep{leung26}, which find a steeper relation indicative of downsizing at these redshifts, with more massive MQGs forming earlier than their lower-mass counterparts.}

Selections at higher redshift correspond to later $t_{50}$ due to the limited cosmic time available, but also to earlier median $z_{50}$ values: $z_{50}=3.40$ at $z=2$, $z_{50}=4.47$ at $z=3$, $z_{50}=5.25$ at $z=4$, $z_{50}=6.08$ at $z=5$, and $z_{50}=7.28$ at $z=6$ (for 2 galaxies), consistent with expectations. The scatter decreases with redshift, which may reflect smaller sample sizes at higher redshifts. In L400m7 (thin lines), $t_{50}$ interestingly peaks at $M_{\star}\approx10^{10.2}\,\rm M_{\odot}$ and increases again at higher masses.

Qualitatively, the simulation values in the top panel are consistent with the observational sample. The \citet{nanayakkara25} data show shorter formation timescales around $M_{\star}\approx10^{11}\rm\,M_{\odot}$, indicating later mass assembly and burstier SFHs than in {\sc COLIBRE}, a trend also seen across the full mass range in \citet{baker25b}. However, as discussed in Section \ref{ssec:res1-ages-discussion}, early-time SFHs are observationally uncertain because old stellar populations provide limited constraints. The same applies to the few longer timescales inferred from $z\approx3$ observations. Observed galaxies at the high-mass end ($M_{\star}\gtrsim10^{11.2}\,\rm M_{\odot}$) are not well matched by the simulation medians either\textcolor{black}{, although we require at least 5 galaxies per stellar mass bin to compute the medians, and thus galaxies of these masses may exist but with insufficient statistics}. The \citet{carnall24} observations display a different behaviour compared to the other two studies. This is likely due to their distinct SFH parametrisation using a double power-law in {\sc BAGPIPES} \citep{carnall18} (parametric form as opposed to the non-parametric of the other two works, discussed in \S~\ref{ssec:res1-ages-discussion}), which tends to produce SFHs resembling a single burst with minimal star formation outside that period.

\subsubsection{Quenching timescale}
\label{ssec:res1-ages-results-tq} 

In the middle panel of Fig.~\ref{fig:ages}, we examine the quenching timescale (as defined in lookback time from the time of selection in \S~\ref{ssec:prop-sfh}). We find no strong trend with $M_{\star}$ for L200m6. Higher-redshift systems exhibit lower $t_{\rm q}$ values, although this trend is less pronounced than for the formation times. Galaxies at earlier times quenched more recently relative to their selection redshift, as expected, since the available cosmic time is shorter. The median values are $z_{\rm q}=2.36$ for selection at $z=2$; $z_{\rm q}=3.26$ at $z=3$; $z_{\rm q}=4.20$ at $z=4$; $z_{\rm q}=5.27$ at $z=5$; and $z_{\rm q}=6.14$ at $z=6$ (for 2 galaxies).

The L400m7 simulation (thin lines) exhibits the same peak in $t_{\rm q}$ as in $t_{\rm 50}$, around $M_{\star} \approx 10^{10.2}\,\rm M_{\odot}$. Although not shown, the scatter at $z=2$ is comparable to that in L200m6, despite the larger number of galaxies. L200m7h shows significantly shorter quenching timescales ($t_{\rm q} \lesssim 0.4\rm\,Gyr$) at $z=2$ for low-mass systems ($M_{\star} < 10^{10.5},\rm M_{\odot}$), producing a non-flat positive correlation between quenching time and stellar mass, where more massive galaxies quench earlier. \textcolor{black}{This may hint at the observational trend that more massive systems at $z\gtrsim3$ tend to have longer quenching timescales \citep{merlin25}, a trend that is also weakly present in L200m6.}

{\sc COLIBRE} predictions broadly agree with observations, indicative of recent quenching $t_{\rm q}\lesssim0.6\,\mathrm{Gyr}$. The simulated data recover the \citet{baker25b} data well. The \citet{nanayakkara25} data show shorter quenching timescales around $M_{\star}\approx10^{11}\rm\,M_{\odot}$, consistent with the later formation timescales found previously due to burstier SFHs\textcolor{black}{. The high-mass end is not well reproduced either, as the simulations lack a statistically significant population of systems in that regime}.
\citet{carnall24} uses a different definition for $t_{\rm q}$ (the earliest lookback time at which the galaxy first crosses $\mathrm{sSFR}<0.2/t_{\rm age}$) yielding longer values, although they follow the same overall trend as $t_{50}$ because of the double power-law SFH parametrisation (parametric form). 

\subsubsection{Peak of SFR}
\label{ssec:res1-ages-results-sfrpeak} 

Finally, we show the SFR peak, $\rm SFR_{max}$, in the bottom panel of Fig.~\ref{fig:ages}. The lookback time of the SFR peak, $t_{\rm SFR_{max}}$, follows a similar trend to that of $t_{50}$ (as shown in Fig.~\ref{fig:sfh-example}). $\rm SFR_{max}$ increases with galaxy mass, reflecting the stronger SF episodes required to assemble more stellar mass. The peak also decreases toward lower redshift, indicating that stellar mass can be built up more gradually over time, rather than through a single intense starburst as seen for MQGs selected at $z=4$. Observational results exhibit considerable scatter but generally follow a similar trend with selection redshift. Values in \citet{nanayakkara25} and \citet{baker25b} are quantitatively reproduced\textcolor{black}{, although the simulated values are systematically higher than the observational estimates. This difference is driven by the broader lookback-time bins used in observational reconstructions based on non-parametric SED fitting (see Fig.~\ref{fig:sfh}). We verify that adopting coarser time bins of $100\,\mathrm{Myr}$ (comparable to observational constraints) reduces the inferred $\rm SFR_{max}$ values.}
In addition, a fraction of galaxies experience very strong starbursts in L200m6 ($\rm SFR_{max}>300\, M_{\odot}\,\mathrm{yr^{-1}}$): 10\% at $z=2$, 14\% at $z=3$, 23\% at $z=4$, 27\% at $z=5$ (3 galaxies), and 50\% at $z=6$ (1 galaxy). This may point to an evolutionary link with highly star-forming galaxies.

\subsubsection{Discussion}
\label{ssec:res1-ages-discussion} 

The level of agreement between simulations and observations is highly sensitive to sample selection and the assumptions made during SED fitting, even when spectroscopic data are available. Choices made during SED fitting can introduce systematic biases that affect the inferred galaxy properties. In particular, the complex SFHs involving multiple bursts, described in \S~\ref{ssec:prop-sfh}, are challenging to recover via SED fitting \citep{suess22, wangb25}.


For example, \citet{nanayakkara25} compares two different codes: {\sc Prospector} and {\sc FAST++} \citep{kriek09}. While both produce qualitatively consistent average SFHs for large samples, individual SFHs differ due to distinct assumptions regarding metallicity, SFHs, SSPs, and other parameters. In particular, the parametric nature of {\sc FAST++} \citep[the {\sc BAGPIPES} results from][]{carnall24} favours less gradual mass build-up and narrower bursts than the more flexible SFH parametrisation of {\sc Prospector}; this results in stronger early-time enhancement \citep[fig.~11 in][]{nanayakkara25}. Some scatter in derived properties such as stellar masses is therefore expected \citep{leja19}, and according to \citet{nanayakkara25} could be up to $\approx 0.4$~dex.

Even within {\sc Prospector}, choices such as the selection of SSP models and the priors used in non-parametric SFH reconstructions can influence the inferred properties of individual galaxies. These effects are typically minor, but they become more pronounced during the first $\approx$250 Myr of cosmic history, when old stellar populations dominate the observed spectrum and time resolution is inherently limited, likely contributing to the shorter observational timescales. Incorporating rest-frame NIR data helps mitigate these uncertainties. In this work, we compare to the results of \citet{nanayakkara25}, who use an exponentially increasing continuity prior between SFH bins, motivated by the average halo mass accretion rate over cosmic time \citep{dekel13}, which is consistent with expectations from a $\Lambda$CDM framework. By contrast, \citet{baker25b} also employ {\sc Prospector} with a non-parametric SFH, but adopt a flat continuity prior \citep{leja19b}, use different lookback time binning, and rely on distinct SSP models: C3K \citep{conroy09} in \citet{nanayakkara25} versus MILES \citep{falcon-barroso11} in \citet{baker25b}; among other methodological differences. Appendix~\ref{appendix:sfh} illustrates how these choices affect the recovered median SFH shapes of observed galaxies.

We emphasise that, although observational estimates of these SFH properties carry significant uncertainties, and we have not applied the $0.3$~dex convolution error for $M_{\star}$ and SFR when selecting our sample for this analysis, there is still overall broad agreement between our {\sc COLIBRE} predictions and the observational estimates across all measured properties. Performing SED-fitting of {\sc COLIBRE} MQGs is left for future work and will help clarify these comparisons.

\subsection{Dust and molecular hydrogen fractions of MQGs}
\label{ssec:res1-dust-gas} 

Dust and molecular gas are fundamental components in quenched galaxies, closely tied to their SFH. Molecular gas, the primary fuel for star formation, is expected to be depleted in MQGs due to quenching mechanisms such as gas ejection or thermal heating, resulting in low gas reservoirs. Dust exhibits a similar decline: it originates from star formation processes, grows within the ISM, and can be destroyed or expelled by feedback. Importantly, dust grains facilitate the formation of molecular hydrogen and their emission is commonly used as a proxy for molecular gas content \citep{tacconi20}. As such, molecular gas, dust, and star formation are intimately connected, and their co-evolution provides key insights into the quenching process.

\begin{figure}
\centering
\includegraphics[width=0.48\textwidth]{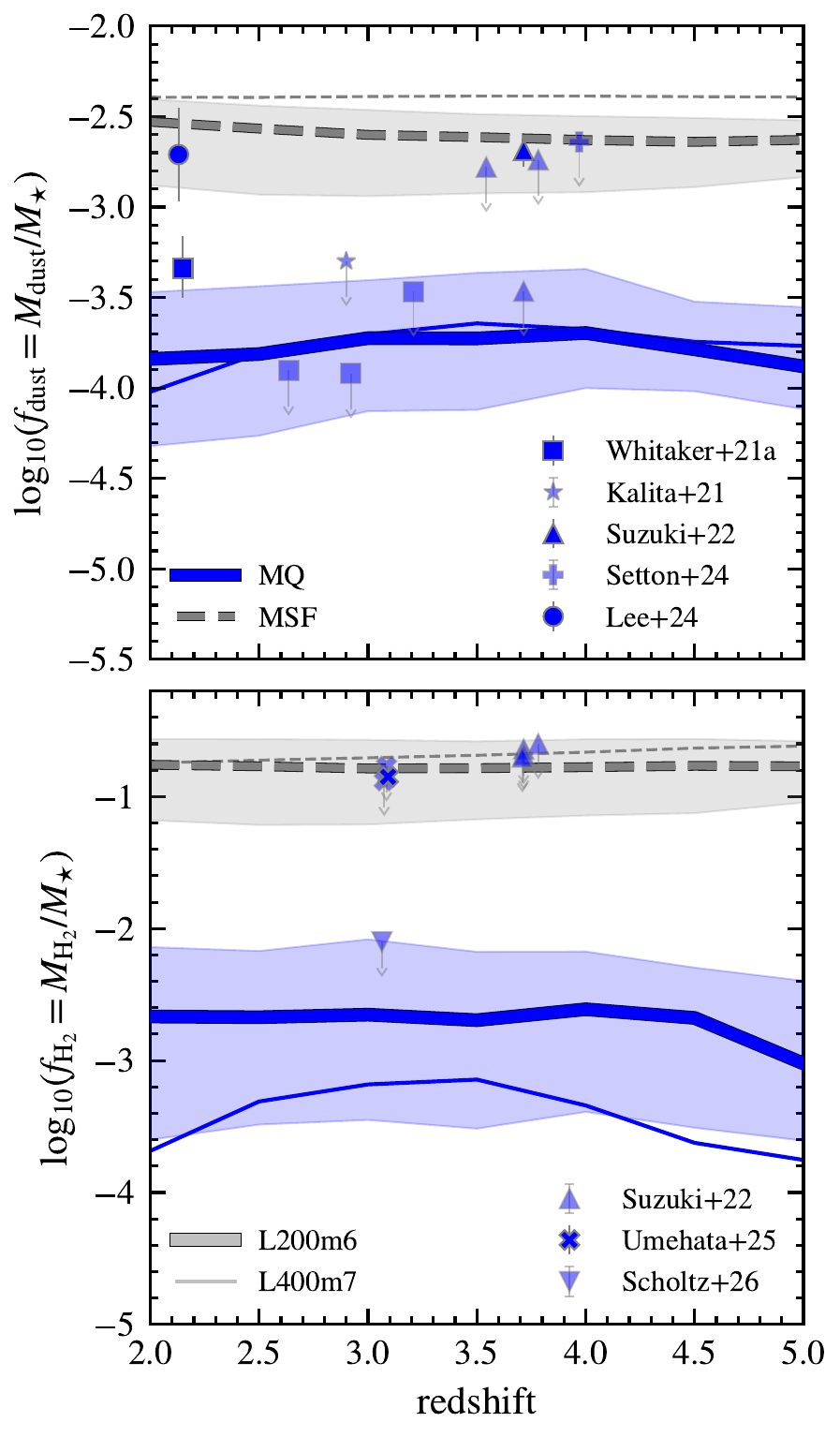}
\caption{Dust fraction, $f_{\rm dust}=M_{\rm dust}/M_{\star}$, (\textit{top panel}) and molecular hydrogen fraction, $f_{\rm H_{2}}=M_{\rm H_{2}}/M_{\star}$, (\textit{bottom panel}) as a function of redshift at $2\le z\le 5$ for L200m6. Median values for the MQGs (thick solid blue) and MSFGs (thick dashed grey) with the corresponding 16th and 84th percentile range at each selection redshift. At $z=6$, MQGs are too few to define a median (2 galaxies), since we consider here all galaxies together rather than splitting by stellar mass bins. Thin lines show the corresponding median from L400m7. Observational data of MQGs are taken from the literature: for dust fractions from \citet{whitaker21, kalita21, suzuki22, setton24, lee24}, and for molecular hydrogen fractions from \citet{suzuki22, umehata25, scholtz25}. Arrows indicate upper limits from non-detections (also shown with a more faded colour).}
\label{fig:dust-z} 
\end{figure}

Observations at lower redshift indicate that the dust \citep{gobat18, magdis21} and molecular gas \citep{spilker18} content of MQGs can evolve significantly. However, constraints for high-redshift quiescent galaxies remain limited: their emission is intrinsically faint, and until recently, telescopes lacked the sensitivity to detect them. Consequently, most available measurements are either upper limits from non-detections or are biased toward quiescent systems that are unusually gas- and dust-rich. Upcoming observations with instruments such as \textit{JWST} MIRI and NIRSpec, deep ALMA surveys, and the future \textit{PRIMA} space-based observatory \citep[][covering the wavelength gap between the previous two]{faisst25} are expected to improve these measurements substantially, making theoretical predictions from simulations particularly valuable. {\sc COLIBRE} provides an ideal testbed for such predictions, as it models dust self-consistently, and captures a cold gas phase, with temperatures as low as $\sim\! 10\,\text{K}$ (\S~\ref{sec:colibre}). This enables a relatively realistic treatment of both components and a direct investigation of their content.

\subsubsection{Dust fraction evolution}
\label{sssec:res1-dust-gas-dustev} 

Fig.~\ref{fig:dust-z} shows the evolution of the dust fraction, $f_{\rm dust}=M_{\rm dust}/M_{\star}$, for MQGs as a function of selection redshift. The blue line indicates the median value, with shaded regions representing the 16th and 84th percentile range. We find that the dust fraction remains roughly constant at $f_{\rm dust} \sim 10^{-4}$ across $2 \le z \le 6$ for both L200m6 (thick) and L400m7 (thin lines) (beyond $z>5$ there are only 2 MQGs for L200m6). For comparison, the dust fractions of massive star-forming galaxies (MSFGs; selected using the same stellar mass cut and sSFR threshold as in \S~\ref{sec:prop}) are shown (dashed grey line). MSFGs exhibit dust fractions at least two orders of magnitude higher than MQGs, consistent with the expectation that ISM material in passive systems is removed by the feedback processes responsible for quenching, even at high redshift and over short timescales \citep{lesniewska25}.

\begin{figure*}
\centering
\includegraphics[width=\textwidth]{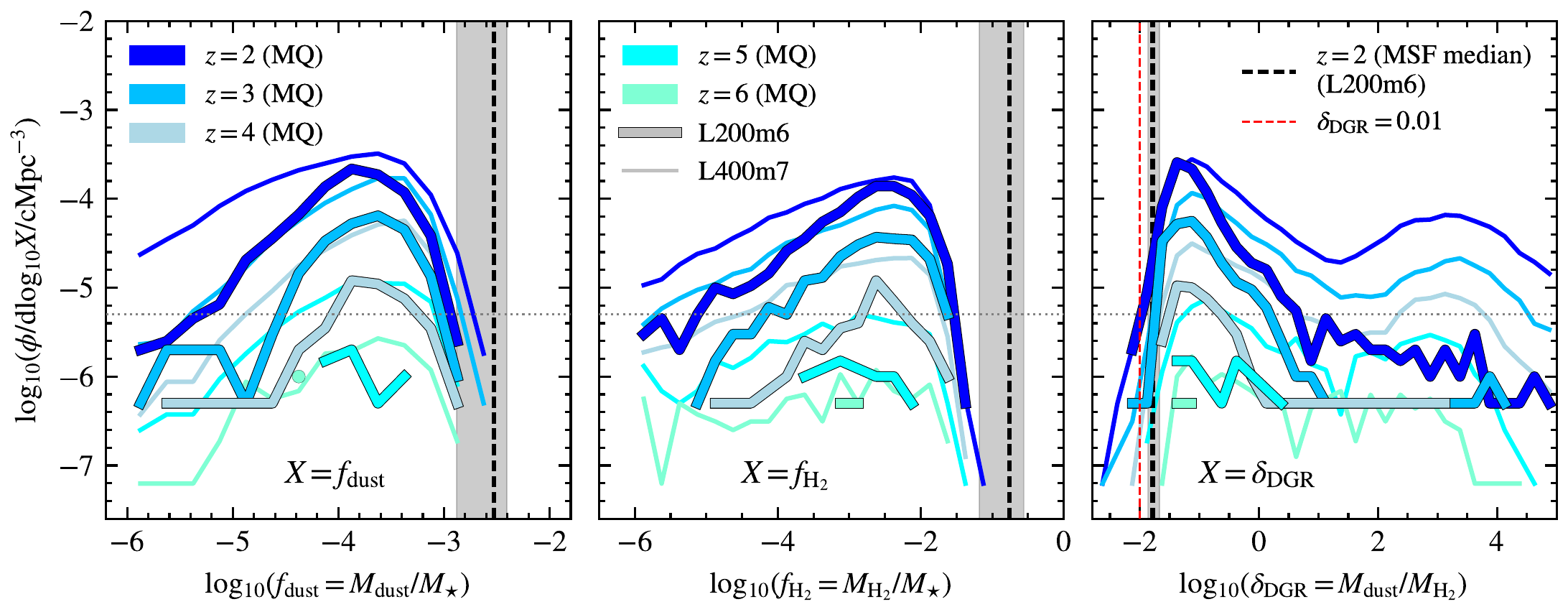}
\caption{Distribution functions of dust fraction, $f_{\rm dust}=M_{\rm dust}/M_{\star}$ (\textit{left panel}), molecular gas fraction, $f_{\rm H_{2}}=M_{\rm H_{2}}/M_{\star}$ (\textit{middle panel}), and dust-to-gas ratio, $\delta_{\rm DGR}=M_{\rm dust}/M_{\rm H_{2}}$ (\textit{right panel}) for L200m6. Solid lines show the distributions for MQGs, colour-coded by selection redshift ($2\le z\le 6$). Vertical dashed lines indicate the median values for MSFGs at $z=2$, with shaded regions marking the corresponding 16th and 84th percentile range. Thin lines show the corresponding medians for MQGs from L400m7. In the right panel, the red vertical dashed line marks the canonical value $\delta_{\rm DGR}=0.01$.
The horizontal dotted line denotes the threshold of 10 galaxies for L200m6, below which the statistics become unreliable.}
\label{fig:dust-gas} 
\end{figure*}

Observational constraints on dust content are typically obtained from far-infrared/sub-mm continuum data by fitting a modified blackbody spectrum to the SED, assuming a fixed dust temperature and emissivity. These assumptions, however, introduce significant uncertainties, especially at high redshift \citep{somovigo25}. The top panel of Fig.~\ref{fig:dust-z} compiles current dust measurements and upper limits for spectroscopically confirmed quenched \textcolor{black}{galaxies}\footnotemark. \citet{lee24} report a single source with unusually high dust content, clearly at odds with the MQGs in our simulations and likely representing a rare dust-rich quenched system, same as one galaxy in \citet{suzuki22}, and another one in \citet{whitaker21} (although the dust fraction is slightly lower). These systems are rare in {\sc COLIBRE} (Fig.~\ref{fig:dust-gas}), but there is overlap with the \citet{whitaker21} system. Other observations shown with arrows and a more faded colour \citep{whitaker21, kalita21, suzuki22, setton24} are upper limits, which are consistent with our predictions. \textcolor{black}{Additional measurements from \citet{siegel25} and \citet{chang26} are not included in the figure. Most correspond to non-detections, except for one (likely dust-rich) system reported by \citet{chang26}, with $f_{\rm dust}\approx10^{-3}$ at $z\approx3.5$, which also lies above the median trend predicted by {\sc COLIBRE}.}

\footnotetext{\citet{lee24, suzuki22} only account for photometric errors (the latter one not including $M_{\star}$ uncertainties); while \citet{whitaker21} consider uncertainties additionally when varying the dust temperature.}

\subsubsection{Molecular hydrogen fraction evolution}
\label{sssec:res1-dust-gas-gasev} 

Molecular gas observations are less reliable due to the absence of strong emission lines or direct $\rm H_{2}$ detections. Molecular gas masses are therefore typically inferred either from dust continuum measurements, assuming a dust-to-gas ratio (DGR) \citep[e.g.][]{whitaker21b, chang25}, or from emission of tracers such as CO, [CII], or [CI] \citep[e.g.][]{suzuki22}, which rely on uncertain conversion factors. The bottom panel of Fig.~\ref{fig:dust-z} shows the median molecular gas fraction, here referring exclusively to $\rm H_{2}$, and its scatter in these systems. For L200m6, the molecular hydrogen fraction remains roughly constant over time and is about two orders of magnitude lower in MQGs than in star-forming counterparts, consistent with $\rm H_{2}$ photodissociation driven by feedback processes. In L400m7, the molecular hydrogen fraction is about $1$~dex lower. \textcolor{black}{Visual inspection of individual evolutionary histories suggests} a more efficient destruction of molecular gas by a more bursty feedback while the dust content remains comparable. \textcolor{black}{A detailed investigation of this difference is beyond the scope of this paper.}

Observational estimates not derived from dust continuum, and thus not requiring corrections, as they trace $\rm H_{2}$ alone; are included: \textcolor{black}{most} of them are upper limits either from [CI] \citep{suzuki22} or CO \citep{scholtz25, umehata25}, illustrating the difficulty of measuring molecular gas in these systems due to their low reservoirs. \textcolor{black}{An exception is one of the galaxies in \citet{umehata25}, which represents the first CO detection of an early MQG}. The system in \citet{scholtz25} hosts a neutral outflow that efficiently removed gas, indicative of ejective AGN feedback. \textcolor{black}{An additional non-detection for a $z\approx3.25$ MQG was reported by \citet{wang26}, with $f_{\rm H_2}\lesssim10^{-1}$}. The {\sc COLIBRE} predictions are consistent with all the observational upper limits, \textcolor{black}{although they do not reproduce the CO-detected system from \citet{umehata25}, which likely corresponds to a gas-rich MQG. However, this interpretation depends on the adopted conversion factor between CO and $\mathrm{H_2}$, which itself carries substantial uncertainties \citep{bolatto13}. The inferred gas masses also depend on the tracer used. \citet{deugenio26} analysed the same sample from \citet{umehata25}, together with an extra non-detected system, using [CII] observations and found a difference of up to $\sim0.4$ dex for the CO-detected galaxy.}

\subsubsection{Distribution functions}
\label{sssec:res1-dust-gas-dist} 

In Fig.~\ref{fig:dust-gas}, we show (from left to right) the distributions of the dust fraction, $f_\text{dust}$; molecular hydrogen fraction, $f_{\rm H_2}$; and dust-to-gas ratio (DGR), $\delta_\text{DGR}=M_{\rm dust}/M_{\rm H_2}$ for L200m6 (thick) and L200m7 (thin lines). Each coloured line represents the distribution at a different redshift, while vertical dashed lines indicate the median values for MSFGs of the same stellar mass. 

Focusing on $f_{\rm dust}$ (left panel), for L200m6 we see that typical values cluster around $f_{\rm dust} \sim 10^{-4}$ but the distribution has extended tails. Some galaxies reach $f_{\rm dust} \sim 10^{-3}$, comparable to values found in MSFGs. These likely represent galaxies that have not yet removed their dust, consistent with the fact that systems with $f_{\rm dust} > 10^{-4}$ show shorter median quenching times ($t_{\rm q} \sim \rm 0.4\,Gyr$) than those with $f_{\rm dust} \le 10^{-4}$ ($t_{\rm q} \sim \rm 0.7\,Gyr$) at $z=2$. Appendix~\ref{appendix:corr} shows that this correlation is shallow and exhibits substantial scatter, though; \textcolor{black}{likely reflecting a mixture of systems that are still quenching and others that have begun to reform their dust content. This behaviour is visible in the individual galaxy evolutionary histories and gas/dust images in appendix~B of \citetalias{chandro-gomez_inprep}.}

The L400m7 simulation exhibits a higher distribution, as was seen for the number densities. Interestingly, it also shows an increased number of galaxies with extremely low dust fractions ($f_{\rm dust} < 10^{-4}$), \textcolor{black}{particularly at $z=2$; a distinct feature confirmed by normalising the distribution by the total number of galaxies at each redshift.} 
This may suggest that a more bursty feedback implementation can reduce dust content more dramatically. A similar broadening at the low-dust end is observed in L200m7h. This may occur because jets affect dust more strongly. In {\sc Simba}, the combination of AGN feedback and fast/slow quenching determines the future fate of dust: the jet mode removes dust over short (slow quenchers) or long (fast quenchers) timescales, while radiative-mode systems allow dust regrowth in the ISM \citep{lorenzon25}. Prospective work could explore whether similar correlations exist in {\sc COLIBRE}. 

Dust may regrow in the ISM or through external processes after quenching, potentially linked to AGN feedback. Both scenarios have been proposed in both observations \citep[e.g.][]{donevski23, kakimoto24} and other simulations, including {\sc Simba} \citep{lorenzon25}. {\sc Simba}, like {\sc COLIBRE}, includes explicit ISM dust growth, but following \citet{mckinnon17}. Unlike COLIBRE, however, {\sc Simba}'s dust budget is dominated by direct stellar production, driven by the order-of-magnitude larger stellar dust yields adopted from \citet{dwek98} \citep[see fig.~1 in][]{trayford25}, and it does not model the cold ISM.


Focusing on the molecular hydrogen fraction (middle panel), we see that the peak values for the quenched population are more than an order of magnitude lower than those of star-forming galaxies for L200m6, consistent with gas depletion due to quenching. Nevertheless, a fraction of galaxies retain significant molecular gas reservoirs, with $f_{\rm H_{2}} > 10^{-2}$. These are relatively high-dust content galaxies (see the correlation between $f_{\rm dust}$ and $f_{\rm H_{2}}$ in Appendix~\ref{appendix:corr}) that quench at later times.

L400m7 again shows a larger number of galaxies with extremely low molecular gas fractions ($f_{\rm H_2} < 10^{-4}$), \textcolor{black}{across the entire redshift range analysed, a trend confirmed by normalising the histogram by the number of galaxies at each redshift. This supports the idea that burstier feedback removes molecular gas more efficiently (at least at $z>2$).} In contrast, L200m7h produces results similar to L200m6 for $f_{\rm H_2}$.


The DGR distribution (right panel) reveals that passive galaxies exhibit a wide scatter in this quantity, with median values intriguingly higher ($\delta_{\rm DGR}\sim0.1$) than those of star-forming galaxies. Gas metallicity (measured within the $50\,\rm pkpc$ fiducial aperture as the \textcolor{black}{gas-mass-weighted} linear sum of the diffuse oxygen-to-hydrogen ratio, \textcolor{black}{excluding metals locked into dust}) for MQGs is lower and yet, they have a higher DGR (e.g. the median values for $12+\mathrm{log_{10}(O/H)}$ at $z=2$ are $\approx8.14$ for MQGs and $\approx8.49$ for MSFGs). This suggests that molecular gas is destroyed more rapidly than dust. L400m7 (and similarly L200m7h) predicts a pronounced local peak in DGR around $\delta_{\rm DGR} \sim 10^{3}$, corresponding to galaxies that retain significant dust, \textcolor{black}{potentially linked to the more efficient feedback and destruction of $\rm H_2$ in these simulations}.

Here we omit galaxies with $M_{\rm H_{2}}=0$. In L200m6, only $\approx 2\%$ of MQGs at $z=2$ fall into this category (and 90\% of those also have $M_{\rm dust}=0$), with the fraction decreasing toward higher redshifts. The $z=2$ median value for MSFGs is computed including the $M_{\rm H_{2}}=0$ cases, although in practice none occur.

The DGR results support the idea that dust content may more closely reflect the nature of the quenching feedback (e.g. AGN-driven), with molecular gas and dust evolving on different timescales. 
This behaviour has been reported using {\sc Simba} \citep{lorenzon25}. However, in {\sc Simba} the $\mathrm{H_2}$ content is computed using the \citet{krumholz11} prescription rather than by resolving the cold ISM, and its DGRs are systematically lower than in {\sc COLIBRE}. The latter difference arises from the contrasting dust models: in {\sc Simba}, stellar yields play the dominant role, while ISM dust growth is insufficient to offset them, resulting in systematically lower dust masses and DGRs. \textcolor{black}{In contrast, \citet{trayford25} show that dust growth via accretion in the cold, dense ISM (along with grain size transfer processes) dominates in {\sc COLIBRE}. Removing this channel reduces the cosmic dust density by more than 0.5 dex relative to the fiducial model (see their section 4.2).}

Moreover, the common observational assumption of a constant DGR of $\sim0.01$ (red vertical dashed line) for the total ISM gas content \citep[not only for the dense, cold $\rm H_{2}$; see, e.g.][]{whitaker21, setton24} may be an oversimplification, as even when computing the ratio as $M_{\rm dust}/(M_{\rm HI}+M_{\rm H_2})$ including neutral hydrogen, the model predicts substantially lower DGRs, with median values of $\sim0.001$ for MQGs.



More observational data are needed to robustly test these predictions. While recent programs are beginning to address this, current samples remain small. Future surveys will increase the number of detected objects and may enable stacking analyses for dust and molecular gas measurements, including non-detections, following the approach of \citet{adscheid25}, but for spectroscopic samples. This will be crucial since, as noted earlier, existing observations are biased toward dust- and gas-rich systems.

\subsection{Sizes of MQGs}
\label{ssec:res1-size}

Quenched galaxies are known to evolve in size, with trends differing from those of their star-forming counterparts at $z \lesssim 3$. \citet{vanderwel14} find that quenched systems are more compact than star-forming galaxies and exhibit a more dramatic evolution, reaching much smaller sizes near $z = 3$ than their descendants at $z \approx 0$ \citep{buitrago08, vandokkum08, cutler22}. Thanks to the high resolution of \textit{JWST}, direct measurements of rest-frame NIR and optical sizes are now possible up to $z \approx 5$. Observations reveal that high-$z$ MQGs are very compact, with typical effective radii below 1 kpc \citep{wrightl24, ito24, williams24, kawinwanichakij25}. These studies also report negative colour gradients \citep{wrightl24, ito24}, indicative of older stellar populations in the centre, which supports an inside-out growth scenario.

In simulations, reproducing such compact sizes can be challenging because of resolution limitations \citep{remus25}. It is therefore interesting to examine the sizes of MQGs in {\sc COLIBRE}.

\subsubsection{Results \& discussion}
\label{ssec:res1-size-results} 

\begin{figure}
\centering
\includegraphics[width=0.48\textwidth]{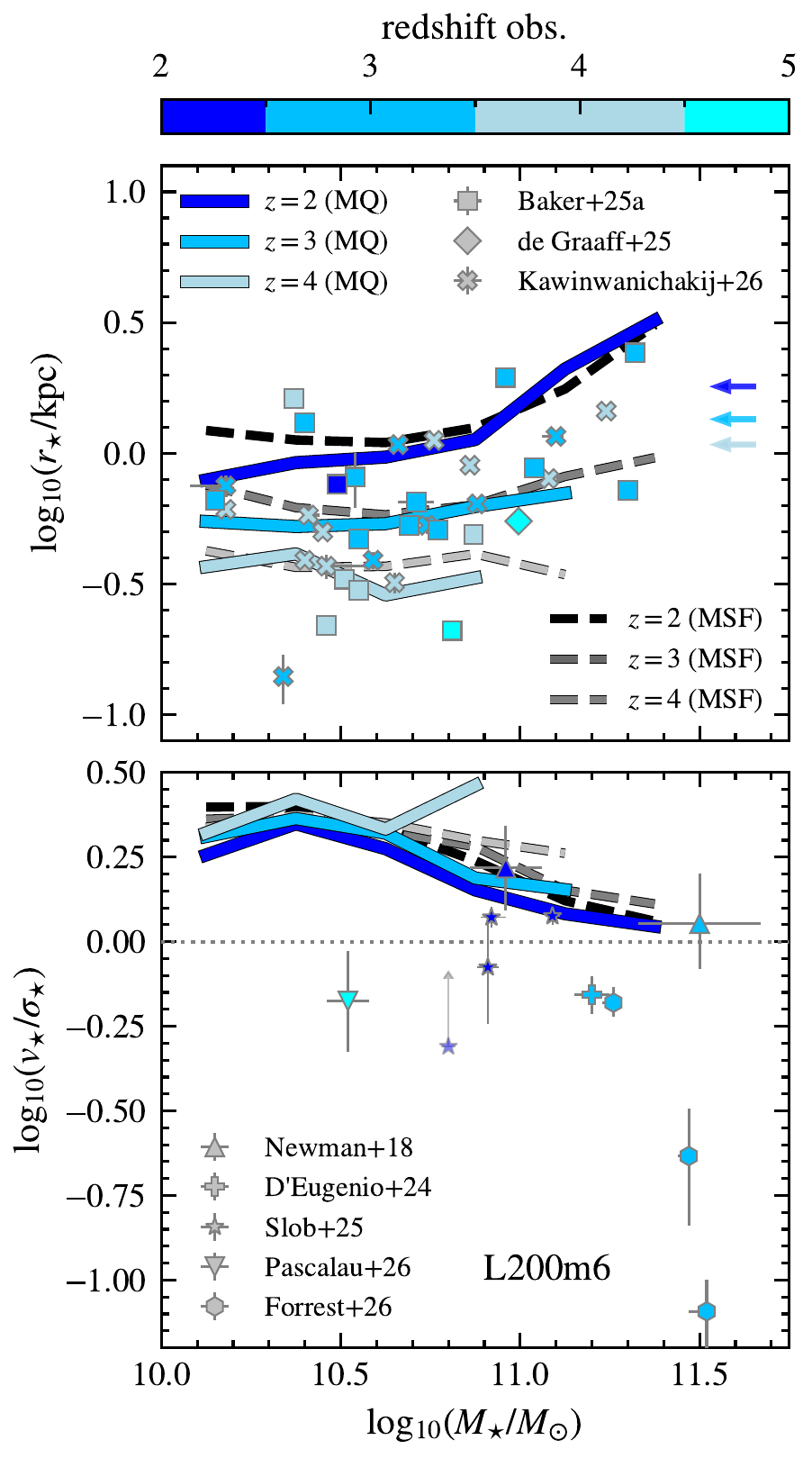}
\caption{Stellar size-mass relation, with $r_{\star}$ defined as the half-mass radius (\textit{top panel}); and $v_{\star}/\sigma_{\star}$ versus stellar mass relation (\textit{bottom panel}) for L200m6. Solid blue-palette lines show the median predictions for MQGs, and dashed grey-palette lines for MSFGs at different selection redshifts, $2 \le z \le 4$. At $z>4$, the number of massive quenched systems is too small to define a median split in stellar mass bins (9 galaxies at $z=5$ and 2 at $z=6$). Observational data of MQGs are from \citet{baker25b, degraaff25, kawinwanichakij25} for the size-mass relation (defined as half-light radius), and \citet{newman18, d'eugenio24, slob25, pascalau25, forrest25} for the $v_{\star}/\sigma_{\star}$-mass relation. Arrows in the error bars indicate lower limits for galaxies where rotational velocity could not be constrained (also shown with a more faded colour). Blue arrows in the size–mass relation mark the gravitational softening length, $\epsilon_{\rm prop}$, at each redshift.}
\label{fig:sizes} 
\end{figure}

The top panel of Fig.~\ref{fig:sizes} shows the galaxy mass–size relation for MQGs from the fiducial L200m6 simulation, where sizes are defined by the stellar half-mass radius averaged over three random projections ($r_{\star}$; see \S~\ref{sec:prop}). These are shown as solid coloured lines for $2 \le z \le 4$ for the fiducial simulation. Dashed lines indicate the corresponding relation for MSFGs at the same redshifts. Overall, we find no significant differences between the two populations, although MQGs appear slightly more compact at the low-mass end ($M_{\star} \lesssim 10^{10.5}\,\rm M_{\odot}$) at $z=2$. This result is consistent with \citet{ormerod24}, who report no distinction between both populations at $z>3$; but contrasts with \citet{vanderwel14}, who find that quenched systems are more compact even at $z\approx 2-3$\textcolor{black}{, as well as \citet{yang25size}, where this trend is extended to $z\approx4$}.

Blue arrows indicate the gravitational softening length, $\epsilon_{\rm prop}$, at different redshifts (see Table~\ref{tab:runs}). \textcolor{black}{Gravity is softened below $3\epsilon_{\rm prop}$.} \textcolor{black}{The
flattening of the median sizes of low-mass galaxies occurs on scales of order $\epsilon_{\rm prop}$, suggesting that caution is
needed when interpreting predictions on these scales. However, the flattening at our low-mass end (when including all systems, MQGs and MSFGs) is also present in L400m7 (see Appendix~\ref{appendix:l400m7-size-kin}), which has a larger softening length, as well as in L025m5 and L050m5, which have a smaller softening length \citep[fig.~10 in][]{ludlow26}. This suggests that the shape of the size-mass relation is not dictated by softening. \citet{ludlow23} reported a similar result for {\sc EAGLE} (see their fig.~8), where median sizes flatten at low masses on a scale comparable to the softening length, a result they attributed to poor DM mass resolution. Given that the size-mass relations in {\sc COLIBRE} converge well with respect to both $3\epsilon_{\rm prop}$ and $m_{\rm DM}$, their shapes appear robust to numerical effects. Furthermore, \citet{ludlow26} found that the median stellar surface density profiles in {\sc COLIBRE} are independent of resolution down to radii substantially smaller than $\epsilon_{\rm prop}$, again suggesting that softening length itself does not limit our ability to resolve galaxy structure.}


We find that galaxies become larger with decreasing redshift, highlighting their size evolution. The most massive systems are also more extended, and the slope of the mass–size relation appears to change around $M_{\star} \sim 10^{10.8}\,\mathrm{M_{\odot}}$, at least in the $z=2$ results. This trend is consistent with \citet{ji24}, who identify a similar transitional mass scale for MQGs, where more massive objects evolve more rapidly with likely higher ex-situ stellar mass contributions. We also find that the most compact galaxies tend to have later formation times (with median $t_{\rm 50} \approx \rm 1.3\,Gyr$ for $r_{\star} \le 1\,\rm kpc$, while $t_{\rm 50} \approx \rm 1.5\,Gyr$ for $r_{\star} > 1\,\rm kpc$ at $z=2$, and similar at other redshifts as shown in Appendix~\ref{appendix:corr}) in agreement with observational results \citep[see fig.~11 in][]{kawinwanichakij25}.

The galaxy sizes in our analysis are defined as the stellar half-mass radius. Observational studies, however, typically use half-light radii, and are influenced by projection effects (although in our case, averaging over three random projections helps to minimise these). Observationally, galaxy sizes are typically measured by fitting a \citet{sersic68} profile to the surface brightness distribution (including point spread function (PSF) convolution) and deriving the half-light effective radius. The inferred sizes depend not only on the fitting algorithm used \citep{haussler07} but also on the wavelength probed \citep{suess19}, with NIR measurements offering the most reliable tracer of stellar mass \citep{bell01}. For this reason, we introduce NIR-based observations in Fig.~\ref{fig:sizes}, despite the methodological and sample selection differences with simulations. Taking these caveats into account, the agreement with \citet{degraaff25, kawinwanichakij25, baker25b} appears to be of a qualitatively good nature.

\subsection{Kinematics of MQGs}
\label{ssec:res1-kin} 

We also examine the kinematics of MQGs, characterised by the stellar rotation-to-dispersion velocity ratio, $v_{\star}/\sigma_{\star}$. At $z \approx 0$, MQGs are primarily dispersion dominated, but their rotational support increases at higher redshift \citep{vandokkum08, vanderwel11, belli17}. Measuring stellar kinematics in these compact systems is challenging and, before \textit{JWST}, was only feasible in strongly lensed cases \citep{newman18}. Observationally, $v_{\star}/\sigma_{\star}$ is derived from spectral absorption features via line fitting, which requires very high signal-to-noise ratios and is particularly demanding at high redshift. Additional systematics arise from wavelength coverage, fitting codes, and modelling assumptions \citep{belli17}. \textit{JWST} has now opened a new window: with NIRSpec/IFU and NIRSpec/MSA, spatially resolved kinematics can be obtained even for faint, compact galaxies, complemented by forward kinematic modelling \citep{pascalau25, slob25}. These advances highlight the importance of simulation-based predictions for interpreting observations.

\subsubsection{Results and discussion}
\label{ssec:res1-kin-results} 

The bottom panel of Fig.~\ref{fig:sizes} shows {\sc COLIBRE} predictions for $v_{\star}/\sigma_{\star}$ (computed as in \S~\ref{ssec:prop-vsigma}) as a function of stellar mass in the range $2 \le z \le 4$ for L200m6. Solid blue-palette lines indicate the median values for MQGs, while dashed grey-palette lines show MSFGs. Observational results from lensed systems \citep{newman18} and \textit{JWST} \citep{d'eugenio24, pascalau25, slob25, forrest25} are included for comparison.

The horizontal dotted line ($v_{\star}/\sigma_{\star} = 1$) serves as a reference. There is a clear offset between the simulation predictions and the observations: most {\sc COLIBRE} MQGs lie above this threshold, appearing more rotationally supported than the observed galaxies. However, it is important to recall that the simulation values represent upper limits relative to the observations, due to differences in definitions and the caveats (\S~\ref{ssec:prop-vsigma}). In addition, some observational values (e.g. \citealt{forrest25}) are reported using the spin parameter $\lambda_{r}$ and have been converted to $v_{\star}/\sigma_{\star}$ using the relation in \citet{cappellari16}; apart from differences in sample selection. A more detailed comparison that closely mimics the observational analysis is left for future work.

Current observations suggest MQGs are commonly fast rotators at high redshift \citep{pascalau25}. One exception is the slow rotator reported by \citet[][the lowest datapoint]{forrest25}, likely the result of a major merger or isotropic gas infall \citep[values around $v_{\star}/\sigma_{\star} = 0.4$ are generally taken to indicate dispersion-dominated systems;][]{cappellari16}. In {\sc COLIBRE}, \textcolor{black}{individual $z=2$ MQGs show values as low as $v_{\star}/\sigma_{\star} \approx 0.6$ in the \textcolor{black}{bottom right} panel of Fig.~\ref{fig:corr-t}. This would not correspond to slow rotators under the observational definition. However, we cannot draw this conclusion because the observational and simulation quantities are defined differently.}

Both MQGs and MSFGs in {\sc COLIBRE} become more dispersion-supported towards lower redshift, with the most massive systems showing the strongest bulge dominance by $z \lesssim 3$. \textcolor{black}{At $z=4$, there is a tentative increase in the rotational support of the most massive MQGs. Although this trend is based on a small number of objects, it may indicate that strongly rotationally supported discs are required to assemble such massive systems at very early cosmic times.} At $z \gtrsim 2$, however, MQGs and MSFGs exhibit similar kinematics, implying that quenching preserves discs and precedes morphological transformation. \textcolor{black}{This is consistent with the findings of \citet{kimmig25}, who also report that the majority of systems are fast rotators, with a much smaller population of slow rotators formed by $z>4$}. Predictions for L400m7 are provided in Appendix~\ref{appendix:l400m7-size-kin}, where the trend reverses, with morphological transformation occurring before quenching. This is likely a consequence of its larger softening length, limiting the ability to resolve the internal structure of galaxies.

In the {\sc EAGLE} simulations, \citet{lagos22} showed this behaviour with galaxies quenching first, and only later ($\approx$2 Gyr) evolving into slow rotators through mergers (particularly dry mergers) that increase masses, sizes, and erode rotational support \citep[see also][]{lagos18}. Feedback mechanisms do not appear to affect kinematics directly, as seen in \citet{d'eugenio24}, where a rotation-supported system showed clear evidence of ejective feedback. The absence of star-forming gas in quenched galaxies likely makes mergers (particularly dry mergers) more effective at disrupting rotational support. This may explain why MQGs in {\sc COLIBRE} show slightly lower $v_{\star}/\sigma_{\star}$ than coeval MSFGs by $z \approx 2$, in agreement with \citet{pascalau25}. However, this contrasts with the findings of \citet{shuntov25}, who report that most high-$z$ MQGs are bulge-dominated, implying that a morphological transformation occurred before quenching.

This emerging population of slow rotators is found primarily in dense environments or recently quenched systems, consistent with our finding of a correlation between dispersion and environment (Appendix~\ref{appendix:corr}). More extended objects are also more dispersion-supported (\textcolor{black}{lower} left panel of Fig.~\ref{fig:corr-t}), supporting the merger-driven scenario, since mergers increase size.





\section{Conclusions}
\label{sec:conclusions}

We have analysed predictions from the new {\sc COLIBRE} hydrodynamical simulations in its multiple volumes, resolutions, and AGN feedback models for the population of Massive Quiescent Galaxies (MQGs; defined as galaxies with $M_{\star}>10^{10}\,\mathrm{M_{\odot}}$ and $\mathrm{sSFR}<0.2/t_{\rm age}$) at high-redshift ($z\geq2$). Recent \textit{JWST} surveys have suggested that MQGs are more abundant than previously observed. Given the poorly understood nature of these MQGs and current observational limitations, we investigated what the state-of-the-art {\sc COLIBRE} galaxy formation model predicts for a wide range of their properties. Our main findings are as follows:

\begin{enumerate}
    \item \textbf{Number densities}: We studied how MQG number densities change across the {\sc COLIBRE} simulations (\S~\ref{sssec:res1-ndens-results}). We find a clear resolution effect: higher-resolution runs produce less quenching; while the hybrid AGN feedback model yields even fewer quenched systems, underpredicting the observational estimates. The fiducial L200m6 simulation yields results broadly consistent with the most recent and robust observational estimates (those covering larger areas or larger spectroscopic samples), especially after we include  uncertainties 
    to the simulation values that exemplify observational uncertainties. Compared to most galaxy formation models that also reproduce observations at $z=0$, {\sc COLIBRE} predicts higher number densities (Fig.~\ref{fig:ndens-models}), \textcolor{black}{driven directly by the importance of super-Eddington accretion at high redshift \citep{chaikin26} and indirectly thanks to the explicit modelling of the cold gas phase.}
    \item \textbf{Stellar mass functions}: MQG stellar mass functions in {\sc COLIBRE} agree with the latest observations when uncertainties are considered, and outperform other models (Fig.~\ref{fig:smf}).
    \item \textbf{Star formation histories}: MQGs show good qualitative agreement with observationally inferred SFHs (\S~\ref{ssec:res1-ages}), with extended formation times (median $t_{50}\approx0.5$–$1.5\,\rm Gyr$), rapid quenching (median $t_{\rm q}\lesssim0.6\,\rm Gyr$), and evidence of strong starbursts.
    \item \textbf{Dust and molecular hydrogen}: With its explicit modelling of dust and a cold ISM phase, {\sc COLIBRE} predicts that MQGs have much lower dust and molecular hydrogen fractions (medians $f_{\rm dust}\sim10^{-4}$ and $f_{\rm H_2}\sim10^{-2.5}$) than massive star-forming counterparts, with \textcolor{black}{potential} evidence for dust regrowth (\S~\ref{ssec:res1-dust-gas}). Although current observations are limited in sample size and biased toward dust- and gas-rich systems, they generally agree with these predictions, with only a few cases showing very high dust \textcolor{black}{and gas} fractions (Fig.~\ref{fig:dust-z}).
    \item \textbf{Stellar sizes and kinematics}: MQG sizes (in terms of the half-mass radius) and kinematics are broadly similar to those of coeval MSFGs (Fig.~\ref{fig:sizes}), suggesting a decoupling between quenching and morphological transformation at $z\gtrsim2$, with the latter likely driven by subsequent mergers given the subtle signs of increased dispersion support by $z\approx2$.
\end{enumerate}

Taken together, the comparison between {\sc COLIBRE} and current observations shows encouraging qualitative and often quantitative agreement across a wide range of properties. Although systematic differences remain, likely in some part attributable to observational uncertainties, sample selection, and measurement techniques; the level of tension previously reported between models and observations has decreased. We conclude that this convergence reflects progress in both simulations and observations. This also highlights the need for more accurate modelling of the physical processes that shape MQGs, as well as larger, deeper spectroscopic samples to further reduce systematics. We note that this observational field is rapidly evolving, so measurements may continue to change. Closing this gap will be key to advancing our understanding of the earliest quenched galaxies and their role in galaxy evolution.

Several caveats must be considered. From the simulation side, galaxy formation models are calibrated in the local Universe, and results at high-$z$ remain sensitive to choices in e.g. AGN feedback modelling and uncertain parameters tied to BH physics, relevant for MQGs \citep{lagos25, husko25}, \textcolor{black}{as we further discuss in \citetalias{chandro-gomez_inprep}}. From the observational side, caveats related to sample selection, wavelength coverage, and measurement techniques (discussed throughout this paper) must be borne in mind when interpreting results. Nevertheless, when these factors are accounted for, this work, \textcolor{black}{together with \citetalias{chandro-gomez_inprep}}, provides robust predictions within the {\sc COLIBRE} model for the population of MQGs, offering insights into the physical mechanisms that shape their origin and evolution, and in the process easing the tension between observations and simulations. Its predictions also offer guidance for upcoming surveys, given current observational limitations.



\section*{Acknowledgements}

\textcolor{black}{We thank the anonymous referee for their constructive feedback.}ACG acknowledges Research Training Program and ICRAR scholarships. ACG acknowledges support for this project by the University of Western Australia via a Research Collaboration Award. This work used the DiRAC@Durham facility managed by the Institute for Computational Cosmology on behalf of the STFC DiRAC HPC Facility (\url{www.dirac.ac.uk}). The equipment was funded by BEIS capital funding via STFC capital grants ST/K00042X/1, ST/P002293/1, ST/R002371/1 and ST/S002502/1, Durham University and STFC operations grant ST/R000832/1. DiRAC is part of the National e-Infrastructure. This project has received funding from the Netherlands Organization for Scientific Research (NWO) through research programme Athena 184.034.002. WMB would like to acknowledge support from DARK via the DARK Fellowship. This work was supported by a research grant (VIL54489) from VILLUM FONDEN. ABL acknowledges support by the Italian Ministry for Universities (MUR) program “Dipartimenti di Eccellenza 2023-2027” within the Centro Bicocca di Cosmologia Quantitativa (BiCoQ), and support by UNIMIB’s Fondo Di Ateneo Quota Competitiva (project 2024-ATEQC-0050). \textcolor{black}{EC acknowledges support from STFC consolidated grant ST/X001075/1.} SP acknowledges support by the Austrian Science Fund (FWF) through grant-DOI: 10.55776/V982. JT acknowledges support of a STFC Early Stage Research and Development grant (ST/X004651/1). Minor typos, grammar and spelling mistakes were identified with the assistance of ChatGPT-4o\footnote{\url{openai.com}} when preparing this document. No passages of text or structural outlines for this paper were created with the help of any large language models.

\section*{Data Availability}

The data underlying this article will be provided upon reasonable request to the corresponding author.



\bibliographystyle{mnras}
\bibliography{colibre-mqg} 

@ARTICLE{lagos25,
       author = {{Lagos}, Claudia del P. and {Valentino}, Francesco and {Wright}, Ruby J. and {de Graaff}, Anna and {Glazebrook}, Karl and {De Lucia}, Gabriella and {Robotham}, Aaron S.~G. and {Nanayakkara}, Themiya and {Chandro-Gomez}, Angel and {Bravo}, Mat{\'\i}as and {Baugh}, Carlton M. and {Harborne}, Katherine E. and {Hirschmann}, Michaela and {Fontanot}, Fabio and {Xie}, Lizhi and {Chittenden}, Harry},
        title = "{The diverse star formation histories of early massive, quenched galaxies in modern galaxy formation simulations}",
      journal = {\mnras},
         year = 2025,
        month = jan,
       volume = {536},
       number = {3},
        pages = {2324-2354},
          doi = {10.1093/mnras/stae2626},
archivePrefix = {arXiv},
       eprint = {2409.16916},
 primaryClass = {astro-ph.GA},
       adsurl = {https://ui.adsabs.harvard.edu/abs/2025MNRAS.536.2324L}
}

@ARTICLE{carnall23,
       author = {{Carnall}, A.~C. and {McLeod}, D.~J. and {McLure}, R.~J. and {Dunlop}, J.~S. and {Begley}, R. and {Cullen}, F. and {Donnan}, C.~T. and {Hamadouche}, M.~L. and {Jewell}, S.~M. and {Jones}, E.~W. and {Pollock}, C.~L. and {Wild}, V.},
        title = "{A surprising abundance of massive quiescent galaxies at 3 < z < 5 in the first data from JWST CEERS}",
      journal = {\mnras},
         year = 2023,
        month = apr,
       volume = {520},
       number = {3},
        pages = {3974-3985},
          doi = {10.1093/mnras/stad369},
archivePrefix = {arXiv},
       eprint = {2208.00986},
 primaryClass = {astro-ph.GA},
       adsurl = {https://ui.adsabs.harvard.edu/abs/2023MNRAS.520.3974C}
}

@ARTICLE{baker25,
       author = {{Baker}, William M. and {Valentino}, Francesco and {Lagos}, Claudia del P. and {Ito}, Kei and {Jespersen}, Christian Kragh and {Gottumukkala}, Rashmi and {Hjorth}, Jens and {Langeroodi}, Danial and {Sedgewick}, Aidan},
        title = "{Exploring over 700 massive quiescent galaxies at z = 2─7: Demographics and stellar mass functions}",
      journal = {\aap},
         year = 2025,
        month = oct,
       volume = {702},
          eid = {A270},
        pages = {A270},
          doi = {10.1051/0004-6361/202555829},
archivePrefix = {arXiv},
       eprint = {2506.04119},
 primaryClass = {astro-ph.GA},
       adsurl = {https://ui.adsabs.harvard.edu/abs/2025A&A...702A.270B}
}

@ARTICLE{alberts24,
       author = {{Alberts}, Stacey and {Williams}, Christina C. and {Helton}, Jakob M. and {Suess}, Katherine A. and {Ji}, Zhiyuan and {Shivaei}, Irene and {Lyu}, Jianwei and {Rieke}, George and {Baker}, William M. and {Bonaventura}, Nina and {Bunker}, Andrew J. and {Carniani}, Stefano and {Charlot}, Stephane and {Curtis-Lake}, Emma and {D'Eugenio}, Francesco and {Eisenstein}, Daniel J. and {de Graaff}, Anna and {Hainline}, Kevin N. and {Hausen}, Ryan and {Johnson}, Benjamin D. and {Maiolino}, Roberto and {Parlanti}, Eleonora and {Rieke}, Marcia J. and {Robertson}, Brant E. and {Sun}, Yang and {Tacchella}, Sandro and {Willmer}, Christopher N.~A. and {Willott}, Chris J.},
        title = "{To High Redshift and Low Mass: Exploring the Emergence of Quenched Galaxies and Their Environments at 3 < z < 6 in the Ultra-deep JADES MIRI F770W Parallel}",
      journal = {\apj},
         year = 2024,
        month = nov,
       volume = {975},
       number = {1},
          eid = {85},
        pages = {85},
          doi = {10.3847/1538-4357/ad66cc},
archivePrefix = {arXiv},
       eprint = {2312.12207},
 primaryClass = {astro-ph.GA},
       adsurl = {https://ui.adsabs.harvard.edu/abs/2024ApJ...975...85A}
}

@ARTICLE{valentino23,
       author = {{Valentino}, Francesco and {Brammer}, Gabriel and {Gould}, Katriona M.~L. and {Kokorev}, Vasily and {Fujimoto}, Seiji and {Jespersen}, Christian Kragh and {Vijayan}, Aswin P. and {Weaver}, John R. and {Ito}, Kei and {Tanaka}, Masayuki and {Ilbert}, Olivier and {Magdis}, Georgios E. and {Whitaker}, Katherine E. and {Faisst}, Andreas L. and {Gallazzi}, Anna and {Gillman}, Steven and {Gim{\'e}nez-Arteaga}, Clara and {G{\'o}mez-Guijarro}, Carlos and {Kubo}, Mariko and {Heintz}, Kasper E. and {Hirschmann}, Michaela and {Oesch}, Pascal and {Onodera}, Masato and {Rizzo}, Francesca and {Lee}, Minju and {Strait}, Victoria and {Toft}, Sune},
        title = "{An Atlas of Color-selected Quiescent Galaxies at z > 3 in Public JWST Fields}",
      journal = {\apj},
         year = 2023,
        month = apr,
       volume = {947},
       number = {1},
          eid = {20},
        pages = {20},
          doi = {10.3847/1538-4357/acbefa},
archivePrefix = {arXiv},
       eprint = {2302.10936},
 primaryClass = {astro-ph.GA},
       adsurl = {https://ui.adsabs.harvard.edu/abs/2023ApJ...947...20V}
}

@ARTICLE{nanayakkara25,
       author = {{Nanayakkara}, Themiya and {Glazebrook}, Karl and {Schreiber}, Corentin and {Chittenden}, Harry and {Brammer}, Gabriel and {Esdaile}, James and {Jacobs}, Colin and {Kacprzak}, Glenn G. and {Kawinwanichakij}, Lalitwadee and {Kimmig}, Lucas C. and {Labbe}, Ivo and {Lagos}, Claudia and {Marchesini}, Danilo and {Mart{\`\i}nez-Mar{\`\i}n}, M. and {Marsan}, Z. Cemile and {Oesch}, Pascal A. and {Papovich}, Casey and {Remus}, Rhea-Silvia and {Tran}, Kim-Vy H.},
        title = "{The Formation Histories of Massive and Quiescent Galaxies in the 3 < z < 4.5 Universe}",
      journal = {\apj},
         year = 2025,
        month = mar,
       volume = {981},
       number = {1},
          eid = {78},
        pages = {78},
          doi = {10.3847/1538-4357/ada6ac},
archivePrefix = {arXiv},
       eprint = {2410.02076},
 primaryClass = {astro-ph.GA},
       adsurl = {https://ui.adsabs.harvard.edu/abs/2025ApJ...981...78N}
}

@ARTICLE{weibel25,
       author = {{Weibel}, Andrea and {de Graaff}, Anna and {Setton}, David J. and {Miller}, Tim B. and {Oesch}, Pascal A. and {Brammer}, Gabriel and {Lagos}, Claudia D.~P. and {Whitaker}, Katherine E. and {Williams}, Christina C. and {Baggen}, Josephine F.~W. and {Bezanson}, Rachel and {Boogaard}, Leindert A. and {Cleri}, Nikko J. and {Greene}, Jenny E. and {Hirschmann}, Michaela and {Hviding}, Raphael E. and {Kuruvanthodi}, Adarsh and {Labb{\'e}}, Ivo and {Leja}, Joel and {Maseda}, Michael V. and {Matthee}, Jorryt and {McConachie}, Ian and {Naidu}, Rohan P. and {Roberts-Borsani}, Guido and {Schaerer}, Daniel and {Suess}, Katherine A. and {Valentino}, Francesco and {van Dokkum}, Pieter and {Wang}, Bingjie},
        title = "{RUBIES Reveals a Massive Quiescent Galaxy at z = 7.3}",
      journal = {\apj},
         year = 2025,
        month = apr,
       volume = {983},
       number = {1},
          eid = {11},
        pages = {11},
          doi = {10.3847/1538-4357/adab7a},
archivePrefix = {arXiv},
       eprint = {2409.03829},
 primaryClass = {astro-ph.GA},
       adsurl = {https://ui.adsabs.harvard.edu/abs/2025ApJ...983...11W}
}

@ARTICLE{carnall24,
       author = {{Carnall}, A.~C. and {Cullen}, F. and {McLure}, R.~J. and {McLeod}, D.~J. and {Begley}, R. and {Donnan}, C.~T. and {Dunlop}, J.~S. and {Shapley}, A.~E. and {Rowlands}, K. and {Almaini}, O. and {Arellano-C{\'o}rdova}, K.~Z. and {Barrufet}, L. and {Cimatti}, A. and {Ellis}, R.~S. and {Grogin}, N.~A. and {Hamadouche}, M.~L. and {Illingworth}, G.~D. and {Koekemoer}, A.~M. and {Leung}, H. -H. and {Lovell}, C.~C. and {P{\'e}rez-Gonz{\'a}lez}, P.~G. and {Santini}, P. and {Stanton}, T.~M. and {Wild}, V.},
        title = "{The JWST EXCELS survey: too much, too young, too fast? Ultra-massive quiescent galaxies at 3 < z < 5}",
      journal = {\mnras},
         year = 2024,
        month = oct,
       volume = {534},
       number = {1},
        pages = {325-348},
          doi = {10.1093/mnras/stae2092},
archivePrefix = {arXiv},
       eprint = {2405.02242},
 primaryClass = {astro-ph.GA},
       adsurl = {https://ui.adsabs.harvard.edu/abs/2024MNRAS.534..325C}
}

@ARTICLE{baker25b,
       author = {{Baker}, William M. and {Lim}, Seunghwan and {D'Eugenio}, Francesco and {Maiolino}, Roberto and {Ji}, Zhiyuan and {Arribas}, Santiago and {Bunker}, Andrew J. and {Carniani}, Stefano and {Charlot}, Stephane and {de Graaff}, Anna and {Hainline}, Kevin and {Looser}, Tobias J. and {Lyu}, Jianwei and {Rinaldi}, Pierluigi and {Robertson}, Brant and {Schaller}, Matthieu and {Schaye}, Joop and {Scholtz}, Jan and {{\"U}bler}, Hannah and {Williams}, Christina C. and {Willmer}, Christopher N.~A. and {Willott}, Chris and {Zhu}, Yongda},
        title = "{The abundance and nature of high-redshift quiescent galaxies from JADES spectroscopy and the FLAMINGO simulations}",
      journal = {\mnras},
         year = 2025,
        month = may,
       volume = {539},
       number = {1},
        pages = {557-589},
          doi = {10.1093/mnras/staf475},
archivePrefix = {arXiv},
       eprint = {2410.14773},
 primaryClass = {astro-ph.GA},
       adsurl = {https://ui.adsabs.harvard.edu/abs/2025MNRAS.539..557B}
}

@ARTICLE{zhang25,
       author = {{Zhang}, Yunchong and {de Graaff}, Anna and {Setton}, David J. and {Price}, Sedona H. and {Bezanson}, Rachel and {del P. Lagos}, Claudia and {Cutler}, Sam E. and {McConachie}, Ian and {Cleri}, Nikko J. and {Cooper}, Olivia R. and {Gottumukkala}, Rashmi and {Greene}, Jenny E. and {Hirschmann}, Michaela and {Khullar}, Gourav and {Labbe}, Ivo and {Leja}, Joel and {Maseda}, Michael V. and {Matthee}, Jorryt and {Miller}, Tim B. and {Nanayakkara}, Themiya and {Suess}, Katherine A. and {Wang}, Bingjie and {Whitaker}, Katherine E. and {Williams}, Christina C.},
        title = "{RUBIES Spectroscopically Confirms the High Number Density of Quiescent Galaxies from 2 < z< 5}",
      journal = {\apj},
         year = 2026,
        month = feb,
       volume = {997},
       number = {2},
          eid = {252},
        pages = {252},
          doi = {10.3847/1538-4357/ae24e1},
archivePrefix = {arXiv},
       eprint = {2508.08577},
 primaryClass = {astro-ph.GA},
       adsurl = {https://ui.adsabs.harvard.edu/abs/2026ApJ...997..252Z}
}

@ARTICLE{bellstedt25,
       author = {{Bellstedt}, Sabine and {Robotham}, Aaron S.~G.},
        title = "{PROGENY II: the impact of libraries and model configurations on inferred galaxy properties in SED fitting}",
      journal = {\mnras},
         year = 2025,
        month = jul,
       volume = {540},
       number = {3},
        pages = {2703-2726},
          doi = {10.1093/mnras/staf889},
archivePrefix = {arXiv},
       eprint = {2410.17698},
 primaryClass = {astro-ph.GA},
       adsurl = {https://ui.adsabs.harvard.edu/abs/2025MNRAS.540.2703B}
}

@ARTICLE{conroy09,
       author = {{Conroy}, Charlie and {Gunn}, James E. and {White}, Martin},
        title = "{The Propagation of Uncertainties in Stellar Population Synthesis Modeling. I. The Relevance of Uncertain Aspects of Stellar Evolution and the Initial Mass Function to the Derived Physical Properties of Galaxies}",
      journal = {\apj},
         year = 2009,
        month = jul,
       volume = {699},
       number = {1},
        pages = {486-506},
          doi = {10.1088/0004-637X/699/1/486},
archivePrefix = {arXiv},
       eprint = {0809.4261},
 primaryClass = {astro-ph},
       adsurl = {https://ui.adsabs.harvard.edu/abs/2009ApJ...699..486C}
}

@ARTICLE{cochrane25,
       author = {{Cochrane}, R.~K. and {Katz}, H. and {Begley}, R. and {Hayward}, C.~C. and {Best}, P.~N.},
        title = "{High-z Stellar Masses Can Be Recovered Robustly with JWST Photometry}",
      journal = {\apjl},
         year = 2025,
        month = jan,
       volume = {978},
       number = {2},
          eid = {L42},
        pages = {L42},
          doi = {10.3847/2041-8213/ad9a4d},
archivePrefix = {arXiv},
       eprint = {2412.02622},
 primaryClass = {astro-ph.GA},
       adsurl = {https://ui.adsabs.harvard.edu/abs/2025ApJ...978L..42C}
}

@ARTICLE{robotham20,
       author = {{Robotham}, A.~S.~G. and {Bellstedt}, S. and {Lagos}, C. del P. and {Thorne}, J.~E. and {Davies}, L.~J. and {Driver}, S.~P. and {Bravo}, M.},
        title = "{ProSpect: generating spectral energy distributions with complex star formation and metallicity histories}",
      journal = {\mnras},
         year = 2020,
        month = jun,
       volume = {495},
       number = {1},
        pages = {905-931},
          doi = {10.1093/mnras/staa1116},
archivePrefix = {arXiv},
       eprint = {2002.06980},
 primaryClass = {astro-ph.GA},
       adsurl = {https://ui.adsabs.harvard.edu/abs/2020MNRAS.495..905R}
}

@ARTICLE{kennicutt12,
       author = {{Kennicutt}, Robert C. and {Evans}, Neal J.},
        title = "{Star Formation in the Milky Way and Nearby Galaxies}",
      journal = {\araa},
         year = 2012,
        month = sep,
       volume = {50},
        pages = {531-608},
          doi = {10.1146/annurev-astro-081811-125610},
archivePrefix = {arXiv},
       eprint = {1204.3552},
 primaryClass = {astro-ph.GA},
       adsurl = {https://ui.adsabs.harvard.edu/abs/2012ARA&A..50..531K}
}

@ARTICLE{forrest24,
       author = {{Forrest}, Ben and {Cooper}, M.~C. and {Muzzin}, Adam and {Wilson}, Gillian and {Marchesini}, Danilo and {McConachie}, Ian and {Gomez}, Percy and {Annunziatella}, Marianna and {Marsan}, Z. Cemile and {Braspenning}, Joey and {Chang}, Wenjun and {de Lucia}, Gabriella and {Fontanot}, Fabio and {Hirschmann}, Michaela and {Nelson}, Dylan and {Pillepich}, Annalisa and {Schaye}, Joop and {Urbano Stawinski}, Stephanie M. and {Stefanon}, Mauro and {Xie}, Lizhi},
        title = "{MAGAZ3NE: Massive, Extremely Dusty Galaxies at z {\ensuremath{\sim}} 2 Lead to Photometric Overestimation of Number Densities of the Most Massive Galaxies at 3 < z < 4}",
      journal = {\apj},
         year = 2024,
        month = dec,
       volume = {977},
       number = {1},
          eid = {51},
        pages = {51},
          doi = {10.3847/1538-4357/ad8b1c},
archivePrefix = {arXiv},
       eprint = {2404.19018},
 primaryClass = {astro-ph.GA},
       adsurl = {https://ui.adsabs.harvard.edu/abs/2024ApJ...977...51F}
}

@ARTICLE{schreiber18,
       author = {{Schreiber}, C. and {Glazebrook}, K. and {Nanayakkara}, T. and {Kacprzak}, G.~G. and {Labb{\'e}}, I. and {Oesch}, P. and {Yuan}, T. and {Tran}, K. -V. and {Papovich}, C. and {Spitler}, L. and {Straatman}, C.},
        title = "{Near infrared spectroscopy and star-formation histories of 3 {\ensuremath{\leq}} z {\ensuremath{\leq}} 4 quiescent galaxies}",
      journal = {\aap},
         year = 2018,
        month = oct,
       volume = {618},
          eid = {A85},
        pages = {A85},
          doi = {10.1051/0004-6361/201833070},
archivePrefix = {arXiv},
       eprint = {1807.02523},
 primaryClass = {astro-ph.GA},
       adsurl = {https://ui.adsabs.harvard.edu/abs/2018A&A...618A..85S}
}

@ARTICLE{wangt25,
       author = {{Wang}, Tao and {Sun}, Hanwen and {Zhou}, Luwenjia and {Xu}, Ke and {Cheng}, Cheng and {Li}, Zhaozhou and {Chen}, Yangyao and {Mo}, H.~J. and {Dekel}, Avishai and {Yang}, Tiancheng and {Wang}, Yijun and {Chen}, Longyue and {Zheng}, Xianzhong and {Cai}, Zheng and {Elbaz}, David and {Dai}, Y. -S. and {Huang}, J. -S.},
        title = "{JWST/MIRI Reveals the True Number Density of Massive Galaxies in the Early Universe}",
      journal = {\apjl},
         year = 2025,
        month = jul,
       volume = {988},
       number = {1},
          eid = {L35},
        pages = {L35},
          doi = {10.3847/2041-8213/adebe7},
archivePrefix = {arXiv},
       eprint = {2403.02399},
 primaryClass = {astro-ph.GA},
       adsurl = {https://ui.adsabs.harvard.edu/abs/2025ApJ...988L..35W}
}

@ARTICLE{ploeckinger25,
       author = {{Ploeckinger}, Sylvia and {Richings}, Alexander J. and {Schaye}, Joop and {Trayford}, James W. and {Schaller}, Matthieu and {Chaikin}, Evgenii},
        title = "{HYBRID-CHIMES: a model for radiative cooling and the abundances of ions and molecules in simulations of galaxy formation}",
      journal = {\mnras},
         year = 2025,
        month = oct,
       volume = {543},
       number = {2},
        pages = {891-916},
          doi = {10.1093/mnras/staf1402},
archivePrefix = {arXiv},
       eprint = {2506.15773},
 primaryClass = {astro-ph.GA},
       adsurl = {https://ui.adsabs.harvard.edu/abs/2025MNRAS.543..891P}
}

@ARTICLE{richings14a,
       author = {{Richings}, A.~J. and {Schaye}, J. and {Oppenheimer}, B.~D.},
        title = "{Non-equilibrium chemistry and cooling in the diffuse interstellar medium - I. Optically thin regime}",
      journal = {\mnras},
         year = 2014,
        month = jun,
       volume = {440},
       number = {4},
        pages = {3349-3369},
          doi = {10.1093/mnras/stu525},
archivePrefix = {arXiv},
       eprint = {1401.4719},
 primaryClass = {astro-ph.GA},
       adsurl = {https://ui.adsabs.harvard.edu/abs/2014MNRAS.440.3349R}
}

@ARTICLE{richings14b,
       author = {{Richings}, A.~J. and {Schaye}, J. and {Oppenheimer}, B.~D.},
        title = "{Non-equilibrium chemistry and cooling in the diffuse interstellar medium - II. Shielded gas}",
      journal = {\mnras},
         year = 2014,
        month = aug,
       volume = {442},
       number = {3},
        pages = {2780-2796},
          doi = {10.1093/mnras/stu1046},
archivePrefix = {arXiv},
       eprint = {1403.6155},
 primaryClass = {astro-ph.GA},
       adsurl = {https://ui.adsabs.harvard.edu/abs/2014MNRAS.442.2780R}
}

@ARTICLE{trayford25,
       author = {{Trayford}, James W. and {Schaye}, Joop and {Correa}, Camila and {Ploeckinger}, Sylvia and {Richings}, Alexander J. and {Chaikin}, Evgenii and {Schaller}, Matthieu and {Ben{\'\i}tez-Llambay}, Alejandro and {Frenk}, Carlos and {Hu{\v{s}}ko}, Filip},
        title = "{Modelling the evolution and influence of dust in cosmological simulations that include the cold phase of the interstellar medium}",
      journal = {\mnras},
         year = 2026,
        month = feb,
       volume = {545},
       number = {4},
          eid = {staf2040},
        pages = {staf2040},
          doi = {10.1093/mnras/staf2040},
archivePrefix = {arXiv},
       eprint = {2505.13056},
 primaryClass = {astro-ph.GA},
       adsurl = {https://ui.adsabs.harvard.edu/abs/2026MNRAS.545f2040T}
}

@ARTICLE{zhukovska08,
       author = {{Zhukovska}, S. and {Gail}, H. -P. and {Trieloff}, M.},
        title = "{Evolution of interstellar dust and stardust in the solar neighbourhood}",
      journal = {\aap},
         year = 2008,
        month = feb,
       volume = {479},
       number = {2},
        pages = {453-480},
          doi = {10.1051/0004-6361:20077789},
archivePrefix = {arXiv},
       eprint = {0706.1155},
 primaryClass = {astro-ph},
       adsurl = {https://ui.adsabs.harvard.edu/abs/2008A&A...479..453Z}
}

@ARTICLE{hirashita14,
       author = {{Hirashita}, Hiroyuki and {Voshchinnikov}, Nikolai V.},
        title = "{Effects of grain growth mechanisms on the extinction curve and the metal depletion in the interstellar medium}",
      journal = {\mnras},
         year = 2014,
        month = jan,
       volume = {437},
       number = {2},
        pages = {1636-1645},
          doi = {10.1093/mnras/stt1997},
archivePrefix = {arXiv},
       eprint = {1310.4679},
 primaryClass = {astro-ph.GA},
       adsurl = {https://ui.adsabs.harvard.edu/abs/2014MNRAS.437.1636H}
}

@ARTICLE{tsai95,
       author = {{Tsai}, John C. and {Mathews}, William G.},
        title = "{Interstellar Grains in Elliptical Galaxies: Grain Evolution}",
      journal = {\apj},
         year = 1995,
        month = jul,
       volume = {448},
        pages = {84},
          doi = {10.1086/175943},
archivePrefix = {arXiv},
       eprint = {astro-ph/9502053},
 primaryClass = {astro-ph},
       adsurl = {https://ui.adsabs.harvard.edu/abs/1995ApJ...448...84T}
}

@ARTICLE{nobels24,
       author = {{Nobels}, Folkert S.~J. and {Schaye}, Joop and {Schaller}, Matthieu and {Ploeckinger}, Sylvia and {Chaikin}, Evgenii and {Richings}, Alexander J.},
        title = "{Tests of subgrid models for star formation using simulations of isolated disc galaxies}",
      journal = {\mnras},
         year = 2024,
        month = aug,
       volume = {532},
       number = {3},
        pages = {3299-3321},
          doi = {10.1093/mnras/stae1390},
archivePrefix = {arXiv},
       eprint = {2309.13750},
 primaryClass = {astro-ph.GA},
       adsurl = {https://ui.adsabs.harvard.edu/abs/2024MNRAS.532.3299N}
}

@ARTICLE{kennicutt98,
       author = {{Kennicutt}, Jr., Robert C.},
        title = "{The Global Schmidt Law in Star-forming Galaxies}",
      journal = {\apj},
         year = 1998,
        month = may,
       volume = {498},
       number = {2},
        pages = {541-552},
          doi = {10.1086/305588},
archivePrefix = {arXiv},
       eprint = {astro-ph/9712213},
 primaryClass = {astro-ph},
       adsurl = {https://ui.adsabs.harvard.edu/abs/1998ApJ...498..541K}
}

@ARTICLE{dalla-vecchia&schaye12,
       author = {{Dalla Vecchia}, Claudio and {Schaye}, Joop},
        title = "{Simulating galactic outflows with thermal supernova feedback}",
      journal = {\mnras},
         year = 2012,
        month = oct,
       volume = {426},
       number = {1},
        pages = {140-158},
          doi = {10.1111/j.1365-2966.2012.21704.x},
archivePrefix = {arXiv},
       eprint = {1203.5667},
 primaryClass = {astro-ph.GA},
       adsurl = {https://ui.adsabs.harvard.edu/abs/2012MNRAS.426..140D}
}

@ARTICLE{chaikin23,
       author = {{Chaikin}, Evgenii and {Schaye}, Joop and {Schaller}, Matthieu and {Ben{\'\i}tez-Llambay}, Alejandro and {Nobels}, Folkert S.~J. and {Ploeckinger}, Sylvia},
        title = "{A thermal-kinetic subgrid model for supernova feedback in simulations of galaxy formation}",
      journal = {\mnras},
         year = 2023,
        month = aug,
       volume = {523},
       number = {3},
        pages = {3709-3731},
          doi = {10.1093/mnras/stad1626},
archivePrefix = {arXiv},
       eprint = {2211.04619},
 primaryClass = {astro-ph.GA},
       adsurl = {https://ui.adsabs.harvard.edu/abs/2023MNRAS.523.3709C}
}

@ARTICLE{booth&schaye09,
       author = {{Booth}, C.~M. and {Schaye}, Joop},
        title = "{Cosmological simulations of the growth of supermassive black holes and feedback from active galactic nuclei: method and tests}",
      journal = {\mnras},
         year = 2009,
        month = sep,
       volume = {398},
       number = {1},
        pages = {53-74},
          doi = {10.1111/j.1365-2966.2009.15043.x},
archivePrefix = {arXiv},
       eprint = {0904.2572},
 primaryClass = {astro-ph.CO},
       adsurl = {https://ui.adsabs.harvard.edu/abs/2009MNRAS.398...53B}
}

@ARTICLE{michaux21,
       author = {{Michaux}, Micha{\"e}l and {Hahn}, Oliver and {Rampf}, Cornelius and {Angulo}, Raul E.},
        title = "{Accurate initial conditions for cosmological N-body simulations: minimizing truncation and discreteness errors}",
      journal = {\mnras},
         year = 2021,
        month = jan,
       volume = {500},
       number = {1},
        pages = {663-683},
          doi = {10.1093/mnras/staa3149},
archivePrefix = {arXiv},
       eprint = {2008.09588},
 primaryClass = {astro-ph.CO},
       adsurl = {https://ui.adsabs.harvard.edu/abs/2021MNRAS.500..663M}
}

@ARTICLE{schaller24,
       author = {{Schaller}, Matthieu and {Borrow}, Josh and {Draper}, Peter W. and {Ivkovic}, Mladen and {McAlpine}, Stuart and {Vandenbroucke}, Bert and {Bah{\'e}}, Yannick and {Chaikin}, Evgenii and {Chalk}, Aidan B.~G. and {Chan}, Tsang Keung and {Correa}, Camila and {van Daalen}, Marcel and {Elbers}, Willem and {Gonnet}, Pedro and {Hausammann}, Lo{\"\i}c and {Helly}, John and {Hu{\v{s}}ko}, Filip and {Kegerreis}, Jacob A. and {Nobels}, Folkert S.~J. and {Ploeckinger}, Sylvia and {Revaz}, Yves and {Roper}, William J. and {Ruiz-Bonilla}, Sergio and {Sandnes}, Thomas D. and {Uyttenhove}, Yolan and {Willis}, James S. and {Xiang}, Zhen},
        title = "{SWIFT: A modern highly-parallel gravity and smoothed particle hydrodynamics solver for astrophysical and cosmological applications}",
      journal = {\mnras},
         year = 2024,
        month = may,
       volume = {530},
       number = {2},
        pages = {2378-2419},
          doi = {10.1093/mnras/stae922},
archivePrefix = {arXiv},
       eprint = {2305.13380},
 primaryClass = {astro-ph.IM},
       adsurl = {https://ui.adsabs.harvard.edu/abs/2024MNRAS.530.2378S}
}

@ARTICLE{ludlow19,
       author = {{Ludlow}, Aaron D. and {Schaye}, Joop and {Schaller}, Matthieu and {Richings}, Jack},
        title = "{Energy equipartition between stellar and dark matter particles in cosmological simulations results in spurious growth of galaxy sizes}",
      journal = {\mnras},
         year = 2019,
        month = sep,
       volume = {488},
       number = {1},
        pages = {L123-L128},
          doi = {10.1093/mnrasl/slz110},
archivePrefix = {arXiv},
       eprint = {1903.10110},
 primaryClass = {astro-ph.GA},
       adsurl = {https://ui.adsabs.harvard.edu/abs/2019MNRAS.488L.123L}
}

@ARTICLE{dehnen14,
       author = {{Dehnen}, Walter},
        title = "{A fast multipole method for stellar dynamics}",
      journal = {Computational Astrophysics and Cosmology},
         year = 2014,
        month = sep,
       volume = {1},
          eid = {1},
        pages = {1},
          doi = {10.1186/s40668-014-0001-7},
archivePrefix = {arXiv},
       eprint = {1405.2255},
 primaryClass = {astro-ph.IM},
       adsurl = {https://ui.adsabs.harvard.edu/abs/2014ComAC...1....1D}
}

@ARTICLE{borrow22,
       author = {{Borrow}, Josh and {Schaller}, Matthieu and {Bower}, Richard G. and {Schaye}, Joop},
        title = "{SPHENIX: smoothed particle hydrodynamics for the next generation of galaxy formation simulations}",
      journal = {\mnras},
         year = 2022,
        month = apr,
       volume = {511},
       number = {2},
        pages = {2367-2389},
          doi = {10.1093/mnras/stab3166},
archivePrefix = {arXiv},
       eprint = {2012.03974},
 primaryClass = {astro-ph.GA},
       adsurl = {https://ui.adsabs.harvard.edu/abs/2022MNRAS.511.2367B}
}

@ARTICLE{abbott22,
       author = {{Abbott}, T.~M.~C. and {Aguena}, M. and {Alarcon}, A. and {Allam}, S. and {Alves}, O. and {Amon}, A. and {Andrade-Oliveira}, F. and {Annis}, J. and {Avila}, S. and {Bacon}, D. and {Baxter}, E. and {Bechtol}, K. and {Becker}, M.~R. and {Bernstein}, G.~M. and {Bhargava}, S. and {Birrer}, S. and {Blazek}, J. and {Brandao-Souza}, A. and {Bridle}, S.~L. and {Brooks}, D. and {Buckley-Geer}, E. and {Burke}, D.~L. and {Camacho}, H. and {Campos}, A. and {Carnero Rosell}, A. and {Carrasco Kind}, M. and {Carretero}, J. and {Castander}, F.~J. and {Cawthon}, R. and {Chang}, C. and {Chen}, A. and {Chen}, R. and {Choi}, A. and {Conselice}, C. and {Cordero}, J. and {Costanzi}, M. and {Crocce}, M. and {da Costa}, L.~N. and {da Silva Pereira}, M.~E. and {Davis}, C. and {Davis}, T.~M. and {De Vicente}, J. and {DeRose}, J. and {Desai}, S. and {Di Valentino}, E. and {Diehl}, H.~T. and {Dietrich}, J.~P. and {Dodelson}, S. and {Doel}, P. and {Doux}, C. and {Drlica-Wagner}, A. and {Eckert}, K. and {Eifler}, T.~F. and {Elsner}, F. and {Elvin-Poole}, J. and {Everett}, S. and {Evrard}, A.~E. and {Fang}, X. and {Farahi}, A. and {Fernandez}, E. and {Ferrero}, I. and {Fert{\'e}}, A. and {Fosalba}, P. and {Friedrich}, O. and {Frieman}, J. and {Garc{\'\i}a-Bellido}, J. and {Gatti}, M. and {Gaztanaga}, E. and {Gerdes}, D.~W. and {Giannantonio}, T. and {Giannini}, G. and {Gruen}, D. and {Gruendl}, R.~A. and {Gschwend}, J. and {Gutierrez}, G. and {Harrison}, I. and {Hartley}, W.~G. and {Herner}, K. and {Hinton}, S.~R. and {Hollowood}, D.~L. and {Honscheid}, K. and {Hoyle}, B. and {Huff}, E.~M. and {Huterer}, D. and {Jain}, B. and {James}, D.~J. and {Jarvis}, M. and {Jeffrey}, N. and {Jeltema}, T. and {Kovacs}, A. and {Krause}, E. and {Kron}, R. and {Kuehn}, K. and {Kuropatkin}, N. and {Lahav}, O. and {Leget}, P. -F. and {Lemos}, P. and {Liddle}, A.~R. and {Lidman}, C. and {Lima}, M. and {Lin}, H. and {MacCrann}, N. and {Maia}, M.~A.~G. and {Marshall}, J.~L. and {Martini}, P. and {McCullough}, J. and {Melchior}, P. and {Mena-Fern{\'a}ndez}, J. and {Menanteau}, F. and {Miquel}, R. and {Mohr}, J.~J. and {Morgan}, R. and {Muir}, J. and {Myles}, J. and {Nadathur}, S. and {Navarro-Alsina}, A. and {Nichol}, R.~C. and {Ogando}, R.~L.~C. and {Omori}, Y. and {Palmese}, A. and {Pandey}, S. and {Park}, Y. and {Paz-Chinch{\'o}n}, F. and {Petravick}, D. and {Pieres}, A. and {Plazas Malag{\'o}n}, A.~A. and {Porredon}, A. and {Prat}, J. and {Raveri}, M. and {Rodriguez-Monroy}, M. and {Rollins}, R.~P. and {Romer}, A.~K. and {Roodman}, A. and {Rosenfeld}, R. and {Ross}, A.~J. and {Rykoff}, E.~S. and {Samuroff}, S. and {S{\'a}nchez}, C. and {Sanchez}, E. and {Sanchez}, J. and {Sanchez Cid}, D. and {Scarpine}, V. and {Schubnell}, M. and {Scolnic}, D. and {Secco}, L.~F. and {Serrano}, S. and {Sevilla-Noarbe}, I. and {Sheldon}, E. and {Shin}, T. and {Smith}, M. and {Soares-Santos}, M. and {Suchyta}, E. and {Swanson}, M.~E.~C. and {Tabbutt}, M. and {Tarle}, G. and {Thomas}, D. and {To}, C. and {Troja}, A. and {Troxel}, M.~A. and {Tucker}, D.~L. and {Tutusaus}, I. and {Varga}, T.~N. and {Walker}, A.~R. and {Weaverdyck}, N. and {Wechsler}, R. and {Weller}, J. and {Yanny}, B. and {Yin}, B. and {Zhang}, Y. and {Zuntz}, J. and {DES Collaboration}},
        title = "{Dark Energy Survey Year 3 results: Cosmological constraints from galaxy clustering and weak lensing}",
      journal = {\prd},
         year = 2022,
        month = jan,
       volume = {105},
       number = {2},
          eid = {023520},
        pages = {023520},
          doi = {10.1103/PhysRevD.105.023520},
archivePrefix = {arXiv},
       eprint = {2105.13549},
 primaryClass = {astro-ph.CO},
       adsurl = {https://ui.adsabs.harvard.edu/abs/2022PhRvD.105b3520A}
}

@ARTICLE{forouhar25,
       author = {{Forouhar Moreno}, Victor J. and {Helly}, John and {McGibbon}, Robert and {Schaye}, Joop and {Schaller}, Matthieu and {Han}, Jiaxin and {Kugel}, Roi and {Bah{\'e}}, Yannick M.},
        title = "{Assessing subhalo finders in cosmological hydrodynamical simulations}",
      journal = {\mnras},
         year = 2025,
        month = oct,
       volume = {543},
       number = {2},
        pages = {1339-1372},
          doi = {10.1093/mnras/staf1478},
archivePrefix = {arXiv},
       eprint = {2502.06932},
 primaryClass = {astro-ph.CO},
       adsurl = {https://ui.adsabs.harvard.edu/abs/2025MNRAS.543.1339F}
}

@ARTICLE{han18,
       author = {{Han}, Jiaxin and {Cole}, Shaun and {Frenk}, Carlos S. and {Benitez-Llambay}, Alejandro and {Helly}, John},
        title = "{HBT+: an improved code for finding subhaloes and building merger trees in cosmological simulations}",
      journal = {\mnras},
         year = 2018,
        month = feb,
       volume = {474},
       number = {1},
        pages = {604-617},
          doi = {10.1093/mnras/stx2792},
archivePrefix = {arXiv},
       eprint = {1708.03646},
 primaryClass = {astro-ph.CO},
       adsurl = {https://ui.adsabs.harvard.edu/abs/2018MNRAS.474..604H}
}

@ARTICLE{chandro-gomez25,
       author = {{Chandro-G{\'o}mez}, {\'A}ngel and {Lagos}, Claudia del P. and {Power}, Chris and {Moreno}, Victor J. Forouhar and {Helly}, John C. and {Lacey}, Cedric G. and {McGibbon}, Robert J. and {Schaller}, Matthieu and {Schaye}, Joop},
        title = "{On the accuracy of dark matter halo merger trees and the consequences for semi-analytic models of galaxy formation}",
      journal = {\mnras},
         year = 2025,
        month = may,
       volume = {539},
       number = {2},
        pages = {776-807},
          doi = {10.1093/mnras/staf519},
archivePrefix = {arXiv},
       eprint = {2501.07677},
 primaryClass = {astro-ph.GA},
       adsurl = {https://ui.adsabs.harvard.edu/abs/2025MNRAS.539..776C}
}

@ARTICLE{mcgibbon25,
       author = {{McGibbon}, Robert and {Helly}, John and {Schaye}, Joop and {Schaller}, Matthieu and {Vandenbroucke}, Bert},
        title = "{SOAP: A Python Package for Calculating the Properties of Galaxies and Halos Formed in Cosmological Simulations}",
      journal = {The Journal of Open Source Software},
         year = 2025,
        month = jul,
       volume = {10},
       number = {111},
          eid = {8252},
        pages = {8252},
          doi = {10.21105/joss.08252},
archivePrefix = {arXiv},
       eprint = {2507.22669},
 primaryClass = {astro-ph.IM},
       adsurl = {https://ui.adsabs.harvard.edu/abs/2025JOSS...10.8252M}
}

@ARTICLE{bahe22,
       author = {{Bah{\'e}}, Yannick M. and {Schaye}, Joop and {Schaller}, Matthieu and {Bower}, Richard G. and {Borrow}, Josh and {Chaikin}, Evgenii and {Kugel}, Roi and {Nobels}, Folkert and {Ploeckinger}, Sylvia},
        title = "{The importance of black hole repositioning for galaxy formation simulations}",
      journal = {\mnras},
         year = 2022,
        month = oct,
       volume = {516},
       number = {1},
        pages = {167-184},
          doi = {10.1093/mnras/stac1339},
archivePrefix = {arXiv},
       eprint = {2109.01489},
 primaryClass = {astro-ph.GA},
       adsurl = {https://ui.adsabs.harvard.edu/abs/2022MNRAS.516..167B}
}

@ARTICLE{cappellari16,
       author = {{Cappellari}, Michele},
        title = "{Structure and Kinematics of Early-Type Galaxies from Integral Field Spectroscopy}",
      journal = {\araa},
         year = 2016,
        month = sep,
       volume = {54},
        pages = {597-665},
          doi = {10.1146/annurev-astro-082214-122432},
archivePrefix = {arXiv},
       eprint = {1602.04267},
 primaryClass = {astro-ph.GA},
       adsurl = {https://ui.adsabs.harvard.edu/abs/2016ARA&A..54..597C}
}

@ARTICLE{pascalau25,
       author = {{Pascalau}, Robert G. and {D'Eugenio}, Francesco and {Tacchella}, Sandro and {Maiolino}, Roberto and {Cappellari}, Michele and {Duan}, Qiao and {Lagos}, Claudia del P. and {Bunker}, Andrew J. and {Jones}, Gareth C. and {Scholtz}, Jan and {{\"U}bler}, Hannah and {Cresci}, Giovanni and {Arribas}, Santiago and {Perna}, Michele and {van der Wel}, Arjen and {Danhaive}, A. Lola and {McClymont}, William and {Williams}, Christina C. and {de Graaff}, Anna and {Vani}, Akash and {Maseda}, Michael V. and {Carnall}, Adam C. and {Charlot}, St{\'e}phane and {Carniani}, Stefano and {Goh}, Tze P. and {Ji}, Zhiyuan and {P{\'e}rez Gonz{\'a}lez}, Pablo},
        title = "{When relics were made: vigorous stellar rotation and low dark matter content in the massive ultra-compact galaxy GS-9209 at z = 4.66}",
      journal = {\mnras},
         year = 2026,
        month = mar,
       volume = {547},
       number = {1},
          eid = {stag210},
        pages = {stag210},
          doi = {10.1093/mnras/stag210},
archivePrefix = {arXiv},
       eprint = {2505.06349},
 primaryClass = {astro-ph.GA},
       adsurl = {https://ui.adsabs.harvard.edu/abs/2026MNRAS.547ag210P}
}

@ARTICLE{slob25,
       author = {{Slob}, Martje and {Kriek}, Mariska and {de Graaff}, Anna and {Cheng}, Chloe M. and {Beverage}, Aliza G. and {Bezanson}, Rachel and {F{\"o}rster Schreiber}, Natascha M. and {Lorenz}, Brian and {Mancera Pi{\~n}a}, Pavel E. and {Marchesini}, Danilo and {Muzzin}, Adam and {Newman}, Andrew B. and {Price}, Sedona H. and {Suess}, Katherine A. and {van de Sande}, Jesse and {van Dokkum}, Pieter and {Weisz}, Daniel R.},
        title = "{Fast rotators at cosmic noon: Stellar kinematics for 15 quiescent galaxies from JWST-SUSPENSE}",
      journal = {\aap},
         year = 2025,
        month = oct,
       volume = {702},
          eid = {A110},
        pages = {A110},
          doi = {10.1051/0004-6361/202555812},
archivePrefix = {arXiv},
       eprint = {2506.04310},
 primaryClass = {astro-ph.GA},
       adsurl = {https://ui.adsabs.harvard.edu/abs/2025A&A...702A.110S}
}

@ARTICLE{franx08,
       author = {{Franx}, Marijn and {van Dokkum}, Pieter G. and {F{\"o}rster Schreiber}, Natascha M. and {Wuyts}, Stijn and {Labb{\'e}}, Ivo and {Toft}, Sune},
        title = "{Structure and Star Formation in Galaxies out to z = 3: Evidence for Surface Density Dependent Evolution and Upsizing}",
      journal = {\apj},
         year = 2008,
        month = dec,
       volume = {688},
       number = {2},
        pages = {770-788},
          doi = {10.1086/592431},
archivePrefix = {arXiv},
       eprint = {0808.2642},
 primaryClass = {astro-ph},
       adsurl = {https://ui.adsabs.harvard.edu/abs/2008ApJ...688..770F}
}

@ARTICLE{long24,
       author = {{Long}, Arianna S. and {Antwi-Danso}, Jacqueline and {Lambrides}, Erini L. and {Lovell}, Christopher C. and {de la Vega}, Alexander and {Valentino}, Francesco and {Zavala}, Jorge A. and {Casey}, Caitlin M. and {Wilkins}, Stephen M. and {Yung}, L.~Y. Aaron and {Arrabal Haro}, Pablo and {Bagley}, Micaela B. and {Bisigello}, Laura and {Chworowsky}, Katherine and {Cooper}, M.~C. and {Cooper}, Olivia R. and {Cooray}, Asantha R. and {Croton}, Darren and {Dickinson}, Mark and {Finkelstein}, Steven L. and {Franco}, Maximilien and {Gould}, Katriona M.~L. and {Hirschmann}, Michaela and {Hutchison}, Taylor A. and {Kartaltepe}, Jeyhan S. and {Kocevski}, Dale D. and {Koekemoer}, Anton M. and {Lucas}, Ray A. and {McKinney}, Jed and {Nere}, Rachel and {Papovich}, Casey and {P{\'e}rez-Gonz{\'a}lez}, Pablo G. and {Pirzkal}, Nor and {Santini}, Paola},
        title = "{Efficient NIRCam Selection of Quiescent Galaxies at 3 < z < 6 in CEERS}",
      journal = {\apj},
         year = 2024,
        month = jul,
       volume = {970},
       number = {1},
          eid = {68},
        pages = {68},
          doi = {10.3847/1538-4357/ad4cea},
archivePrefix = {arXiv},
       eprint = {2305.04662},
 primaryClass = {astro-ph.GA},
       adsurl = {https://ui.adsabs.harvard.edu/abs/2024ApJ...970...68L}
}

@ARTICLE{lee24,
       author = {{Lee}, Minju M. and {Steidel}, Charles C. and {Brammer}, Gabriel and {F{\"o}rster-Schreiber}, Natascha and {Renzini}, Alvio and {Liu}, Daizhong and {Herrera-Camus}, Rodrigo and {Naab}, Thorsten and {Price}, Sedona H. and {{\"U}bler}, Hannah and {Arriagada-Neira}, Sebasti{\'a}n and {Magdis}, Georgios},
        title = "{High dust content of a quiescent galaxy at z   2 revealed by deep ALMA observation}",
      journal = {\mnras},
         year = 2024,
        month = feb,
       volume = {527},
       number = {4},
        pages = {9529-9547},
          doi = {10.1093/mnras/stad3718},
archivePrefix = {arXiv},
       eprint = {2311.00023},
 primaryClass = {astro-ph.GA},
       adsurl = {https://ui.adsabs.harvard.edu/abs/2024MNRAS.527.9529L}
}

@ARTICLE{setton24,
       author = {{Setton}, David J. and {Khullar}, Gourav and {Miller}, Tim B. and {Bezanson}, Rachel and {Greene}, Jenny E. and {Suess}, Katherine A. and {Whitaker}, Katherine E. and {Antwi-Danso}, Jacqueline and {Atek}, Hakim and {Brammer}, Gabriel and {Cutler}, Sam E. and {Dayal}, Pratika and {Feldmann}, Robert and {Fujimoto}, Seiji and {Furtak}, Lukas J. and {Glazebrook}, Karl and {Goulding}, Andy D. and {Kokorev}, Vasily and {Labbe}, Ivo and {Leja}, Joel and {Ma}, Yilun and {Marchesini}, Danilo and {Nanayakkara}, Themiya and {Pan}, Richard and {Price}, Sedona H. and {Siegel}, Jared C. and {Shipley}, Heath and {Weaver}, John R. and {van Dokkum}, Pieter and {Wang}, Bingjie and {Williams}, Christina C.},
        title = "{UNCOVER NIRSpec/PRISM Spectroscopy Unveils Evidence of Early Core Formation in a Massive, Centrally Dusty Quiescent Galaxy at z $_{spec}$ = 3.97}",
      journal = {\apj},
         year = 2024,
        month = oct,
       volume = {974},
       number = {1},
          eid = {145},
        pages = {145},
          doi = {10.3847/1538-4357/ad6a18},
archivePrefix = {arXiv},
       eprint = {2402.05664},
 primaryClass = {astro-ph.GA},
       adsurl = {https://ui.adsabs.harvard.edu/abs/2024ApJ...974..145S}
}

@ARTICLE{whitaker21,
       author = {{Whitaker}, Katherine E. and {Williams}, Christina C. and {Mowla}, Lamiya and {Spilker}, Justin S. and {Toft}, Sune and {Narayanan}, Desika and {Pope}, Alexandra and {Magdis}, Georgios E. and {van Dokkum}, Pieter G. and {Akhshik}, Mohammad and {Bezanson}, Rachel and {Brammer}, Gabriel B. and {Leja}, Joel and {Man}, Allison and {Nelson}, Erica J. and {Richard}, Johan and {Pacifici}, Camilla and {Sharon}, Keren and {Valentino}, Francesco},
        title = "{Quenching of star formation from a lack of inflowing gas to galaxies}",
      journal = {\nat},
         year = 2021,
        month = sep,
       volume = {597},
       number = {7877},
        pages = {485-488},
          doi = {10.1038/s41586-021-03806-7},
archivePrefix = {arXiv},
       eprint = {2109.10384},
 primaryClass = {astro-ph.GA},
       adsurl = {https://ui.adsabs.harvard.edu/abs/2021Natur.597..485W}
}

@ARTICLE{suzuki22,
       author = {{Suzuki}, Tomoko L. and {Glazebrook}, Karl and {Schreiber}, Corentin and {Kodama}, Tadayuki and {Kacprzak}, Glenn G. and {Leiton}, Roger and {Nanayakkara}, Themiya and {Oesch}, Pascal A. and {Papovich}, Casey and {Spitler}, Lee and {Straatman}, Caroline M.~S. and {Tran}, Kim-Vy and {Wang}, Tao},
        title = "{Low Star Formation Activity and Low Gas Content of Quiescent Galaxies at z = 3.5-4.0 Constrained with ALMA}",
      journal = {\apj},
         year = 2022,
        month = sep,
       volume = {936},
       number = {1},
          eid = {61},
        pages = {61},
          doi = {10.3847/1538-4357/ac7ce3},
archivePrefix = {arXiv},
       eprint = {2206.14238},
 primaryClass = {astro-ph.GA},
       adsurl = {https://ui.adsabs.harvard.edu/abs/2022ApJ...936...61S}
}

@ARTICLE{kalita21,
       author = {{Kalita}, Boris S. and {Daddi}, Emanuele and {D'Eugenio}, Chiara and {Valentino}, Francesco and {Rich}, R. Michael and {G{\'o}mez-Guijarro}, Carlos and {Coogan}, Rosemary T. and {Delvecchio}, Ivan and {Elbaz}, David and {Neill}, James D. and {Puglisi}, Annagrazia and {Strazzullo}, Veronica},
        title = "{An Ancient Massive Quiescent Galaxy Found in a Gas-rich z   3 Group}",
      journal = {\apjl},
         year = 2021,
        month = aug,
       volume = {917},
       number = {2},
          eid = {L17},
        pages = {L17},
          doi = {10.3847/2041-8213/ac16dc},
archivePrefix = {arXiv},
       eprint = {2107.13241},
 primaryClass = {astro-ph.GA},
       adsurl = {https://ui.adsabs.harvard.edu/abs/2021ApJ...917L..17K}
}

@ARTICLE{whitaker21b,
       author = {{Whitaker}, Katherine E. and {Narayanan}, Desika and {Williams}, Christina C. and {Li}, Qi and {Spilker}, Justin S. and {Dav{\'e}}, Romeel and {Akhshik}, Mohammad and {Akins}, Hollis B. and {Bezanson}, Rachel and {Katz}, Neal and {Leja}, Joel and {Magdis}, Georgios E. and {Mowla}, Lamiya and {Nelson}, Erica J. and {Pope}, Alexandra and {Privon}, George C. and {Toft}, Sune and {Valentino}, Francesco},
        title = "{High Molecular-gas to Dust Mass Ratios Predicted in Most Quiescent Galaxies}",
      journal = {\apjl},
         year = 2021,
        month = dec,
       volume = {922},
       number = {2},
          eid = {L30},
        pages = {L30},
          doi = {10.3847/2041-8213/ac399f},
archivePrefix = {arXiv},
       eprint = {2111.05349},
 primaryClass = {astro-ph.GA},
       adsurl = {https://ui.adsabs.harvard.edu/abs/2021ApJ...922L..30W}
}

@ARTICLE{park24,
       author = {{Park}, Minjung and {Belli}, Sirio and {Conroy}, Charlie and {Johnson}, Benjamin D. and {Davies}, Rebecca L. and {Leja}, Joel and {Tacchella}, Sandro and {Mendel}, J. Trevor and {Benton}, Chlo{\"e} and {Bugiani}, Letizia and {Emami}, Razieh and {Khoram}, Amir H. and {Li}, Yijia and {Maheson}, Gabriel and {Mathews}, Elijah P. and {Naidu}, Rohan P. and {Nelson}, Erica J. and {Terrazas}, Bryan A. and {Weinberger}, Rainer},
        title = "{Widespread Rapid Quenching at Cosmic Noon Revealed by JWST Deep Spectroscopy}",
      journal = {\apj},
         year = 2024,
        month = nov,
       volume = {976},
       number = {1},
          eid = {72},
        pages = {72},
          doi = {10.3847/1538-4357/ad7e15},
archivePrefix = {arXiv},
       eprint = {2404.17945},
 primaryClass = {astro-ph.GA},
       adsurl = {https://ui.adsabs.harvard.edu/abs/2024ApJ...976...72P}
}

@ARTICLE{chang25,
       author = {{Chang}, Wenjun and {Wilson}, Gillian and {Forrest}, Ben and {McConachie}, Ian and {Webb}, Tracy and {Noble}, Allison and {Muzzin}, Adam and {Cooper}, M.~C. and {Marchesini}, Danilo and {Canalizo}, Gabriela and {Battisti}, A.~J. and {Henry}, Aurelien and {Gomez}, Percy and {Urbano Stawinski}, Stephanie M. and {Wisz}, M.~E.},
        title = "{MAGAZ3NE: Far-IR and Radio Insights into the Nature and Properties of Ultramassive Galaxies at z {\ensuremath{\gtrsim}} 3}",
      journal = {\apj},
         year = 2026,
        month = apr,
       volume = {1001},
       number = {2},
          eid = {131},
        pages = {131},
          doi = {10.3847/1538-4357/ae4ede},
archivePrefix = {arXiv},
       eprint = {2508.08460},
 primaryClass = {astro-ph.GA},
       adsurl = {https://ui.adsabs.harvard.edu/abs/2026ApJ..1001..131C}
}

@ARTICLE{scholtz25,
       author = {{Scholtz}, Jan and {D'Eugenio}, Francesco and {Maiolino}, Roberto and {P{\'e}rez-Gonz{\'a}lez}, Pablo G. and {Circosta}, Chiara and {Tacchella}, Sandro and {Williams}, Christina C. and {Alberts}, Stacey and {Arribas}, Santiago and {Baker}, William M. and {Bertola}, Elena and {Bunker}, Andrew J. and {Carniani}, Stefano and {Charlot}, Stephane and {Cresci}, Giovanni and {Jones}, Gareth C. and {Kumari}, Nimisha and {Lamperti}, Isabella and {Looser}, Tobias J. and {Pino}, Bruno Rodr{\'\i}guez Del and {Robertson}, Brant and {Parlanti}, Eleonora and {Perna}, Michele and {{\"U}bler}, Hannah and {Venturi}, Giacomo and {Witstok}, Joris},
        title = "{Measurement of the gas consumption history of a massive quiescent galaxy}",
      journal = {Nature Astronomy},
         year = 2026,
        month = mar,
       volume = {10},
        pages = {431-439},
          doi = {10.1038/s41550-025-02751-z},
archivePrefix = {arXiv},
       eprint = {2405.19401},
 primaryClass = {astro-ph.GA},
       adsurl = {https://ui.adsabs.harvard.edu/abs/2026NatAs..10..431S}
}

@ARTICLE{donevski23,
       author = {{Donevski}, D. and {Damjanov}, I. and {Nanni}, A. and {Man}, A. and {Giulietti}, M. and {Romano}, M. and {Lapi}, A. and {Narayanan}, D. and {Dav{\'e}}, R. and {Shivaei}, I. and {Sohn}, J. and {Junais} and {Pantoni}, L. and {Li}, Q.},
        title = "{In pursuit of giants. II. Evolution of dusty quiescent galaxies over the last six billion years from the hCOSMOS survey}",
      journal = {\aap},
         year = 2023,
        month = oct,
       volume = {678},
          eid = {A35},
        pages = {A35},
          doi = {10.1051/0004-6361/202346066},
archivePrefix = {arXiv},
       eprint = {2304.05842},
 primaryClass = {astro-ph.GA},
       adsurl = {https://ui.adsabs.harvard.edu/abs/2023A&A...678A..35D}
}

@ARTICLE{kakimoto24,
       author = {{Kakimoto}, Takumi and {Tanaka}, Masayuki and {Onodera}, Masato and {Shimakawa}, Rhythm and {Wu}, Po-Feng and {Gould}, Katriona M.~L. and {Ito}, Kei and {Jin}, Shuowen and {Kubo}, Mariko and {Suzuki}, Tomoko L. and {Toft}, Sune and {Valentino}, Francesco and {Yabe}, Kiyoto},
        title = "{A Massive Quiescent Galaxy in a Group Environment at z = 4.53}",
      journal = {\apj},
         year = 2024,
        month = mar,
       volume = {963},
       number = {1},
          eid = {49},
        pages = {49},
          doi = {10.3847/1538-4357/ad1ff1},
archivePrefix = {arXiv},
       eprint = {2308.15011},
 primaryClass = {astro-ph.GA},
       adsurl = {https://ui.adsabs.harvard.edu/abs/2024ApJ...963...49K}
}

@ARTICLE{lorenzon25,
       author = {{Lorenzon}, G. and {Donevski}, D. and {Lisiecki}, K. and {Lovell}, C. and {Romano}, M. and {Narayanan}, D. and {Dav{\'e}}, R. and {Man}, A. and {Whitaker}, K.~E. and {Nanni}, A. and {Long}, A. and {Lee}, M.~M. and {Junais} and {Ma{\l}ek}, K. and {Rodighiero}, G. and {Li}, Q.},
        title = "{Tracing the evolutionary pathways of dust and cold gas in high-z quiescent galaxies with SIMBA}",
      journal = {\aap},
         year = 2025,
        month = jan,
       volume = {693},
          eid = {A118},
        pages = {A118},
          doi = {10.1051/0004-6361/202450393},
archivePrefix = {arXiv},
       eprint = {2404.10568},
 primaryClass = {astro-ph.GA},
       adsurl = {https://ui.adsabs.harvard.edu/abs/2025A&A...693A.118L}
}

@ARTICLE{driver22,
       author = {{Driver}, Simon P. and {Bellstedt}, Sabine and {Robotham}, Aaron S.~G. and {Baldry}, Ivan K. and {Davies}, Luke J. and {Liske}, Jochen and {Obreschkow}, Danail and {Taylor}, Edward N. and {Wright}, Angus H. and {Alpaslan}, Mehmet and {Bamford}, Steven P. and {Bauer}, Amanda E. and {Bland-Hawthorn}, Joss and {Bilicki}, Maciej and {Bravo}, Mat{\'\i}as and {Brough}, Sarah and {Casura}, Sarah and {Cluver}, Michelle E. and {Colless}, Matthew and {Conselice}, Christopher J. and {Croom}, Scott M. and {de Jong}, Jelte and {D'Eugenio}, Franceso and {De Propris}, Roberto and {Dogruel}, Burak and {Drinkwater}, Michael J. and {Dvornik}, Andrej and {Farrow}, Daniel J. and {Frenk}, Carlos S. and {Giblin}, Benjamin and {Graham}, Alister W. and {Grootes}, Meiert W. and {Gunawardhana}, Madusha L.~P. and {Hashemizadeh}, Abdolhosein and {H{\"a}u{\ss}ler}, Boris and {Heymans}, Catherine and {Hildebrandt}, Hendrik and {Holwerda}, Benne W. and {Hopkins}, Andrew M. and {Jarrett}, Tom H. and {Heath Jones}, D. and {Kelvin}, Lee S. and {Koushan}, Soheil and {Kuijken}, Konrad and {Lara-L{\'o}pez}, Maritza A. and {Lange}, Rebecca and {L{\'o}pez-S{\'a}nchez}, {\'A}ngel R. and {Loveday}, Jon and {Mahajan}, Smriti and {Meyer}, Martin and {Moffett}, Amanda J. and {Napolitano}, Nicola R. and {Norberg}, Peder and {Owers}, Matt S. and {Radovich}, Mario and {Raouf}, Mojtaba and {Peacock}, John A. and {Phillipps}, Steven and {Pimbblet}, Kevin A. and {Popescu}, Cristina and {Said}, Khaled and {Sansom}, Anne E. and {Seibert}, Mark and {Sutherland}, Will J. and {Thorne}, Jessica E. and {Tuffs}, Richard J. and {Turner}, Ryan and {van der Wel}, Arjen and {van Kampen}, Eelco and {Wilkins}, Steve M.},
        title = "{Galaxy And Mass Assembly (GAMA): Data Release 4 and the z < 0.1 total and z < 0.08 morphological galaxy stellar mass functions}",
      journal = {\mnras},
         year = 2022,
        month = jun,
       volume = {513},
       number = {1},
        pages = {439-467},
          doi = {10.1093/mnras/stac472},
archivePrefix = {arXiv},
       eprint = {2203.08539},
 primaryClass = {astro-ph.GA},
       adsurl = {https://ui.adsabs.harvard.edu/abs/2022MNRAS.513..439D}
}

@ARTICLE{hardwick22,
       author = {{Hardwick}, Jennifer A. and {Cortese}, Luca and {Obreschkow}, Danail and {Catinella}, Barbara and {Cook}, Robin H.~W.},
        title = "{xGASS: characterizing the slope and scatter of the stellar mass-angular momentum relation for nearby galaxies}",
      journal = {\mnras},
         year = 2022,
        month = jan,
       volume = {509},
       number = {3},
        pages = {3751-3763},
          doi = {10.1093/mnras/stab3261},
archivePrefix = {arXiv},
       eprint = {2111.15048},
 primaryClass = {astro-ph.GA},
       adsurl = {https://ui.adsabs.harvard.edu/abs/2022MNRAS.509.3751H}
}

@ARTICLE{graham23,
       author = {{Graham}, Alister W. and {Sahu}, Nandini},
        title = "{Appreciating mergers for understanding the non-linear M$_{bh}$-M$_{*,spheroid}$ and M$_{bh}$-M$_{*, galaxy}$ relations, updated herein, and the implications for the (reduced) role of AGN feedback}",
      journal = {\mnras},
         year = 2023,
        month = jan,
       volume = {518},
       number = {2},
        pages = {2177-2200},
          doi = {10.1093/mnras/stac2019},
archivePrefix = {arXiv},
       eprint = {2209.14526},
 primaryClass = {astro-ph.GA},
       adsurl = {https://ui.adsabs.harvard.edu/abs/2023MNRAS.518.2177G}
}

@ARTICLE{schaye15,
       author = {{Schaye}, Joop and {Crain}, Robert A. and {Bower}, Richard G. and {Furlong}, Michelle and {Schaller}, Matthieu and {Theuns}, Tom and {Dalla Vecchia}, Claudio and {Frenk}, Carlos S. and {McCarthy}, I.~G. and {Helly}, John C. and {Jenkins}, Adrian and {Rosas-Guevara}, Y.~M. and {White}, Simon D.~M. and {Baes}, Maarten and {Booth}, C.~M. and {Camps}, Peter and {Navarro}, Julio F. and {Qu}, Yan and {Rahmati}, Alireza and {Sawala}, Till and {Thomas}, Peter A. and {Trayford}, James},
        title = "{The EAGLE project: simulating the evolution and assembly of galaxies and their environments}",
      journal = {\mnras},
         year = 2015,
        month = jan,
       volume = {446},
       number = {1},
        pages = {521-554},
          doi = {10.1093/mnras/stu2058},
archivePrefix = {arXiv},
       eprint = {1407.7040},
 primaryClass = {astro-ph.GA},
       adsurl = {https://ui.adsabs.harvard.edu/abs/2015MNRAS.446..521S}
}

@ARTICLE{pilepich18,
       author = {{Pillepich}, Annalisa and {Springel}, Volker and {Nelson}, Dylan and {Genel}, Shy and {Naiman}, Jill and {Pakmor}, R{\"u}diger and {Hernquist}, Lars and {Torrey}, Paul and {Vogelsberger}, Mark and {Weinberger}, Rainer and {Marinacci}, Federico},
        title = "{Simulating galaxy formation with the IllustrisTNG model}",
      journal = {\mnras},
         year = 2018,
        month = jan,
       volume = {473},
       number = {3},
        pages = {4077-4106},
          doi = {10.1093/mnras/stx2656},
archivePrefix = {arXiv},
       eprint = {1703.02970},
 primaryClass = {astro-ph.GA},
       adsurl = {https://ui.adsabs.harvard.edu/abs/2018MNRAS.473.4077P}
}

@ARTICLE{dave19,
       author = {{Dav{\'e}}, Romeel and {Angl{\'e}s-Alc{\'a}zar}, Daniel and {Narayanan}, Desika and {Li}, Qi and {Rafieferantsoa}, Mika H. and {Appleby}, Sarah},
        title = "{SIMBA: Cosmological simulations with black hole growth and feedback}",
      journal = {\mnras},
         year = 2019,
        month = jun,
       volume = {486},
       number = {2},
        pages = {2827-2849},
          doi = {10.1093/mnras/stz937},
archivePrefix = {arXiv},
       eprint = {1901.10203},
 primaryClass = {astro-ph.GA},
       adsurl = {https://ui.adsabs.harvard.edu/abs/2019MNRAS.486.2827D}
}

@ARTICLE{lagos24,
       author = {{Lagos}, Claudia del P. and {Bravo}, Mat{\'\i}as and {Tobar}, Rodrigo and {Obreschkow}, Danail and {Power}, Chris and {Robotham}, Aaron S.~G. and {Proctor}, Katy L. and {Hansen}, Samuel and {Chandro-G{\'o}mez}, {\'A}ngel and {Carrivick}, Julian},
        title = "{Quenching massive galaxies across cosmic time with the semi-analytic model SHARK V2.0}",
      journal = {\mnras},
         year = 2024,
        month = jul,
       volume = {531},
       number = {3},
        pages = {3551-3578},
          doi = {10.1093/mnras/stae1024},
archivePrefix = {arXiv},
       eprint = {2309.02310},
 primaryClass = {astro-ph.GA},
       adsurl = {https://ui.adsabs.harvard.edu/abs/2024MNRAS.531.3551L}
}

@ARTICLE{de-lucia24,
       author = {{De Lucia}, Gabriella and {Fontanot}, Fabio and {Xie}, Lizhi and {Hirschmann}, Michaela},
        title = "{Tracing the quenching journey across cosmic time}",
      journal = {\aap},
         year = 2024,
        month = jul,
       volume = {687},
          eid = {A68},
        pages = {A68},
          doi = {10.1051/0004-6361/202349045},
archivePrefix = {arXiv},
       eprint = {2401.06211},
 primaryClass = {astro-ph.GA},
       adsurl = {https://ui.adsabs.harvard.edu/abs/2024A&A...687A..68D}
}

@ARTICLE{lacey16,
       author = {{Lacey}, Cedric G. and {Baugh}, Carlton M. and {Frenk}, Carlos S. and {Benson}, Andrew J. and {Bower}, Richard G. and {Cole}, Shaun and {Gonzalez-Perez}, Violeta and {Helly}, John C. and {Lagos}, Claudia D.~P. and {Mitchell}, Peter D.},
        title = "{A unified multiwavelength model of galaxy formation}",
      journal = {\mnras},
         year = 2016,
        month = nov,
       volume = {462},
       number = {4},
        pages = {3854-3911},
          doi = {10.1093/mnras/stw1888},
archivePrefix = {arXiv},
       eprint = {1509.08473},
 primaryClass = {astro-ph.GA},
       adsurl = {https://ui.adsabs.harvard.edu/abs/2016MNRAS.462.3854L}
}

@ARTICLE{schaye25,
       author = {{Schaye}, Joop and {Chaikin}, Evgenii and {Schaller}, Matthieu and {Ploeckinger}, Sylvia and {Hu{\v{s}}ko}, Filip and {McGibbon}, Robert J. and {Trayford}, James W. and {Ben{\'\i}tez-Llambay}, Alejandro and {Correa}, Camila and {Frenk}, Carlos S. and {Richings}, Alexander J. and {Forouhar Moreno}, Victor J. and {Bah{\'e}}, Yannick M. and {Borrow}, Josh and {Durrant}, Anna and {Gebek}, Andrea and {Helly}, John C. and {Jenkins}, Adrian and {Lacey}, Cedric G. and {Ludlow}, Aaron and {Nobels}, Folkert S.~J.},
        title = "{The COLIBRE project: cosmological hydrodynamical simulations of galaxy formation and evolution}",
      journal = {\mnras},
         year = 2026,
        month = may,
       volume = {548},
       number = {1},
          eid = {stag375},
        pages = {stag375},
          doi = {10.1093/mnras/stag375},
archivePrefix = {arXiv},
       eprint = {2508.21126},
 primaryClass = {astro-ph.GA},
       adsurl = {https://ui.adsabs.harvard.edu/abs/2026MNRAS.548ag375S}
}

@ARTICLE{carnall18,
       author = {{Carnall}, A.~C. and {McLure}, R.~J. and {Dunlop}, J.~S. and {Dav{\'e}}, R.},
        title = "{Inferring the star formation histories of massive quiescent galaxies with BAGPIPES: evidence for multiple quenching mechanisms}",
      journal = {\mnras},
         year = 2018,
        month = nov,
       volume = {480},
       number = {4},
        pages = {4379-4401},
          doi = {10.1093/mnras/sty2169},
archivePrefix = {arXiv},
       eprint = {1712.04452},
 primaryClass = {astro-ph.GA},
       adsurl = {https://ui.adsabs.harvard.edu/abs/2018MNRAS.480.4379C}
}

@ARTICLE{kriek09,
       author = {{Kriek}, Mariska and {van Dokkum}, Pieter G. and {Labb{\'e}}, Ivo and {Franx}, Marijn and {Illingworth}, Garth D. and {Marchesini}, Danilo and {Quadri}, Ryan F.},
        title = "{An Ultra-Deep Near-Infrared Spectrum of a Compact Quiescent Galaxy at z = 2.2}",
      journal = {\apj},
         year = 2009,
        month = jul,
       volume = {700},
       number = {1},
        pages = {221-231},
          doi = {10.1088/0004-637X/700/1/221},
archivePrefix = {arXiv},
       eprint = {0905.1692},
 primaryClass = {astro-ph.CO},
       adsurl = {https://ui.adsabs.harvard.edu/abs/2009ApJ...700..221K}
}

@ARTICLE{dekel13,
       author = {{Dekel}, A. and {Zolotov}, A. and {Tweed}, D. and {Cacciato}, M. and {Ceverino}, D. and {Primack}, J.~R.},
        title = "{Toy models for galaxy formation versus simulations}",
      journal = {\mnras},
         year = 2013,
        month = oct,
       volume = {435},
       number = {2},
        pages = {999-1019},
          doi = {10.1093/mnras/stt1338},
archivePrefix = {arXiv},
       eprint = {1303.3009},
 primaryClass = {astro-ph.CO},
       adsurl = {https://ui.adsabs.harvard.edu/abs/2013MNRAS.435..999D}
}

@ARTICLE{leja19,
       author = {{Leja}, Joel and {Johnson}, Benjamin D. and {Conroy}, Charlie and {van Dokkum}, Pieter and {Speagle}, Joshua S. and {Brammer}, Gabriel and {Momcheva}, Ivelina and {Skelton}, Rosalind and {Whitaker}, Katherine E. and {Franx}, Marijn and {Nelson}, Erica J.},
        title = "{An Older, More Quiescent Universe from Panchromatic SED Fitting of the 3D-HST Survey}",
      journal = {\apj},
         year = 2019,
        month = jun,
       volume = {877},
       number = {2},
          eid = {140},
        pages = {140},
          doi = {10.3847/1538-4357/ab1d5a},
archivePrefix = {arXiv},
       eprint = {1812.05608},
 primaryClass = {astro-ph.GA},
       adsurl = {https://ui.adsabs.harvard.edu/abs/2019ApJ...877..140L}
}

@ARTICLE{leja19b,
       author = {{Leja}, Joel and {Carnall}, Adam C. and {Johnson}, Benjamin D. and {Conroy}, Charlie and {Speagle}, Joshua S.},
        title = "{How to Measure Galaxy Star Formation Histories. II. Nonparametric Models}",
      journal = {\apj},
         year = 2019,
        month = may,
       volume = {876},
       number = {1},
          eid = {3},
        pages = {3},
          doi = {10.3847/1538-4357/ab133c},
archivePrefix = {arXiv},
       eprint = {1811.03637},
 primaryClass = {astro-ph.GA},
       adsurl = {https://ui.adsabs.harvard.edu/abs/2019ApJ...876....3L}
}

@ARTICLE{paxton15,
       author = {{Paxton}, Bill and {Marchant}, Pablo and {Schwab}, Josiah and {Bauer}, Evan B. and {Bildsten}, Lars and {Cantiello}, Matteo and {Dessart}, Luc and {Farmer}, R. and {Hu}, H. and {Langer}, N. and {Townsend}, R.~H.~D. and {Townsley}, Dean M. and {Timmes}, F.~X.},
        title = "{Modules for Experiments in Stellar Astrophysics (MESA): Binaries, Pulsations, and Explosions}",
      journal = {\apjs},
         year = 2015,
        month = sep,
       volume = {220},
       number = {1},
          eid = {15},
        pages = {15},
          doi = {10.1088/0067-0049/220/1/15},
archivePrefix = {arXiv},
       eprint = {1506.03146},
 primaryClass = {astro-ph.SR},
       adsurl = {https://ui.adsabs.harvard.edu/abs/2015ApJS..220...15P}
}

@ARTICLE{chabrier03,
       author = {{Chabrier}, Gilles},
        title = "{Galactic Stellar and Substellar Initial Mass Function}",
      journal = {\pasp},
         year = 2003,
        month = jul,
       volume = {115},
       number = {809},
        pages = {763-795},
          doi = {10.1086/376392},
archivePrefix = {arXiv},
       eprint = {astro-ph/0304382},
 primaryClass = {astro-ph},
       adsurl = {https://ui.adsabs.harvard.edu/abs/2003PASP..115..763C}
}

@ARTICLE{calzetti00,
       author = {{Calzetti}, Daniela and {Armus}, Lee and {Bohlin}, Ralph C. and {Kinney}, Anne L. and {Koornneef}, Jan and {Storchi-Bergmann}, Thaisa},
        title = "{The Dust Content and Opacity of Actively Star-forming Galaxies}",
      journal = {\apj},
         year = 2000,
        month = apr,
       volume = {533},
       number = {2},
        pages = {682-695},
          doi = {10.1086/308692},
archivePrefix = {arXiv},
       eprint = {astro-ph/9911459},
 primaryClass = {astro-ph},
       adsurl = {https://ui.adsabs.harvard.edu/abs/2000ApJ...533..682C}
}

@ARTICLE{falcon-barroso11,
       author = {{Falc{\'o}n-Barroso}, J. and {S{\'a}nchez-Bl{\'a}zquez}, P. and {Vazdekis}, A. and {Ricciardelli}, E. and {Cardiel}, N. and {Cenarro}, A.~J. and {Gorgas}, J. and {Peletier}, R.~F.},
        title = "{An updated MILES stellar library and stellar population models}",
      journal = {\aap},
         year = 2011,
        month = aug,
       volume = {532},
          eid = {A95},
        pages = {A95},
          doi = {10.1051/0004-6361/201116842},
archivePrefix = {arXiv},
       eprint = {1107.2303},
 primaryClass = {astro-ph.CO},
       adsurl = {https://ui.adsabs.harvard.edu/abs/2011A&A...532A..95F}
}

@ARTICLE{johnson21,
       author = {{Johnson}, Benjamin D. and {Leja}, Joel and {Conroy}, Charlie and {Speagle}, Joshua S.},
        title = "{Stellar Population Inference with Prospector}",
      journal = {\apjs},
         year = 2021,
        month = jun,
       volume = {254},
       number = {2},
          eid = {22},
        pages = {22},
          doi = {10.3847/1538-4365/abef67},
archivePrefix = {arXiv},
       eprint = {2012.01426},
 primaryClass = {astro-ph.GA},
       adsurl = {https://ui.adsabs.harvard.edu/abs/2021ApJS..254...22J}
}

@ARTICLE{tacconi20,
       author = {{Tacconi}, Linda J. and {Genzel}, Reinhard and {Sternberg}, Amiel},
        title = "{The Evolution of the Star-Forming Interstellar Medium Across Cosmic Time}",
      journal = {\araa},
         year = 2020,
        month = aug,
       volume = {58},
        pages = {157-203},
          doi = {10.1146/annurev-astro-082812-141034},
archivePrefix = {arXiv},
       eprint = {2003.06245},
 primaryClass = {astro-ph.GA},
       adsurl = {https://ui.adsabs.harvard.edu/abs/2020ARA&A..58..157T}
}

@ARTICLE{spilker18,
       author = {{Spilker}, Justin and {Bezanson}, Rachel and {Bari{\v{s}}i{\'c}}, Ivana and {Bell}, Eric and {Lagos}, Claudia del P. and {Maseda}, Michael and {Muzzin}, Adam and {Pacifici}, Camilla and {Sobral}, David and {Straatman}, Caroline and {van der Wel}, Arjen and {van Dokkum}, Pieter and {Weiner}, Benjamin and {Whitaker}, Katherine and {Williams}, Christina C. and {Wu}, Po-Feng},
        title = "{Molecular Gas Contents and Scaling Relations for Massive, Passive Galaxies at Intermediate Redshifts from the LEGA-C Survey}",
      journal = {\apj},
         year = 2018,
        month = jun,
       volume = {860},
       number = {2},
          eid = {103},
        pages = {103},
          doi = {10.3847/1538-4357/aac438},
archivePrefix = {arXiv},
       eprint = {1805.02667},
 primaryClass = {astro-ph.GA},
       adsurl = {https://ui.adsabs.harvard.edu/abs/2018ApJ...860..103S}
}

@ARTICLE{magdis21,
       author = {{Magdis}, Georgios E. and {Gobat}, Raphael and {Valentino}, Francesco and {Daddi}, Emanuele and {Zanella}, Anita and {Kokorev}, Vasily and {Toft}, Sune and {Jin}, Shuowen and {Whitaker}, Katherine E.},
        title = "{The interstellar medium of quiescent galaxies and its evolution with time}",
      journal = {\aap},
         year = 2021,
        month = mar,
       volume = {647},
          eid = {A33},
        pages = {A33},
          doi = {10.1051/0004-6361/202039280},
archivePrefix = {arXiv},
       eprint = {2101.04700},
 primaryClass = {astro-ph.GA},
       adsurl = {https://ui.adsabs.harvard.edu/abs/2021A&A...647A..33M}
}

@ARTICLE{dwek98,
       author = {{Dwek}, Eli},
        title = "{The Evolution of the Elemental Abundances in the Gas and Dust Phases of the Galaxy}",
      journal = {\apj},
         year = 1998,
        month = jul,
       volume = {501},
        pages = {643},
          doi = {10.1086/305829},
archivePrefix = {arXiv},
       eprint = {astro-ph/9707024},
 primaryClass = {astro-ph},
       adsurl = {https://ui.adsabs.harvard.edu/abs/1998ApJ...501..643D}
}

@ARTICLE{adscheid25,
       author = {{Adscheid}, Sylvia and {Magnelli}, Benjamin and {Ciesla}, Laure and {Liu}, Daizhong and {Schinnerer}, Eva and {Bertoldi}, Frank},
        title = "{A$^{3}$COSMOS: The dust content of massive quiescent galaxies and its evolution with cosmic time}",
      journal = {\aap},
         year = 2025,
        month = oct,
       volume = {702},
          eid = {A186},
        pages = {A186},
          doi = {10.1051/0004-6361/202554400},
archivePrefix = {arXiv},
       eprint = {2508.18097},
 primaryClass = {astro-ph.GA},
       adsurl = {https://ui.adsabs.harvard.edu/abs/2025A&A...702A.186A}
}

@ARTICLE{muzzin25,
       author = {{Muzzin}, Adam and {Suess}, Katherine A. and {Marchesini}, Danilo and {Robbins}, Luke and {Willott}, Chris J. and {Alberts}, Stacey and {Antwi-Danso}, Jacqueline and {Asada}, Yoshihisa and {Brammer}, Gabriel and {Cutler}, Sam E. and {Iyer}, Kartheik G. and {Labbe}, Ivo and {Martis}, Nicholas S. and {Miller}, Tim B. and {Mitsuhashi}, Ikki and {Pope}, Alexandra and {Sajina}, Anna and {Sarrouh}, Ghassan T.~E. and {Sharma}, Monu and {Stefanon}, Mauro and {Whitaker}, Katherine E. and {Abraham}, Roberto and {Atek}, Hakim and {Bradac}, Marusa and {Berek}, Samantha and {Bezanson}, Rachel and {Brown}, Westley and {Burgasser}, Adam J. and {Chicoine}, Nathalie and {Cloonan}, Aidan P. and {Cooper}, Olivia R. and {Dayal}, Pratika and {de Graaff}, Anna and {Desprez}, Guillaume and {Feldmann}, Robert and {Forrest}, Ben and {Franx}, Marijn and {Fudamoto}, Yoshinobu and {Fujimoto}, Seiji and {Furtak}, Lukas J. and {Glazebrook}, Karl and {Goovaerts}, Ilias and {Greene}, Jenny E. and {Jagga}, Naadiyah and {Jarvis}, William W.~H. and {Kriek}, Mariska and {Khullar}, Gourav and {La Torre}, Valentina and {Leja}, Joel and {Lin}, Jamie and {Lorenz}, Brian and {Lyon}, Daniel and {Markov}, Vladan and {Maseda}, Michael V. and {McConachie}, Ian and {Merchant}, Maya and {Merida}, Rosa M. and {Mowla}, Lamiya and {Myers}, Katherine and {Naidu}, Rohan P. and {Nanayakkara}, Themiya and {Nelson}, Erica J. and {Noirot}, Gael and {Oesch}, Pascal A. and {Omori}, Kiyoaki C. and {Pan}, Richard and {Porraz Barrera}, Natalia and {Price}, Sedona H. and {Ravindranath}, Swara and {Sawicki}, Marcin and {Setton}, David J. and {Smit}, Renske and {Sok}, Visal and {Speagle}, Joshua S. and {Taylor}, Edward N. and {Tan}, Vivian Yun Yan and {Tripodi}, Roberta and {van der Wel}, Arjen and {Perez Vidal}, Edgar and {Wang}, Bingjie and {Weaver}, John R. and {Williams}, Christina C. and {Withers}, Sunna and {Zaidi}, Kumail},
        title = "{MINERVA: A NIRCam Medium Band and MIRI Imaging Survey to Unlock the Hidden Gems of the Distant Universe}",
      journal = {arXiv e-prints},
         year = 2025,
        month = jul,
          eid = {arXiv:2507.19706},
        pages = {arXiv:2507.19706},
          doi = {10.48550/arXiv.2507.19706},
archivePrefix = {arXiv},
       eprint = {2507.19706},
 primaryClass = {astro-ph.GA},
       adsurl = {https://ui.adsabs.harvard.edu/abs/2025arXiv250719706M}
}

@ARTICLE{faisst25,
       author = {{Faisst}, Andreas L. and {Chen}, Chian-Chou and {Ciesla}, Laure and {Gruppioni}, Carlotta},
        title = "{Understanding the evolution of black hole accretion and dust out to z = 4 with a deep imaging extragalactic survey with PRIMA}",
      journal = {Journal of Astronomical Telescopes, Instruments, and Systems},
         year = 2025,
        month = jul,
       volume = {11},
          eid = {031630},
        pages = {031630},
          doi = {10.1117/1.JATIS.11.3.031630},
archivePrefix = {arXiv},
       eprint = {2509.01674},
 primaryClass = {astro-ph.GA},
       adsurl = {https://ui.adsabs.harvard.edu/abs/2025JATIS..11c1630F}
}

@ARTICLE{kawinwanichakij25,
       author = {{Kawinwanichakij}, Lalitwadee and {Glazebrook}, Karl and {Nanayakkara}, Themiya and {Kacprzak}, Glenn G. and {Chittenden}, Harry George and {Jacobs}, Colin and {Chandro-G{\'o}mez}, {\'A}ngel and {Lagos}, Claudia and {Marchesini}, Danilo and {Mart{\`\i}nez-Mar{\`\i}n}, M. and {Oesch}, Pascal A. and {Remus}, Rhea-Silvia},
        title = "{Connecting Environment, Star Formation History, and Morphology of Massive Quiescent Galaxies at 3 < z < 4 with JWST}",
      journal = {\apj},
         year = 2026,
        month = jan,
       volume = {997},
       number = {1},
          eid = {29},
        pages = {29},
          doi = {10.3847/1538-4357/ae0a18},
archivePrefix = {arXiv},
       eprint = {2505.03089},
 primaryClass = {astro-ph.GA},
       adsurl = {https://ui.adsabs.harvard.edu/abs/2026ApJ...997...29K}
}

@ARTICLE{vanderwel14,
       author = {{van der Wel}, A. and {Franx}, M. and {van Dokkum}, P.~G. and {Skelton}, R.~E. and {Momcheva}, I.~G. and {Whitaker}, K.~E. and {Brammer}, G.~B. and {Bell}, E.~F. and {Rix}, H. -W. and {Wuyts}, S. and {Ferguson}, H.~C. and {Holden}, B.~P. and {Barro}, G. and {Koekemoer}, A.~M. and {Chang}, Yu-Yen and {McGrath}, E.~J. and {H{\"a}ussler}, B. and {Dekel}, A. and {Behroozi}, P. and {Fumagalli}, M. and {Leja}, J. and {Lundgren}, B.~F. and {Maseda}, M.~V. and {Nelson}, E.~J. and {Wake}, D.~A. and {Patel}, S.~G. and {Labb{\'e}}, I. and {Faber}, S.~M. and {Grogin}, N.~A. and {Kocevski}, D.~D.},
        title = "{3D-HST+CANDELS: The Evolution of the Galaxy Size-Mass Distribution since z = 3}",
      journal = {\apj},
         year = 2014,
        month = jun,
       volume = {788},
       number = {1},
          eid = {28},
        pages = {28},
          doi = {10.1088/0004-637X/788/1/28},
archivePrefix = {arXiv},
       eprint = {1404.2844},
 primaryClass = {astro-ph.GA},
       adsurl = {https://ui.adsabs.harvard.edu/abs/2014ApJ...788...28V}
}

@ARTICLE{cutler22,
       author = {{Cutler}, Sam E. and {Whitaker}, Katherine E. and {Mowla}, Lamiya A. and {Brammer}, Gabriel B. and {van der Wel}, Arjen and {Marchesini}, Danilo and {van Dokkum}, Pieter G. and {Momcheva}, Ivelina G. and {Song}, Mimi and {Akhshik}, Mohammad and {Nelson}, Erica J. and {Bezanson}, Rachel and {Franx}, Marijn and {Kriek}, Mariska and {Lange-Vagle}, Daniel and {Leja}, Joel and {MacKenty}, John W. and {Muzzin}, Adam and {Shipley}, Heath},
        title = "{Diagnosing DASH: A Catalog of Structural Properties for the COSMOS-DASH Survey}",
      journal = {\apj},
         year = 2022,
        month = jan,
       volume = {925},
       number = {1},
          eid = {34},
        pages = {34},
          doi = {10.3847/1538-4357/ac341c},
archivePrefix = {arXiv},
       eprint = {2111.14848},
 primaryClass = {astro-ph.GA},
       adsurl = {https://ui.adsabs.harvard.edu/abs/2022ApJ...925...34C}
}

@ARTICLE{wrightl24,
       author = {{Wright}, Lillian and {Whitaker}, Katherine E. and {Weaver}, John R. and {Cutler}, Sam E. and {Wang}, Bingjie and {Carnall}, Adam and {Suess}, Katherine A. and {Bezanson}, Rachel and {Nelson}, Erica and {Miller}, Tim B. and {Ito}, Kei and {Valentino}, Francesco},
        title = "{Remarkably Compact Quiescent Candidates at 3 < z < 5 in JWST-CEERS}",
      journal = {\apjl},
         year = 2024,
        month = mar,
       volume = {964},
       number = {1},
          eid = {L10},
        pages = {L10},
          doi = {10.3847/2041-8213/ad2b6d},
archivePrefix = {arXiv},
       eprint = {2311.05394},
 primaryClass = {astro-ph.GA},
       adsurl = {https://ui.adsabs.harvard.edu/abs/2024ApJ...964L..10W}
}

@ARTICLE{ito24,
       author = {{Ito}, Kei and {Valentino}, Francesco and {Brammer}, Gabriel and {Faisst}, Andreas L. and {Gillman}, Steven and {G{\'o}mez-Guijarro}, Carlos and {Gould}, Katriona M.~L. and {Heintz}, Kasper E. and {Ilbert}, Olivier and {Jespersen}, Christian Kragh and {Kokorev}, Vasily and {Kubo}, Mariko and {Magdis}, Georgios E. and {McPartland}, Conor J.~R. and {Onodera}, Masato and {Rizzo}, Francesca and {Tanaka}, Masayuki and {Toft}, Sune and {Vijayan}, Aswin P. and {Weaver}, John R. and {Whitaker}, Katherine E. and {Wright}, Lillian},
        title = "{Size{\textendash}Stellar Mass Relation and Morphology of Quiescent Galaxies at z {\ensuremath{\geq}} 3 in Public JWST Fields}",
      journal = {\apj},
         year = 2024,
        month = apr,
       volume = {964},
       number = {2},
          eid = {192},
        pages = {192},
          doi = {10.3847/1538-4357/ad2512},
archivePrefix = {arXiv},
       eprint = {2307.06994},
 primaryClass = {astro-ph.GA},
       adsurl = {https://ui.adsabs.harvard.edu/abs/2024ApJ...964..192I}
}

@ARTICLE{williams24,
       author = {{Williams}, Christina C. and {Alberts}, Stacey and {Ji}, Zhiyuan and {Hainline}, Kevin N. and {Lyu}, Jianwei and {Rieke}, George and {Endsley}, Ryan and {Suess}, Katherine A. and {Sun}, Fengwu and {Johnson}, Benjamin D. and {Florian}, Michael and {Shivaei}, Irene and {Rujopakarn}, Wiphu and {Baker}, William M. and {Bhatawdekar}, Rachana and {Boyett}, Kristan and {Bunker}, Andrew J. and {Cameron}, Alex J. and {Carniani}, Stefano and {Charlot}, Stephane and {Curtis-Lake}, Emma and {DeCoursey}, Christa and {de Graaff}, Anna and {Egami}, Eiichi and {Eisenstein}, Daniel J. and {Gibson}, Justus L. and {Hausen}, Ryan and {Helton}, Jakob M. and {Maiolino}, Roberto and {Maseda}, Michael V. and {Nelson}, Erica J. and {P{\'e}rez-Gonz{\'a}lez}, Pablo G. and {Rieke}, Marcia J. and {Robertson}, Brant E. and {Saxena}, Aayush and {Tacchella}, Sandro and {Willmer}, Christopher N.~A. and {Willott}, Chris J.},
        title = "{The Galaxies Missed by Hubble and ALMA: The Contribution of Extremely Red Galaxies to the Cosmic Census at 3 < z < 8}",
      journal = {\apj},
         year = 2024,
        month = jun,
       volume = {968},
       number = {1},
          eid = {34},
        pages = {34},
          doi = {10.3847/1538-4357/ad3f17},
archivePrefix = {arXiv},
       eprint = {2311.07483},
 primaryClass = {astro-ph.GA},
       adsurl = {https://ui.adsabs.harvard.edu/abs/2024ApJ...968...34W}
}

@ARTICLE{remus25,
       author = {{Remus}, Rhea-Silvia and {Kimmig}, Lucas C.},
        title = "{Relight the Candle: What Happens to High-redshift Massive Quenched Galaxies}",
      journal = {\apj},
         year = 2025,
        month = mar,
       volume = {982},
       number = {1},
          eid = {30},
        pages = {30},
          doi = {10.3847/1538-4357/ad8b4b},
archivePrefix = {arXiv},
       eprint = {2310.16089},
 primaryClass = {astro-ph.GA},
       adsurl = {https://ui.adsabs.harvard.edu/abs/2025ApJ...982...30R}
}

@ARTICLE{ormerod24,
       author = {{Ormerod}, K. and {Conselice}, C.~J. and {Adams}, N.~J. and {Harvey}, T. and {Austin}, D. and {Trussler}, J. and {Ferreira}, L. and {Caruana}, J. and {Lucatelli}, G. and {Li}, Q. and {Roper}, W.~J.},
        title = "{EPOCHS VI: the size and shape evolution of galaxies since z   8 with JWST Observations}",
      journal = {\mnras},
         year = 2024,
        month = jan,
       volume = {527},
       number = {3},
        pages = {6110-6125},
          doi = {10.1093/mnras/stad3597},
archivePrefix = {arXiv},
       eprint = {2309.04377},
 primaryClass = {astro-ph.GA},
       adsurl = {https://ui.adsabs.harvard.edu/abs/2024MNRAS.527.6110O}
}

@ARTICLE{ji24,
       author = {{Ji}, Zhiyuan and {Williams}, Christina C. and {Suess}, Katherine A. and {Tacchella}, Sandro and {Johnson}, Benjamin D. and {Robertson}, Brant and {Alberts}, Stacey and {Baker}, William M. and {Baum}, Stefi and {Bhatawdekar}, Rachana and {Bonaventura}, Nina and {Boyett}, Kristan and {Bunker}, Andrew J. and {Carniani}, Stefano and {Charlot}, Stephane and {Chen}, Zuyi and {Chevallard}, Jacopo and {Curtis-Lake}, Emma and {D'Eugenio}, Francesco and {de Graaff}, Anna and {DeCoursey}, Christa and {Egami}, Eiichi and {Eisenstein}, Daniel J. and {Hainline}, Kevin and {Hausen}, Ryan and {Helton}, Jakob M. and {Looser}, Tobias J. and {Lyu}, Jianwei and {Maiolino}, Roberto and {Maseda}, Michael V. and {Nelson}, Erica and {Rieke}, George and {Rieke}, Marcia and {Rix}, Hans-Walter and {Sandles}, Lester and {Sun}, Fengwu and {{\"U}bler}, Hannah and {Willmer}, Christopher N.~A. and {Willott}, Chris and {Witstok}, Joris},
        title = "{JADES: Rest-frame UV-to-NIR Size Evolution of Massive Quiescent Galaxies from Redshift z = 5 to z = 0.5}",
      journal = {\apj},
         year = 2026,
        month = feb,
       volume = {998},
       number = {2},
          eid = {239},
        pages = {239},
          doi = {10.3847/1538-4357/ae3b2a},
archivePrefix = {arXiv},
       eprint = {2401.00934},
 primaryClass = {astro-ph.GA},
       adsurl = {https://ui.adsabs.harvard.edu/abs/2026ApJ...998..239J}
}

@BOOK{sersic68,
       author = {{Sersic}, Jose Luis},
        title = "{Atlas de Galaxias Australes}",
         year = 1968,
       adsurl = {https://ui.adsabs.harvard.edu/abs/1968adga.book.....S}
}

@ARTICLE{haussler07,
       author = {{H{\"a}ussler}, Boris and {McIntosh}, Daniel H. and {Barden}, Marco and {Bell}, Eric F. and {Rix}, Hans-Walter and {Borch}, Andrea and {Beckwith}, Steven V.~W. and {Caldwell}, John A.~R. and {Heymans}, Catherine and {Jahnke}, Knud and {Jogee}, Shardha and {Koposov}, Sergey E. and {Meisenheimer}, Klaus and {S{\'a}nchez}, Sebastian F. and {Somerville}, Rachel S. and {Wisotzki}, Lutz and {Wolf}, Christian},
        title = "{GEMS: Galaxy Fitting Catalogs and Testing Parametric Galaxy Fitting Codes: GALFIT and GIM2D}",
      journal = {\apjs},
         year = 2007,
        month = oct,
       volume = {172},
       number = {2},
        pages = {615-633},
          doi = {10.1086/518836},
archivePrefix = {arXiv},
       eprint = {0704.2601},
 primaryClass = {astro-ph},
       adsurl = {https://ui.adsabs.harvard.edu/abs/2007ApJS..172..615H}
}

@ARTICLE{suess19,
       author = {{Suess}, Katherine A. and {Kriek}, Mariska and {Price}, Sedona H. and {Barro}, Guillermo},
        title = "{Half-mass Radii for {\ensuremath{\sim}}7000 Galaxies at 1.0 {\ensuremath{\leq}} z {\ensuremath{\leq}} 2.5: Most of the Evolution in the Mass-Size Relation Is Due to Color Gradients}",
      journal = {\apj},
         year = 2019,
        month = jun,
       volume = {877},
       number = {2},
          eid = {103},
        pages = {103},
          doi = {10.3847/1538-4357/ab1bda},
archivePrefix = {arXiv},
       eprint = {1904.10992},
 primaryClass = {astro-ph.GA},
       adsurl = {https://ui.adsabs.harvard.edu/abs/2019ApJ...877..103S}
}

@ARTICLE{bell01,
       author = {{Bell}, Eric F. and {de Jong}, Roelof S.},
        title = "{Stellar Mass-to-Light Ratios and the Tully-Fisher Relation}",
      journal = {\apj},
         year = 2001,
        month = mar,
       volume = {550},
       number = {1},
        pages = {212-229},
          doi = {10.1086/319728},
archivePrefix = {arXiv},
       eprint = {astro-ph/0011493},
 primaryClass = {astro-ph},
       adsurl = {https://ui.adsabs.harvard.edu/abs/2001ApJ...550..212B}
}

@ARTICLE{buitrago08,
       author = {{Buitrago}, Fernando and {Trujillo}, Ignacio and {Conselice}, Christopher J. and {Bouwens}, Rychard J. and {Dickinson}, Mark and {Yan}, Haojing},
        title = "{Size Evolution of the Most Massive Galaxies at 1.7 < z < 3 from GOODS NICMOS Survey Imaging}",
      journal = {\apjl},
         year = 2008,
        month = nov,
       volume = {687},
       number = {2},
        pages = {L61},
          doi = {10.1086/592836},
archivePrefix = {arXiv},
       eprint = {0807.4141},
 primaryClass = {astro-ph},
       adsurl = {https://ui.adsabs.harvard.edu/abs/2008ApJ...687L..61B}
}

@ARTICLE{vandokkum08,
       author = {{van Dokkum}, Pieter G. and {Franx}, Marijn and {Kriek}, Mariska and {Holden}, Bradford and {Illingworth}, Garth D. and {Magee}, Daniel and {Bouwens}, Rychard and {Marchesini}, Danilo and {Quadri}, Ryan and {Rudnick}, Greg and {Taylor}, Edward N. and {Toft}, Sune},
        title = "{Confirmation of the Remarkable Compactness of Massive Quiescent Galaxies at z \raisebox{-0.5ex}\textasciitilde 2.3: Early-Type Galaxies Did not Form in a Simple Monolithic Collapse}",
      journal = {\apjl},
         year = 2008,
        month = apr,
       volume = {677},
       number = {1},
        pages = {L5},
          doi = {10.1086/587874},
archivePrefix = {arXiv},
       eprint = {0802.4094},
 primaryClass = {astro-ph},
       adsurl = {https://ui.adsabs.harvard.edu/abs/2008ApJ...677L...5V}
}

@ARTICLE{degraaff25,
       author = {{de Graaff}, Anna and {Setton}, David J. and {Brammer}, Gabriel and {Cutler}, Sam and {Suess}, Katherine A. and {Labb{\'e}}, Ivo and {Leja}, Joel and {Weibel}, Andrea and {Maseda}, Michael V. and {Whitaker}, Katherine E. and {Bezanson}, Rachel and {Boogaard}, Leindert A. and {Cleri}, Nikko J. and {De Lucia}, Gabriella and {Franx}, Marijn and {Greene}, Jenny E. and {Hirschmann}, Michaela and {Matthee}, Jorryt and {McConachie}, Ian and {Naidu}, Rohan P. and {Oesch}, Pascal A. and {Price}, Sedona H. and {Rix}, Hans-Walter and {Valentino}, Francesco and {Wang}, Bingjie and {Williams}, Christina C.},
        title = "{Efficient formation of a massive quiescent galaxy at redshift 4.9}",
      journal = {Nature Astronomy},
         year = 2025,
        month = feb,
       volume = {9},
        pages = {280-292},
          doi = {10.1038/s41550-024-02424-3},
archivePrefix = {arXiv},
       eprint = {2404.05683},
 primaryClass = {astro-ph.GA},
       adsurl = {https://ui.adsabs.harvard.edu/abs/2025NatAs...9..280D}
}

@ARTICLE{vanderwel11,
       author = {{van der Wel}, Arjen and {Rix}, Hans-Walter and {Wuyts}, Stijn and {McGrath}, Elizabeth J. and {Koekemoer}, Anton M. and {Bell}, Eric F. and {Holden}, Bradford P. and {Robaina}, Aday R. and {McIntosh}, Daniel H.},
        title = "{The Majority of Compact Massive Galaxies at z \raisebox{-0.5ex}\textasciitilde 2 are Disk Dominated}",
      journal = {\apj},
         year = 2011,
        month = mar,
       volume = {730},
       number = {1},
          eid = {38},
        pages = {38},
          doi = {10.1088/0004-637X/730/1/38},
archivePrefix = {arXiv},
       eprint = {1101.2423},
 primaryClass = {astro-ph.CO},
       adsurl = {https://ui.adsabs.harvard.edu/abs/2011ApJ...730...38V}
}

@ARTICLE{newman18,
       author = {{Newman}, Andrew B. and {Belli}, Sirio and {Ellis}, Richard S. and {Patel}, Shannon G.},
        title = "{Resolving Quiescent Galaxies at z {\ensuremath{\gtrsim}} 2. II. Direct Measures of Rotational Support}",
      journal = {\apj},
         year = 2018,
        month = aug,
       volume = {862},
       number = {2},
          eid = {126},
        pages = {126},
          doi = {10.3847/1538-4357/aacd4f},
archivePrefix = {arXiv},
       eprint = {1806.06815},
 primaryClass = {astro-ph.GA},
       adsurl = {https://ui.adsabs.harvard.edu/abs/2018ApJ...862..126N}
}

@ARTICLE{forrest25,
       author = {{Forrest}, Ben and {Muzzin}, Adam and {Marchesini}, Danilo and {Pan}, Richard and {Ozden}, Nehir and {Antwi-Danso}, Jacqueline and {Chang}, Wenjun and {Cooper}, M.~C. and {Edward}, Adit H. and {Gomez}, Percy and {Kimmig}, Lucas and {Lemaux}, Brian C. and {McConachie}, Ian and {Noble}, Allison and {Remus}, Rhea-Silvia and {Urbano Stawinski}, Stephanie M. and {Wilson}, Gillian and {Wisz}, M.~E.},
        title = "{A massive and evolved slow-rotating galaxy in the early Universe}",
      journal = {Nature Astronomy},
         year = 2026,
        month = may,
          doi = {10.1038/s41550-026-02855-0},
archivePrefix = {arXiv},
       eprint = {2508.10987},
 primaryClass = {astro-ph.GA},
       adsurl = {https://ui.adsabs.harvard.edu/abs/2026NatAs.tmp...95F}
}

@ARTICLE{belli17,
       author = {{Belli}, Sirio and {Newman}, Andrew B. and {Ellis}, Richard S.},
        title = "{MOSFIRE Spectroscopy of Quiescent Galaxies at 1.5 < z < 2.5. I. Evolution of Structural and Dynamical Properties}",
      journal = {\apj},
         year = 2017,
        month = jan,
       volume = {834},
       number = {1},
          eid = {18},
        pages = {18},
          doi = {10.3847/1538-4357/834/1/18},
archivePrefix = {arXiv},
       eprint = {1608.00608},
 primaryClass = {astro-ph.GA},
       adsurl = {https://ui.adsabs.harvard.edu/abs/2017ApJ...834...18B}
}

@ARTICLE{lagos22,
       author = {{Lagos}, Claudia del P. and {Emsellem}, Eric and {van de Sande}, Jesse and {Harborne}, Katherine E. and {Cortese}, Luca and {Davison}, Thomas and {Foster}, Caroline and {Wright}, Ruby J.},
        title = "{The diverse nature and formation paths of slow rotator galaxies in the EAGLE simulations}",
      journal = {\mnras},
         year = 2022,
        month = jan,
       volume = {509},
       number = {3},
        pages = {4372-4391},
          doi = {10.1093/mnras/stab3128},
archivePrefix = {arXiv},
       eprint = {2012.08060},
 primaryClass = {astro-ph.GA},
       adsurl = {https://ui.adsabs.harvard.edu/abs/2022MNRAS.509.4372L}
}

@ARTICLE{lagos18,
       author = {{Lagos}, Claudia del P. and {Schaye}, Joop and {Bah{\'e}}, Yannick and {van de Sande}, Jesse and {Kay}, Scott T. and {Barnes}, David and {Davis}, Timothy A. and {Dalla Vecchia}, Claudio},
        title = "{The connection between mass, environment, and slow rotation in simulated galaxies}",
      journal = {\mnras},
         year = 2018,
        month = jun,
       volume = {476},
       number = {4},
        pages = {4327-4345},
          doi = {10.1093/mnras/sty489},
archivePrefix = {arXiv},
       eprint = {1712.01398},
 primaryClass = {astro-ph.GA},
       adsurl = {https://ui.adsabs.harvard.edu/abs/2018MNRAS.476.4327L}
}

@ARTICLE{harborne19,
       author = {{Harborne}, K.~E. and {Power}, C. and {Robotham}, A.~S.~G. and {Cortese}, L. and {Taranu}, D.~S.},
        title = "{A numerical twist on the spin parameter, {\ensuremath{\lambda}}$_{R}$}",
      journal = {\mnras},
         year = 2019,
        month = feb,
       volume = {483},
       number = {1},
        pages = {249-262},
          doi = {10.1093/mnras/sty3120},
archivePrefix = {arXiv},
       eprint = {1811.06148},
 primaryClass = {astro-ph.GA},
       adsurl = {https://ui.adsabs.harvard.edu/abs/2019MNRAS.483..249H}
}

@ARTICLE{somovigo25,
       author = {{Sommovigo}, Laura and {Algera}, Hiddo},
        title = "{Realistic multitemperature dust: how well can we constrain the dust properties of high-redshift galaxies?}",
      journal = {\mnras},
         year = 2025,
        month = jul,
       volume = {540},
       number = {4},
        pages = {3693-3708},
          doi = {10.1093/mnras/staf897},
archivePrefix = {arXiv},
       eprint = {2505.20105},
 primaryClass = {astro-ph.GA},
       adsurl = {https://ui.adsabs.harvard.edu/abs/2025MNRAS.540.3693S}
}

@ARTICLE{husko25,
       author = {{Hu{\v{s}}ko}, Filip and {Lacey}, Cedric G. and {Schaye}, Joop and {Schaller}, Matthieu and {Chaikin}, Evgenii and {Ploeckinger}, Sylvia and {Ben{\'\i}tez Llambay}, Alejandro and {Richings}, Alexander J. and {Trayford}, James W.},
        title = "{A hybrid active galactic nucleus feedback model with spinning black holes, winds and jets}",
      journal = {\mnras},
         year = 2026,
        month = apr,
       volume = {547},
       number = {2},
          eid = {stag324},
        pages = {stag324},
          doi = {10.1093/mnras/stag324},
archivePrefix = {arXiv},
       eprint = {2509.05179},
 primaryClass = {astro-ph.GA},
       adsurl = {https://ui.adsabs.harvard.edu/abs/2026MNRAS.547ag324H}
}

@ARTICLE{chaikin25,
       author = {{Chaikin}, Evgenii and {Schaye}, Joop and {Schaller}, Matthieu and {Ploeckinger}, Sylvia and {Bah{\'e}}, Yannick M. and {Ben{\'\i}tez-Llambay}, Alejandro and {Correa}, Camila and {Forouhar Moreno}, Victor J. and {Frenk}, Carlos S. and {Hu{\v{s}}ko}, Filip and {Kugel}, Roi and {McGibbon}, Robert and {Richings}, Alexander J. and {Trayford}, James W. and {Borrow}, Josh and {Crain}, Robert A. and {Helly}, John C. and {Lacey}, Cedric G. and {Ludlow}, Aaron and {Nobels}, Folkert S.~J.},
        title = "{COLIBRE: calibrating subgrid feedback in cosmological simulations that include a cold gas phase}",
      journal = {\mnras},
         year = 2026,
        month = may,
       volume = {548},
       number = {1},
          eid = {stag300},
        pages = {stag300},
          doi = {10.1093/mnras/stag300},
archivePrefix = {arXiv},
       eprint = {2509.04067},
 primaryClass = {astro-ph.GA},
       adsurl = {https://ui.adsabs.harvard.edu/abs/2026MNRAS.548ag300C}
}

@ARTICLE{stevenson25,
       author = {{Stevenson}, Struan D. and {Carnall}, Adam C. and {Leung}, Ho-Hin and {Taylor}, Elizabeth and {Cullen}, Fergus and {Dunlop}, James S. and {McLeod}, Derek J. and {McLure}, Ross J. and {Begley}, Ryan and {Arellano-C{\'o}rdova}, Karla Z. and {Barrufet}, Laia and {Bondestam}, Cecilia and {Donnan}, Callum T. and {Ellis}, Richard S. and {Grogin}, Norman A. and {Koekemoer}, Anton M. and {Liu}, Feng-Yuan and {P{\'e}rez-Gonz{\'a}lez}, Pablo G. and {Rowlands}, Kate and {Sanders}, Ryan L. and {Scholte}, Dirk and {Shapley}, Alice E. and {Skarbinski}, Maya and {Stanton}, Thomas M. and {Wild}, Vivienne},
        title = "{PRIMER and JADES reveal an abundance of massive quiescent galaxies at 2 < z < 5}",
      journal = {\mnras},
         year = 2026,
        month = jan,
       volume = {545},
       number = {3},
          eid = {staf2087},
        pages = {staf2087},
          doi = {10.1093/mnras/staf2087},
archivePrefix = {arXiv},
       eprint = {2509.06913},
 primaryClass = {astro-ph.GA},
       adsurl = {https://ui.adsabs.harvard.edu/abs/2026MNRAS.545f2087S}
}

@ARTICLE{kimmig25,
       author = {{Kimmig}, Lucas C. and {Remus}, Rhea-Silvia and {Seidel}, Benjamin and {Valenzuela}, Lucas M. and {Dolag}, Klaus and {Burkert}, Andreas},
        title = "{Blowing Out the Candle: How to Quench Galaxies at High Redshift{\textemdash}An Ensemble of Rapid Starbursts, AGN Feedback, and Environment}",
      journal = {\apj},
         year = 2025,
        month = jan,
       volume = {979},
       number = {1},
          eid = {15},
        pages = {15},
          doi = {10.3847/1538-4357/ad9472},
archivePrefix = {arXiv},
       eprint = {2310.16085},
 primaryClass = {astro-ph.GA},
       adsurl = {https://ui.adsabs.harvard.edu/abs/2025ApJ...979...15K}
}

@ARTICLE{chittenden25,
       author = {{Chittenden}, Harry George and {Glazebrook}, Karl and {Nanayakkara}, Themiya and {Kawinwanichakij}, Lalitwadee and {Lagos}, Claudia and {Kimmig}, Lucas and {Remus}, Rhea-Silvia},
        title = "{On the unique evolutionary mechanisms of massive quiescent galaxies in the epoch of reionization}",
      journal = {\mnras},
         year = 2026,
        month = apr,
       volume = {547},
       number = {4},
          eid = {stag474},
        pages = {stag474},
          doi = {10.1093/mnras/stag474},
archivePrefix = {arXiv},
       eprint = {2504.19696},
 primaryClass = {astro-ph.GA},
       adsurl = {https://ui.adsabs.harvard.edu/abs/2026MNRAS.547ag474C}
}

@ARTICLE{chaikin25b,
       author = {{Chaikin}, Evgenii and {Schaye}, Joop and {Schaller}, Matthieu and {Ploeckinger}, Sylvia and {Ben{\'\i}tez-Llambay}, Alejandro and {Frenk}, Carlos S. and {Hu{\v{s}}ko}, Filip and {McGibbon}, Robert J. and {Richings}, Alexander J. and {Trayford}, James W.},
        title = "{The evolution of the galaxy stellar mass function and star formation rates in the COLIBRE simulations from redshift 17 to 0}",
      journal = {\mnras},
         year = 2026,
        month = jun,
       volume = {548},
       number = {4},
          eid = {stag740},
        pages = {stag740},
          doi = {10.1093/mnras/stag740},
archivePrefix = {arXiv},
       eprint = {2509.07960},
 primaryClass = {astro-ph.GA},
       adsurl = {https://ui.adsabs.harvard.edu/abs/2026MNRAS.548ag740C}
}

@ARTICLE{suess22,
       author = {{Suess}, Katherine A. and {Leja}, Joel and {Johnson}, Benjamin D. and {Bezanson}, Rachel and {Greene}, Jenny E. and {Kriek}, Mariska and {Lower}, Sidney and {Narayanan}, Desika and {Setton}, David J. and {Spilker}, Justin S.},
        title = "{Recovering the Star Formation Histories of Recently Quenched Galaxies: The Impact of Model and Prior Choices}",
      journal = {\apj},
         year = 2022,
        month = aug,
       volume = {935},
       number = {2},
          eid = {146},
        pages = {146},
          doi = {10.3847/1538-4357/ac82b0},
archivePrefix = {arXiv},
       eprint = {2207.02883},
 primaryClass = {astro-ph.GA},
       adsurl = {https://ui.adsabs.harvard.edu/abs/2022ApJ...935..146S}
}

@ARTICLE{glazebrook24,
       author = {{Glazebrook}, Karl and {Nanayakkara}, Themiya and {Schreiber}, Corentin and {Lagos}, Claudia and {Kawinwanichakij}, Lalitwadee and {Jacobs}, Colin and {Chittenden}, Harry and {Brammer}, Gabriel and {Kacprzak}, Glenn G. and {Labbe}, Ivo and {Marchesini}, Danilo and {Marsan}, Z. Cemile and {Oesch}, Pascal A. and {Papovich}, Casey and {Remus}, Rhea-Silvia and {Tran}, Kim-Vy H. and {Esdaile}, James and {Chandro-Gomez}, Angel},
        title = "{A massive galaxy that formed its stars at z {\ensuremath{\approx}} 11}",
      journal = {\nat},
         year = 2024,
        month = apr,
       volume = {628},
       number = {8007},
        pages = {277-281},
          doi = {10.1038/s41586-024-07191-9},
archivePrefix = {arXiv},
       eprint = {2308.05606},
 primaryClass = {astro-ph.GA},
       adsurl = {https://ui.adsabs.harvard.edu/abs/2024Natur.628..277G}
}

@ARTICLE{kurinchi-vendhan24,
       author = {{Kurinchi-Vendhan}, Shalini and {Farcy}, Marion and {Hirschmann}, Michaela and {Valentino}, Francesco},
        title = "{On the origin of star formation quenching in massive galaxies at z {\ensuremath{\gtrsim}} 3 in the cosmological simulations IllustrisTNG}",
      journal = {\mnras},
         year = 2024,
        month = nov,
       volume = {534},
       number = {4},
        pages = {3974-3988},
          doi = {10.1093/mnras/stae2297},
archivePrefix = {arXiv},
       eprint = {2310.03083},
 primaryClass = {astro-ph.GA},
       adsurl = {https://ui.adsabs.harvard.edu/abs/2024MNRAS.534.3974K}
}

@ARTICLE{looser24,
       author = {{Looser}, Tobias J. and {D'Eugenio}, Francesco and {Maiolino}, Roberto and {Witstok}, Joris and {Sandles}, Lester and {Curtis-Lake}, Emma and {Chevallard}, Jacopo and {Tacchella}, Sandro and {Johnson}, Benjamin D. and {Baker}, William M. and {Suess}, Katherine A. and {Carniani}, Stefano and {Ferruit}, Pierre and {Arribas}, Santiago and {Bonaventura}, Nina and {Bunker}, Andrew J. and {Cameron}, Alex J. and {Charlot}, Stephane and {Curti}, Mirko and {de Graaff}, Anna and {Maseda}, Michael V. and {Rawle}, Tim and {Rix}, Hans-Walter and {Del Pino}, Bruno Rodr{\'\i}guez and {Smit}, Renske and {{\"U}bler}, Hannah and {Willott}, Chris and {Alberts}, Stacey and {Egami}, Eiichi and {Eisenstein}, Daniel J. and {Endsley}, Ryan and {Hausen}, Ryan and {Rieke}, Marcia and {Robertson}, Brant and {Shivaei}, Irene and {Williams}, Christina C. and {Boyett}, Kristan and {Chen}, Zuyi and {Ji}, Zhiyuan and {Jones}, Gareth C. and {Kumari}, Nimisha and {Nelson}, Erica and {Perna}, Michele and {Saxena}, Aayush and {Scholtz}, Jan},
        title = "{A recently quenched galaxy 700 million years after the Big Bang}",
      journal = {\nat},
         year = 2024,
        month = may,
       volume = {629},
       number = {8010},
        pages = {53-57},
          doi = {10.1038/s41586-024-07227-0},
archivePrefix = {arXiv},
       eprint = {2302.14155},
 primaryClass = {astro-ph.GA},
       adsurl = {https://ui.adsabs.harvard.edu/abs/2024Natur.629...53L}
}

@ARTICLE{szpila25,
       author = {{Szpila}, Jakub and {Dav{\'e}}, Romeel and {Rennehan}, Douglas and {Cui}, Weiguang and {Hough}, Renier T.},
        title = "{The nature and evolution of early massive quenched galaxies in the SIMBA-C simulation}",
      journal = {\mnras},
         year = 2025,
        month = feb,
       volume = {537},
       number = {2},
        pages = {1849-1868},
          doi = {10.1093/mnras/staf132},
archivePrefix = {arXiv},
       eprint = {2402.08729},
 primaryClass = {astro-ph.GA},
       adsurl = {https://ui.adsabs.harvard.edu/abs/2025MNRAS.537.1849S}
}

@ARTICLE{weller25,
       author = {{Weller}, Emma Jane and {Pacucci}, Fabio and {Ni}, Yueying and {Hernquist}, Lars and {Park}, Minjung},
        title = "{Discrepancies between JWST Observations and Simulations of Quenched Massive Galaxies at z > 3: A Comparative Study with IllustrisTNG and ASTRID}",
      journal = {\apj},
         year = 2025,
        month = feb,
       volume = {979},
       number = {2},
          eid = {181},
        pages = {181},
          doi = {10.3847/1538-4357/ada360},
archivePrefix = {arXiv},
       eprint = {2406.02664},
 primaryClass = {astro-ph.GA},
       adsurl = {https://ui.adsabs.harvard.edu/abs/2025ApJ...979..181W}
}

@ARTICLE{baker25c,
       author = {{Baker}, William M. and {Ito}, Kei and {Valentino}, Francesco and {Zhu}, Pengpei and {Scarpe}, Gianluca and {Gottumukkala}, Rashmi and {Hjorth}, Jens and {Barrufet}, Laia and {Langeroodi}, Danial},
        title = "{Double trouble: Two spectroscopically confirmed low-mass quiescent galaxies at z > 5 in overdensities}",
      journal = {\aap},
         year = 2026,
        month = feb,
       volume = {706},
          eid = {A91},
        pages = {A91},
          doi = {10.1051/0004-6361/202557207},
archivePrefix = {arXiv},
       eprint = {2509.09761},
 primaryClass = {astro-ph.GA},
       adsurl = {https://ui.adsabs.harvard.edu/abs/2026A&A...706A..91B}
}

@ARTICLE{jin24,
       author = {{Jin}, Shuowen and {Sillassen}, Nikolaj B. and {Magdis}, Georgios E. and {Brinch}, Malte and {Shuntov}, Marko and {Brammer}, Gabriel and {Gobat}, Raphael and {Valentino}, Francesco and {Carnall}, Adam C. and {Lee}, Minju and {Vijayan}, Aswin P. and {Gillman}, Steven and {Kokorev}, Vasily and {Le Bail}, Aur{\'e}lien and {Greve}, Thomas R. and {Gullberg}, Bitten and {Gould}, Katriona M.~L. and {Toft}, Sune},
        title = "{Cosmic Vine: A z = 3.44 large-scale structure hosting massive quiescent galaxies}",
      journal = {\aap},
         year = 2024,
        month = mar,
       volume = {683},
          eid = {L4},
        pages = {L4},
          doi = {10.1051/0004-6361/202348540},
archivePrefix = {arXiv},
       eprint = {2311.04867},
 primaryClass = {astro-ph.GA},
       adsurl = {https://ui.adsabs.harvard.edu/abs/2024A&A...683L...4J}
}

@ARTICLE{adamo25,
       author = {{Adamo}, Angela and {Atek}, Hakim and {Bagley}, Micaela B. and {Ba{\~n}ados}, Eduardo and {Barrow}, Kirk S.~S. and {Berg}, Danielle A. and {Bezanson}, Rachel and {Brada{\v{c}}}, Maru{\v{s}}a and {Brammer}, Gabriel and {Carnall}, Adam C. and {Chisholm}, John and {Coe}, Dan and {Dayal}, Pratika and {Eisenstein}, Daniel J. and {Eldridge}, Jan J. and {Ferrara}, Andrea and {Fujimoto}, Seiji and {Graaff}, Anna de and {Habouzit}, Melanie and {Hutchison}, Taylor A. and {Kartaltepe}, Jeyhan S. and {Kassin}, Susan A. and {Kriek}, Mariska and {Labb{\'e}}, Ivo and {Maiolino}, Roberto and {Marques-Chaves}, Rui and {Maseda}, Michael V. and {Mason}, Charlotte and {Matthee}, Jorryt and {McQuinn}, Kristen B.~W. and {Meynet}, Georges and {Naidu}, Rohan P. and {Oesch}, Pascal A. and {Pentericci}, Laura and {P{\'e}rez-Gonz{\'a}lez}, Pablo G. and {Rigby}, Jane R. and {Roberts-Borsani}, Guido and {Schaerer}, Daniel and {Shapley}, Alice E. and {Stark}, Daniel P. and {Stiavelli}, Massimo and {Strom}, Allison L. and {Vanzella}, Eros and {Wang}, Feige and {Wilkins}, Stephen M. and {Williams}, Christina C. and {Willott}, Chris J. and {Wylezalek}, Dominika and {Nota}, Antonella},
        title = "{The first billion years according to JWST}",
      journal = {Nature Astronomy},
         year = 2025,
        month = aug,
       volume = {9},
        pages = {1134-1147},
          doi = {10.1038/s41550-025-02624-5},
archivePrefix = {arXiv},
       eprint = {2405.21054},
 primaryClass = {astro-ph.GA},
       adsurl = {https://ui.adsabs.harvard.edu/abs/2025NatAs...9.1134A}
}

@ARTICLE{nanayakkara22,
       author = {{Nanayakkara}, Themiya and {Esdaile}, James and {Glazebrook}, Karl and {Espejo Salcedo}, Juan M. and {Durre}, Mark and {Jacobs}, Colin},
        title = "{Massive high-redshift quiescent galaxies with JWST}",
      journal = {\pasa},
         year = 2022,
        month = jan,
       volume = {39},
          eid = {e002},
        pages = {e002},
          doi = {10.1017/pasa.2021.61},
archivePrefix = {arXiv},
       eprint = {2103.01459},
 primaryClass = {astro-ph.GA},
       adsurl = {https://ui.adsabs.harvard.edu/abs/2022PASA...39....2N}
}

@ARTICLE{glazebrook17,
       author = {{Glazebrook}, Karl and {Schreiber}, Corentin and {Labb{\'e}}, Ivo and {Nanayakkara}, Themiya and {Kacprzak}, Glenn G. and {Oesch}, Pascal A. and {Papovich}, Casey and {Spitler}, Lee R. and {Straatman}, Caroline M.~S. and {Tran}, Kim-Vy H. and {Yuan}, Tiantian},
        title = "{A massive, quiescent galaxy at a redshift of 3.717}",
      journal = {\nat},
         year = 2017,
        month = apr,
       volume = {544},
       number = {7648},
        pages = {71-74},
          doi = {10.1038/nature21680},
archivePrefix = {arXiv},
       eprint = {1702.01751},
 primaryClass = {astro-ph.GA},
       adsurl = {https://ui.adsabs.harvard.edu/abs/2017Natur.544...71G}
}

@ARTICLE{weaver22,
       author = {{Weaver}, J.~R. and {Kauffmann}, O.~B. and {Ilbert}, O. and {McCracken}, H.~J. and {Moneti}, A. and {Toft}, S. and {Brammer}, G. and {Shuntov}, M. and {Davidzon}, I. and {Hsieh}, B.~C. and {Laigle}, C. and {Anastasiou}, A. and {Jespersen}, C.~K. and {Vinther}, J. and {Capak}, P. and {Casey}, C.~M. and {McPartland}, C.~J.~R. and {Milvang-Jensen}, B. and {Mobasher}, B. and {Sanders}, D.~B. and {Zalesky}, L. and {Arnouts}, S. and {Aussel}, H. and {Dunlop}, J.~S. and {Faisst}, A. and {Franx}, M. and {Furtak}, L.~J. and {Fynbo}, J.~P.~U. and {Gould}, K.~M.~L. and {Greve}, T.~R. and {Gwyn}, S. and {Kartaltepe}, J.~S. and {Kashino}, D. and {Koekemoer}, A.~M. and {Kokorev}, V. and {Le F{\`e}vre}, O. and {Lilly}, S. and {Masters}, D. and {Magdis}, G. and {Mehta}, V. and {Peng}, Y. and {Riechers}, D.~A. and {Salvato}, M. and {Sawicki}, M. and {Scarlata}, C. and {Scoville}, N. and {Shirley}, R. and {Silverman}, J.~D. and {Sneppen}, A. and {Smolc̆i{\'c}}, V. and {Steinhardt}, C. and {Stern}, D. and {Tanaka}, M. and {Taniguchi}, Y. and {Teplitz}, H.~I. and {Vaccari}, M. and {Wang}, W. -H. and {Zamorani}, G.},
        title = "{COSMOS2020: A Panchromatic View of the Universe to z{\ensuremath{\sim}}10 from Two Complementary Catalogs}",
      journal = {\apjs},
         year = 2022,
        month = jan,
       volume = {258},
       number = {1},
          eid = {11},
        pages = {11},
          doi = {10.3847/1538-4365/ac3078},
archivePrefix = {arXiv},
       eprint = {2110.13923},
 primaryClass = {astro-ph.GA},
       adsurl = {https://ui.adsabs.harvard.edu/abs/2022ApJS..258...11W}
}

@ARTICLE{boylan-kolchin23,
       author = {{Boylan-Kolchin}, Michael},
        title = "{Stress testing {\ensuremath{\Lambda}}CDM with high-redshift galaxy candidates}",
      journal = {Nature Astronomy},
         year = 2023,
        month = jun,
       volume = {7},
        pages = {731-735},
          doi = {10.1038/s41550-023-01937-7},
archivePrefix = {arXiv},
       eprint = {2208.01611},
 primaryClass = {astro-ph.CO},
       adsurl = {https://ui.adsabs.harvard.edu/abs/2023NatAs...7..731B}
}

@ARTICLE{forrest20,
       author = {{Forrest}, Ben and {Marsan}, Z. Cemile and {Annunziatella}, Marianna and {Wilson}, Gillian and {Muzzin}, Adam and {Marchesini}, Danilo and {Cooper}, M.~C. and {Chan}, Jeffrey C.~C. and {McConachie}, Ian and {Gomez}, Percy and {Kado-Fong}, Erin and {La Barbera}, Francesco and {Lange-Vagle}, Daniel and {Nantais}, Julie and {Nonino}, Mario and {Saracco}, Paolo and {Stefanon}, Mauro and {van der Burg}, Remco F.~J.},
        title = "{The Massive Ancient Galaxies at z > 3 NEar-infrared (MAGAZ3NE) Survey: Confirmation of Extremely Rapid Star Formation and Quenching Timescales for Massive Galaxies in the Early Universe}",
      journal = {\apj},
         year = 2020,
        month = nov,
       volume = {903},
       number = {1},
          eid = {47},
        pages = {47},
          doi = {10.3847/1538-4357/abb819},
archivePrefix = {arXiv},
       eprint = {2009.07281},
 primaryClass = {astro-ph.GA},
       adsurl = {https://ui.adsabs.harvard.edu/abs/2020ApJ...903...47F}
}

@ARTICLE{gelli23,
       author = {{Gelli}, Viola and {Salvadori}, Stefania and {Ferrara}, Andrea and {Pallottini}, Andrea and {Carniani}, Stefano},
        title = "{Quiescent Low-mass Galaxies Observed by JWST in the Epoch of Reionization}",
      journal = {\apjl},
         year = 2023,
        month = sep,
       volume = {954},
       number = {1},
          eid = {L11},
        pages = {L11},
          doi = {10.3847/2041-8213/acee80},
archivePrefix = {arXiv},
       eprint = {2303.13574},
 primaryClass = {astro-ph.GA},
       adsurl = {https://ui.adsabs.harvard.edu/abs/2023ApJ...954L..11G}
}

@ARTICLE{benson03,
       author = {{Benson}, A.~J. and {Bower}, R.~G. and {Frenk}, C.~S. and {Lacey}, C.~G. and {Baugh}, C.~M. and {Cole}, S.},
        title = "{What Shapes the Luminosity Function of Galaxies?}",
      journal = {\apj},
         year = 2003,
        month = dec,
       volume = {599},
       number = {1},
        pages = {38-49},
          doi = {10.1086/379160},
archivePrefix = {arXiv},
       eprint = {astro-ph/0302450},
 primaryClass = {astro-ph},
       adsurl = {https://ui.adsabs.harvard.edu/abs/2003ApJ...599...38B}
}

@ARTICLE{man18,
       author = {{Man}, Allison and {Belli}, Sirio},
        title = "{Star formation quenching in massive galaxies}",
      journal = {Nature Astronomy},
         year = 2018,
        month = sep,
       volume = {2},
        pages = {695-697},
          doi = {10.1038/s41550-018-0558-1},
archivePrefix = {arXiv},
       eprint = {1809.00722},
 primaryClass = {astro-ph.GA},
       adsurl = {https://ui.adsabs.harvard.edu/abs/2018NatAs...2..695M}
}

@ARTICLE{somerville15,
       author = {{Somerville}, Rachel S. and {Dav{\'e}}, Romeel},
        title = "{Physical Models of Galaxy Formation in a Cosmological Framework}",
      journal = {\araa},
         year = 2015,
        month = aug,
       volume = {53},
        pages = {51-113},
          doi = {10.1146/annurev-astro-082812-140951},
archivePrefix = {arXiv},
       eprint = {1412.2712},
 primaryClass = {astro-ph.GA},
       adsurl = {https://ui.adsabs.harvard.edu/abs/2015ARA&A..53...51S}
}

@ARTICLE{crain15,
       author = {{Crain}, Robert A. and {Schaye}, Joop and {Bower}, Richard G. and {Furlong}, Michelle and {Schaller}, Matthieu and {Theuns}, Tom and {Dalla Vecchia}, Claudio and {Frenk}, Carlos S. and {McCarthy}, Ian G. and {Helly}, John C. and {Jenkins}, Adrian and {Rosas-Guevara}, Yetli M. and {White}, Simon D.~M. and {Trayford}, James W.},
        title = "{The EAGLE simulations of galaxy formation: calibration of subgrid physics and model variations}",
      journal = {\mnras},
         year = 2015,
        month = jun,
       volume = {450},
       number = {2},
        pages = {1937-1961},
          doi = {10.1093/mnras/stv725},
archivePrefix = {arXiv},
       eprint = {1501.01311},
 primaryClass = {astro-ph.GA},
       adsurl = {https://ui.adsabs.harvard.edu/abs/2015MNRAS.450.1937C}
}

@ARTICLE{Lovell23,
       author = {{Lovell}, Christopher C. and {Roper}, Will and {Vijayan}, Aswin P. and {Seeyave}, Louise and {Irodotou}, Dimitrios and {Wilkins}, Stephen M. and {Conselice}, Christopher J. and {Fortuni}, Flaminia and {Kuusisto}, Jussi K. and {Merlin}, Emiliano and {Santini}, Paola and {Thomas}, Peter},
        title = "{First light and reionisation epoch simulations (FLARES) - VIII. The emergence of passive galaxies at z {\ensuremath{\geq}} 5}",
      journal = {\mnras},
         year = 2023,
        month = nov,
       volume = {525},
       number = {4},
        pages = {5520-5539},
          doi = {10.1093/mnras/stad2550},
archivePrefix = {arXiv},
       eprint = {2211.07540},
 primaryClass = {astro-ph.GA},
       adsurl = {https://ui.adsabs.harvard.edu/abs/2023MNRAS.525.5520L}
}

@ARTICLE{hartley23,
       author = {{Hartley}, Abigail I. and {Nelson}, Erica J. and {Suess}, Katherine A. and {Garcia}, Alex M. and {Park}, Minjung and {Hernquist}, Lars and {Bezanson}, Rachel and {Nevin}, Rebecca and {Pillepich}, Annalisa and {Schechter}, Aimee L. and {Terrazas}, Bryan A. and {Torrey}, Paul and {Wellons}, Sarah and {Whitaker}, Katherine E. and {Williams}, Christina C.},
        title = "{The first quiescent galaxies in TNG300}",
      journal = {\mnras},
         year = 2023,
        month = jun,
       volume = {522},
       number = {2},
        pages = {3138-3144},
          doi = {10.1093/mnras/stad1162},
archivePrefix = {arXiv},
       eprint = {2304.09392},
 primaryClass = {astro-ph.GA},
       adsurl = {https://ui.adsabs.harvard.edu/abs/2023MNRAS.522.3138H}
}

@ARTICLE{lee15,
       author = {{Lee}, Seong-Kook and {Im}, Myungshin and {Kim}, Jae-Woo and {Lotz}, Jennifer and {McPartland}, Conor and {Peth}, Michael and {Koekemoer}, Anton},
        title = "{Evolution of Star-formation Properties of High-redshift Cluster Galaxies since z = 2}",
      journal = {\apj},
         year = 2015,
        month = sep,
       volume = {810},
       number = {2},
          eid = {90},
        pages = {90},
          doi = {10.1088/0004-637X/810/2/90},
archivePrefix = {arXiv},
       eprint = {1508.01294},
 primaryClass = {astro-ph.GA},
       adsurl = {https://ui.adsabs.harvard.edu/abs/2015ApJ...810...90L}
}

@ARTICLE{wendland95,
       author = {{Wendland}, Holger},
        title = "{Piecewise polynomial, positive definite and compactly supported radial functions of minimal degree}",
      journal = {Advances in Computational Mathematics},
        year = 1995,
        month = dec,
       volume = {4},
        pages = {8},
          doi = {10.1007/BF02123482},
}

@ARTICLE{ludlow23,
       author = {{Ludlow}, Aaron D. and {Fall}, S. Michael and {Wilkinson}, Matthew J. and {Schaye}, Joop and {Obreschkow}, Danail},
        title = "{Spurious heating of stellar motions by dark matter particles in cosmological simulations of galaxy formation}",
      journal = {\mnras},
         year = 2023,
        month = nov,
       volume = {525},
       number = {4},
        pages = {5614-5630},
          doi = {10.1093/mnras/stad2615},
archivePrefix = {arXiv},
       eprint = {2306.05753},
 primaryClass = {astro-ph.GA},
       adsurl = {https://ui.adsabs.harvard.edu/abs/2023MNRAS.525.5614L}
}

@ARTICLE{benitez-llambay25,
       author = {{Ben{\'\i}tez-Llambay}, Alejandro and {Ploeckinger}, Sylvia and {Schaye}, Joop and {Richings}, Alexander J. and {Chaikin}, Evgenii and {Schaller}, Matthieu and {Trayford}, James W. and {Frenk}, Carlos S. and {Hu{\v{s}}ko}, Filip and {Correa}, Camila},
        title = "{Non-explosive pre-supernova feedback in the COLIBRE model of galaxy formation}",
      journal = {\mnras},
         year = 2026,
        month = mar,
       volume = {546},
       number = {4},
          eid = {stag268},
        pages = {stag268},
          doi = {10.1093/mnras/stag268},
archivePrefix = {arXiv},
       eprint = {2509.25309},
 primaryClass = {astro-ph.GA},
       adsurl = {https://ui.adsabs.harvard.edu/abs/2026MNRAS.546ag268B}
}

@ARTICLE{valentino20,
       author = {{Valentino}, Francesco and {Tanaka}, Masayuki and {Davidzon}, Iary and {Toft}, Sune and {G{\'o}mez-Guijarro}, Carlos and {Stockmann}, Mikkel and {Onodera}, Masato and {Brammer}, Gabriel and {Ceverino}, Daniel and {Faisst}, Andreas L. and {Gallazzi}, Anna and {Hayward}, Christopher C. and {Ilbert}, Olivier and {Kubo}, Mariko and {Magdis}, Georgios E. and {Selsing}, Jonatan and {Shimakawa}, Rhythm and {Sparre}, Martin and {Steinhardt}, Charles and {Yabe}, Kiyoto and {Zabl}, Johannes},
        title = "{Quiescent Galaxies 1.5 Billion Years after the Big Bang and Their Progenitors}",
      journal = {\apj},
         year = 2020,
        month = feb,
       volume = {889},
       number = {2},
          eid = {93},
        pages = {93},
          doi = {10.3847/1538-4357/ab64dc},
archivePrefix = {arXiv},
       eprint = {1909.10540},
 primaryClass = {astro-ph.GA},
       adsurl = {https://ui.adsabs.harvard.edu/abs/2020ApJ...889...93V}
}

@ARTICLE{merlin18,
       author = {{Merlin}, E. and {Fontana}, A. and {Castellano}, M. and {Santini}, P. and {Torelli}, M. and {Boutsia}, K. and {Wang}, T. and {Grazian}, A. and {Pentericci}, L. and {Schreiber}, C. and {Ciesla}, L. and {McLure}, R. and {Derriere}, S. and {Dunlop}, J.~S. and {Elbaz}, D.},
        title = "{Chasing passive galaxies in the early Universe: a critical analysis in CANDELS GOODS-South}",
      journal = {\mnras},
         year = 2018,
        month = jan,
       volume = {473},
       number = {2},
        pages = {2098-2123},
          doi = {10.1093/mnras/stx2385},
archivePrefix = {arXiv},
       eprint = {1709.00429},
 primaryClass = {astro-ph.GA},
       adsurl = {https://ui.adsabs.harvard.edu/abs/2018MNRAS.473.2098M}
}

@ARTICLE{strait23,
       author = {{Strait}, Victoria and {Brammer}, Gabriel and {Muzzin}, Adam and {Desprez}, Guillaume and {Asada}, Yoshihisa and {Abraham}, Roberto and {Brada{\v{c}}}, Maru{\v{s}}a and {Iyer}, Kartheik G. and {Martis}, Nicholas and {Mowla}, Lamiya and {Noirot}, Ga{\"e}l and {Sarrouh}, Ghassan T.~E. and {Sawicki}, Marcin and {Willott}, Chris and {Gould}, Katriona and {Grindlay}, Tess and {Matharu}, Jasleen and {Rihtar{\v{s}}i{\v{c}}}, Gregor},
        title = "{An Extremely Compact, Low-mass Galaxy on its Way to Quiescence at z = 5.2}",
      journal = {\apjl},
         year = 2023,
        month = jun,
       volume = {949},
       number = {2},
          eid = {L23},
        pages = {L23},
          doi = {10.3847/2041-8213/acd457},
archivePrefix = {arXiv},
       eprint = {2303.11349},
 primaryClass = {astro-ph.GA},
       adsurl = {https://ui.adsabs.harvard.edu/abs/2023ApJ...949L..23S}
}

@ARTICLE{baker25d,
       author = {{Baker}, William M. and {D'Eugenio}, Francesco and {Maiolino}, Roberto and {Bunker}, Andrew J. and {Simmonds}, Charlotte and {Tacchella}, Sandro and {Witstok}, Joris and {Arribas}, Santiago and {Carniani}, Stefano and {Charlot}, St{\'e}phane and {Chevallard}, Jacopo and {Curti}, Mirko and {Curtis-Lake}, Emma and {Jones}, Gareth C. and {Kumari}, Nimisha and {Rinaldi}, Pierluigi and {Robertson}, Brant and {Williams}, Christina C. and {Willott}, Chris and {Zhu}, Yongda},
        title = "{Zapped then napped? A rapidly quenched remnant leaker candidate with a steep spectroscopic {\ensuremath{\beta}}$_{UV}$ slope at z = 8.5}",
      journal = {\aap},
         year = 2025,
        month = may,
       volume = {697},
          eid = {A90},
        pages = {A90},
          doi = {10.1051/0004-6361/202553766},
archivePrefix = {arXiv},
       eprint = {2501.09070},
 primaryClass = {astro-ph.GA},
       adsurl = {https://ui.adsabs.harvard.edu/abs/2025A&A...697A..90B}
}

@ARTICLE{gobat18,
       author = {{Gobat}, R. and {Daddi}, E. and {Magdis}, G. and {Bournaud}, F. and {Sargent}, M. and {Martig}, M. and {Jin}, S. and {Finoguenov}, A. and {B{\'e}thermin}, M. and {Hwang}, H.~S. and {Renzini}, A. and {Wilson}, G.~W. and {Aretxaga}, I. and {Yun}, M. and {Strazzullo}, V. and {Valentino}, F.},
        title = "{The unexpectedly large dust and gas content of quiescent galaxies at z > 1.4}",
      journal = {Nature Astronomy},
         year = 2018,
        month = jan,
       volume = {2},
        pages = {239-246},
          doi = {10.1038/s41550-017-0352-5},
archivePrefix = {arXiv},
       eprint = {1703.02207},
 primaryClass = {astro-ph.GA},
       adsurl = {https://ui.adsabs.harvard.edu/abs/2018NatAs...2..239G}
}

@ARTICLE{white91,
       author = {{White}, Simon D.~M. and {Frenk}, Carlos S.},
        title = "{Galaxy Formation through Hierarchical Clustering}",
      journal = {\apj},
         year = 1991,
        month = sep,
       volume = {379},
        pages = {52},
          doi = {10.1086/170483},
       adsurl = {https://ui.adsabs.harvard.edu/abs/1991ApJ...379...52W}
}

@ARTICLE{springel05,
       author = {{Springel}, Volker and {Di Matteo}, Tiziana and {Hernquist}, Lars},
        title = "{Modelling feedback from stars and black holes in galaxy mergers}",
      journal = {\mnras},
         year = 2005,
        month = aug,
       volume = {361},
       number = {3},
        pages = {776-794},
          doi = {10.1111/j.1365-2966.2005.09238.x},
archivePrefix = {arXiv},
       eprint = {astro-ph/0411108},
 primaryClass = {astro-ph},
       adsurl = {https://ui.adsabs.harvard.edu/abs/2005MNRAS.361..776S}
}

@ARTICLE{aoyama17,
       author = {{Aoyama}, Shohei and {Hou}, Kuan-Chou and {Shimizu}, Ikkoh and {Hirashita}, Hiroyuki and {Todoroki}, Keita and {Choi}, Jun-Hwan and {Nagamine}, Kentaro},
        title = "{Galaxy simulation with dust formation and destruction}",
      journal = {\mnras},
         year = 2017,
        month = apr,
       volume = {466},
       number = {1},
        pages = {105-121},
          doi = {10.1093/mnras/stw3061},
archivePrefix = {arXiv},
       eprint = {1609.07547},
 primaryClass = {astro-ph.GA},
       adsurl = {https://ui.adsabs.harvard.edu/abs/2017MNRAS.466..105A}
}

@software{hahn20,
       author = {{Hahn}, Oliver and {Michaux}, Micha{\"e}l and {Rampf}, Cornelius and {Uhlemann}, Cora and {Angulo}, Raul E.},
        title = "{MUSIC2-monofonIC: 3LPT initial condition generator}",
 howpublished = {Astrophysics Source Code Library, record ascl:2008.024},
         year = 2020,
        month = aug,
          eid = {ascl:2008.024},
archivePrefix = {ascl},
       eprint = {2008.024},
       adsurl = {https://ui.adsabs.harvard.edu/abs/2020ascl.soft08024H}
}

@ARTICLE{lim25,
       author = {{Lim}, Seunghwan and {Tacchella}, Sandro and {Maiolino}, Roberto and {Lovell}, Christopher C. and {Schaye}, Joop},
        title = "{Think inside the box: cosmic variance and large-scale conformity of high-redshift massive galaxies in the FLAMINGO simulations}",
      journal = {arXiv e-prints},
         year = 2025,
        month = nov,
          eid = {arXiv:2511.09618},
        pages = {arXiv:2511.09618},
          doi = {10.48550/arXiv.2511.09618},
archivePrefix = {arXiv},
       eprint = {2511.09618},
 primaryClass = {astro-ph.GA},
       adsurl = {https://ui.adsabs.harvard.edu/abs/2025arXiv251109618L}
}

@ARTICLE{crain23,
       author = {{Crain}, Robert A. and {van de Voort}, Freeke},
        title = "{Hydrodynamical Simulations of the Galaxy Population: Enduring Successes and Outstanding Challenges}",
      journal = {\araa},
         year = 2023,
        month = aug,
       volume = {61},
        pages = {473-515},
          doi = {10.1146/annurev-astro-041923-043618},
archivePrefix = {arXiv},
       eprint = {2309.17075},
 primaryClass = {astro-ph.GA},
       adsurl = {https://ui.adsabs.harvard.edu/abs/2023ARA&A..61..473C}
}

@ARTICLE{martinez-marin24,
       author = {{Mart{\'\i}nez-Mar{\'\i}n}, M. and {Glazebrook}, K. and {Nanayakkara}, T. and {Jacobs}, C. and {Labb{\'e}}, I. and {Kacprzak}, G.~G. and {Papovich}, C. and {Schreiber}, C.},
        title = "{The origin of large emission line widths in massive galaxies at redshifts z   3-4}",
      journal = {\mnras},
         year = 2024,
        month = jul,
       volume = {531},
       number = {3},
        pages = {3187-3202},
          doi = {10.1093/mnras/stae1335},
archivePrefix = {arXiv},
       eprint = {2405.12501},
 primaryClass = {astro-ph.GA},
       adsurl = {https://ui.adsabs.harvard.edu/abs/2024MNRAS.531.3187M}
}

@ARTICLE{chaikin22,
       author = {{Chaikin}, Evgenii and {Schaye}, Joop and {Schaller}, Matthieu and {Bah{\'e}}, Yannick M. and {Nobels}, Folkert S.~J. and {Ploeckinger}, Sylvia},
        title = "{The importance of the way in which supernova energy is distributed around young stellar populations in simulations of galaxies}",
      journal = {\mnras},
         year = 2022,
        month = jul,
       volume = {514},
       number = {1},
        pages = {249-264},
          doi = {10.1093/mnras/stac1132},
archivePrefix = {arXiv},
       eprint = {2203.07134},
 primaryClass = {astro-ph.GA},
       adsurl = {https://ui.adsabs.harvard.edu/abs/2022MNRAS.514..249C}
}

@ARTICLE{wangb25,
       author = {{Wang}, Bingjie and {Leja}, Joel and {Atek}, Hakim and {Bezanson}, Rachel and {Burnham}, Emilie and {Dayal}, Pratika and {Feldmann}, Robert and {Greene}, Jenny E. and {Johnson}, Benjamin D. and {Labb{\'e}}, Ivo and {Maseda}, Michael V. and {Nanayakkara}, Themiya and {Price}, Sedona H. and {Suess}, Katherine A. and {Weaver}, John R. and {Whitaker}, Katherine E.},
        title = "{Population Models for Star Formation Timescales in Early Galaxies: The First Step toward Solving Outshining in Star Formation History Inference}",
      journal = {\apj},
         year = 2025,
        month = jul,
       volume = {987},
       number = {2},
          eid = {184},
        pages = {184},
          doi = {10.3847/1538-4357/adddb8},
archivePrefix = {arXiv},
       eprint = {2504.15255},
 primaryClass = {astro-ph.GA},
       adsurl = {https://ui.adsabs.harvard.edu/abs/2025ApJ...987..184W}
}

@ARTICLE{stefanon13,
       author = {{Stefanon}, Mauro and {Marchesini}, Danilo and {Rudnick}, Gregory H. and {Brammer}, Gabriel B. and {Whitaker}, Katherine E.},
        title = "{What are the Progenitors of Compact, Massive, Quiescent Galaxies at z = 2.3? The Population of Massive Galaxies at z > 3 from NMBS and CANDELS}",
      journal = {\apj},
         year = 2013,
        month = may,
       volume = {768},
       number = {1},
          eid = {92},
        pages = {92},
          doi = {10.1088/0004-637X/768/1/92},
archivePrefix = {arXiv},
       eprint = {1301.7063},
 primaryClass = {astro-ph.CO},
       adsurl = {https://ui.adsabs.harvard.edu/abs/2013ApJ...768...92S}
}

@ARTICLE{schmidt59,
       author = {{Schmidt}, Maarten},
        title = "{The Rate of Star Formation.}",
      journal = {\apj},
         year = 1959,
        month = mar,
       volume = {129},
        pages = {243},
          doi = {10.1086/146614},
       adsurl = {https://ui.adsabs.harvard.edu/abs/1959ApJ...129..243S}
}

@ARTICLE{lesniewska25,
       author = {{Le{\'s}niewska}, A. and {Hjorth}, J. and {Gall}, C.},
        title = "{Dust removal timescale in galaxies across cosmic time}",
      journal = {\aap},
         year = 2025,
        month = jul,
       volume = {699},
          eid = {A352},
        pages = {A352},
          doi = {10.1051/0004-6361/202555007},
archivePrefix = {arXiv},
       eprint = {2505.21492},
 primaryClass = {astro-ph.GA},
       adsurl = {https://ui.adsabs.harvard.edu/abs/2025A&A...699A.352L}
}

@ARTICLE{shuntov25,
       author = {{Shuntov}, Marko and {Ilbert}, Olivier and {Lagos}, Claudia del P. and {Toft}, Sune and {Valentino}, Francesco and {Mercier}, Wilfried and {Akins}, Hollis B. and {Binh}, Nguyen and {Brinch}, Malte and {Casey}, Caitlin M. and {Franco}, Maximilien and {Gentile}, Fabrizio and {Gozaliasl}, Ghassem and {Haghjoo}, Aryana and {Harish}, Santosh and {Hirschmann}, Michaela and {Huertas-Company}, Marc and {Jin}, Shuowen and {Kartaltepe}, Jeyhan S. and {Koekemoer}, Anton M. and {Laigle}, Clotilde and {Lewis}, Joseph S.~W. and {Magdis}, Georgios E. and {Joy McCracken}, Henry and {Mobasher}, Bahram and {Moutard}, Thibaud and {Oesch}, Pascal A. and {Paquereau}, Louise and {Renzini}, Alvio and {Rich}, Michael R. and {Sanders}, David B. and {Toni}, Greta and {Tresse}, Laurence and {Weibel}, Andrea and {Weaver}, John R. and {Yang}, Lilan},
        title = "{The stellar mass function of quiescent and star-forming galaxies and its dependence on morphology in COSMOS-Web}",
      journal = {\aap},
         year = 2026,
        month = mar,
       volume = {707},
          eid = {A391},
        pages = {A391},
          doi = {10.1051/0004-6361/202558022},
archivePrefix = {arXiv},
       eprint = {2511.05259},
 primaryClass = {astro-ph.GA},
       adsurl = {https://ui.adsabs.harvard.edu/abs/2026A&A...707A.391S}
}

@ARTICLE{yang25,
       author = {{Yang}, Tiancheng and {Wang}, Tao and {Xu}, Ke and {Sun}, Hanwen and {Zhou}, Luwenjia and {Xie}, Lizhi and {De Lucia}, Gabriella and {del P. Lagos}, Claudia and {Wang}, Kai and {Fontanot}, Fabio and {Guo}, Qi and {Wu}, Yuxuan and {Lu}, Shiying and {Chen}, Longyue and {Hirschmann}, Michaela},
        title = "{A Census of Quiescent Galaxies across 0.5 < z < 8 with JWST/MIRI: Mass-dependent Number Density Evolution of Quiescent Galaxies in the Early Universe}",
      journal = {\apjl},
         year = 2026,
        month = apr,
       volume = {1000},
       number = {2},
          eid = {L42},
        pages = {L42},
          doi = {10.3847/2041-8213/ae4803},
archivePrefix = {arXiv},
       eprint = {2510.12235},
 primaryClass = {astro-ph.GA},
       adsurl = {https://ui.adsabs.harvard.edu/abs/2026ApJ..1000L..42Y}
}

@ARTICLE{krumholz11,
       author = {{Krumholz}, Mark R. and {Gnedin}, Nickolay Y.},
        title = "{A Comparison of Methods for Determining the Molecular Content of Model Galaxies}",
      journal = {\apj},
         year = 2011,
        month = mar,
       volume = {729},
       number = {1},
          eid = {36},
        pages = {36},
          doi = {10.1088/0004-637X/729/1/36},
archivePrefix = {arXiv},
       eprint = {1011.4065},
 primaryClass = {astro-ph.CO},
       adsurl = {https://ui.adsabs.harvard.edu/abs/2011ApJ...729...36K}
}

@ARTICLE{mckinnon17,
       author = {{McKinnon}, Ryan and {Torrey}, Paul and {Vogelsberger}, Mark and {Hayward}, Christopher C. and {Marinacci}, Federico},
        title = "{Simulating the dust content of galaxies: successes and failures}",
      journal = {\mnras},
         year = 2017,
        month = jun,
       volume = {468},
       number = {2},
        pages = {1505-1521},
          doi = {10.1093/mnras/stx467},
archivePrefix = {arXiv},
       eprint = {1606.02714},
 primaryClass = {astro-ph.GA},
       adsurl = {https://ui.adsabs.harvard.edu/abs/2017MNRAS.468.1505M}
}

@ARTICLE{husko25b,
       author = {{Hu{\v{s}}ko}, Filip and {Lacey}, Cedric G. and {Roper}, William J. and {Schaye}, Joop and {Briggs}, Jemima Mae and {Schaller}, Matthieu},
        title = "{The effects of super-Eddington accretion and feedback on the growth of early supermassive black holes and galaxies}",
      journal = {\mnras},
         year = 2025,
        month = mar,
       volume = {537},
       number = {3},
        pages = {2559-2578},
          doi = {10.1093/mnras/staf146},
archivePrefix = {arXiv},
       eprint = {2410.09450},
 primaryClass = {astro-ph.GA},
       adsurl = {https://ui.adsabs.harvard.edu/abs/2025MNRAS.537.2559H}
}

@ARTICLE{pacifici23,
       author = {{Pacifici}, Camilla and {Iyer}, Kartheik G. and {Mobasher}, Bahram and {da Cunha}, Elisabete and {Acquaviva}, Viviana and {Burgarella}, Denis and {Calistro Rivera}, Gabriela and {Carnall}, Adam C. and {Chang}, Yu-Yen and {Chartab}, Nima and {Cooke}, Kevin C. and {Fairhurst}, Ciaran and {Kartaltepe}, Jeyhan and {Leja}, Joel and {Ma{\l}ek}, Katarzyna and {Salmon}, Brett and {Torelli}, Marianna and {Vidal-Garc{\'\i}a}, Alba and {Boquien}, M{\'e}d{\'e}ric and {Brammer}, Gabriel G. and {Brown}, Michael J.~I. and {Capak}, Peter L. and {Chevallard}, Jacopo and {Circosta}, Chiara and {Croton}, Darren and {Davidzon}, Iary and {Dickinson}, Mark and {Duncan}, Kenneth J. and {Faber}, Sandra M. and {Ferguson}, Harry C. and {Fontana}, Adriano and {Guo}, Yicheng and {Haeussler}, Boris and {Hemmati}, Shoubaneh and {Jafariyazani}, Marziye and {Kassin}, Susan A. and {Larson}, Rebecca L. and {Lee}, Bomee and {Mantha}, Kameswara Bharadwaj and {Marchi}, Francesca and {Nayyeri}, Hooshang and {Newman}, Jeffrey A. and {Pandya}, Viraj and {Pforr}, Janine and {Reddy}, Naveen and {Sanders}, Ryan and {Shah}, Ekta and {Shahidi}, Abtin and {Stevans}, Matthew L. and {Triani}, Dian Puspita and {Tyler}, Krystal D. and {Vanderhoof}, Brittany N. and {de la Vega}, Alexander and {Wang}, Weichen and {Weston}, Madalyn E.},
        title = "{The Art of Measuring Physical Parameters in Galaxies: A Critical Assessment of Spectral Energy Distribution Fitting Techniques}",
      journal = {\apj},
         year = 2023,
        month = feb,
       volume = {944},
       number = {2},
          eid = {141},
        pages = {141},
          doi = {10.3847/1538-4357/acacff},
archivePrefix = {arXiv},
       eprint = {2212.01915},
 primaryClass = {astro-ph.GA},
       adsurl = {https://ui.adsabs.harvard.edu/abs/2023ApJ...944..141P}
}

@ARTICLE{lagos25b,
       author = {{Lagos}, Claudia del P. and {Schaye}, Joop and {Schaller}, Matthieu and {Obreschkow}, Danail and {Bah{\'e}}, Yannick M. and {Ben{\'\i}tez-Llambay}, Alejandro and {Chaikin}, Evgenii and {Correa}, Camila and {Davis}, Timothy A. and {Frenk}, Carlos S. and {Hu{\v{s}}ko}, Filip and {Kaasinen}, Melanie and {McGibbon}, Robert J. and {Oman}, Kyle and {Ploeckinger}, Sylvia and {Richings}, Alexander J. and {Trayford}, James W. and {Wang}, Jing and {Wright}, Ruby J.},
        title = "{Kennicutt─Schmidt relation of galaxies over 13 billion years in the COLIBRE hydrodynamical simulations}",
      journal = {\mnras},
         year = 2026,
        month = jun,
       volume = {549},
       number = {2},
          eid = {stag947},
        pages = {stag947},
          doi = {10.1093/mnras/stag947},
archivePrefix = {arXiv},
       eprint = {2512.11309},
 primaryClass = {astro-ph.GA},
       adsurl = {https://ui.adsabs.harvard.edu/abs/2026MNRAS.549ag947L}
}

@ARTICLE{conroy13,
       author = {{Conroy}, Charlie},
        title = "{Modeling the Panchromatic Spectral Energy Distributions of Galaxies}",
      journal = {\araa},
         year = 2013,
        month = aug,
       volume = {51},
       number = {1},
        pages = {393-455},
          doi = {10.1146/annurev-astro-082812-141017},
archivePrefix = {arXiv},
       eprint = {1301.7095},
 primaryClass = {astro-ph.CO},
       adsurl = {https://ui.adsabs.harvard.edu/abs/2013ARA&A..51..393C}
}

@ARTICLE{jones25,
       author = {{Jones}, Gareth T. and {Byrne}, Conor M. and {Stanway}, Elizabeth R.},
        title = "{Impact of uncertainties in spectral energy distribution modelling on inferred galaxy properties}",
      journal = {\mnras},
         year = 2025,
        month = oct,
       volume = {543},
       number = {1},
        pages = {167-189},
          doi = {10.1093/mnras/staf1462},
archivePrefix = {arXiv},
       eprint = {2509.02741},
 primaryClass = {astro-ph.GA},
       adsurl = {https://ui.adsabs.harvard.edu/abs/2025MNRAS.543..167J}
}

@ARTICLE{merlin25,
       author = {{Merlin}, Emiliano and {Fortuni}, Flaminia and {Calabr{\'o}}, Antonello and {Castellano}, Marco and {Santini}, Paola and {Fontana}, Adriano and {Kimmig}, Lucas C. and {Shankar}, Francesco and {Napolitano}, Lorenzo and {Koekemoer}, Anton M. and {Lucas}, Ray A. and {Pacucci}, Fabio and {Cooper}, Michael C. and {Hirschmann}, Michaela and {P{\'e}rez-Gonz{\'a}lez}, Pablo G. and {Barro}, Guillermo and {Dickinson}, Mark and {Gandolfi}, Giovanni and {Paris}, Diego and {Grogin}, Norman A. and {Wang}, Xin},
        title = "{Witnessing downsizing in the making: quiescent and breathing galaxies at the dawn of the Universe}",
      journal = {The Open Journal of Astrophysics},
         year = 2025,
        month = nov,
       volume = {8},
        pages = {E170},
          doi = {10.33232/001c.147267},
archivePrefix = {arXiv},
       eprint = {2509.09764},
 primaryClass = {astro-ph.GA},
       adsurl = {https://ui.adsabs.harvard.edu/abs/2025OJAp....8E.170M}
}

@ARTICLE{turner25,
       author = {{Turner}, Crispin and {Tacchella}, Sandro and {D'Eugenio}, Francesco and {Carniani}, Stefano and {Curti}, Mirko and {Glazebrook}, Karl and {Johnson}, Benjamin D. and {Lim}, Seunghwan and {Looser}, Tobias and {Maiolino}, Roberto and {Nanayakkara}, Themiya and {Wan}, Jenny},
        title = "{Age-dating early quiescent galaxies: high star formation efficiency, but consistent with direct, higher-redshift observations}",
      journal = {\mnras},
         year = 2025,
        month = feb,
       volume = {537},
       number = {2},
        pages = {1826-1848},
          doi = {10.1093/mnras/staf128},
archivePrefix = {arXiv},
       eprint = {2410.05377},
 primaryClass = {astro-ph.GA},
       adsurl = {https://ui.adsabs.harvard.edu/abs/2025MNRAS.537.1826T}
}

@ARTICLE{yang25size,
       author = {{Yang}, Lilan and {Kartaltepe}, Jeyhan S. and {Franco}, Maximilien and {Ding}, Xuheng and {Achenbach}, Mark J. and {Arango-Toro}, Rafael C. and {Casey}, Caitlin M. and {Drakos}, Nicole E. and {Faisst}, Andreas L. and {Gillman}, Steven and {Gozaliasl}, Ghassem and {Huertas-Company}, Marc and {Jin}, Shuowen and {Liu}, Daizhong and {Magdis}, Georgios and {Massey}, Richard and {Silverman}, John D. and {Tanaka}, Takumi S. and {Yu}, Si-Yue and {Akins}, Hollis B. and {Allen}, Natalie and {Ilbert}, Olivier and {Koekemoer}, Anton M. and {McCracken}, Henry Joy and {Paquereau}, Louise and {Rhodes}, Jason and {Robertson}, Brant E. and {Shuntov}, Marko and {Toft}, Sune},
        title = "{COSMOS-Web: Unraveling the Evolution of Galaxy Size and Related Properties at 2 < z < 10}",
      journal = {\apjs},
         year = 2025,
        month = dec,
       volume = {281},
       number = {2},
          eid = {68},
        pages = {68},
          doi = {10.3847/1538-4365/ae0e1b},
archivePrefix = {arXiv},
       eprint = {2504.07185},
 primaryClass = {astro-ph.GA},
       adsurl = {https://ui.adsabs.harvard.edu/abs/2025ApJS..281...68Y}
}

@ARTICLE{lustig23,
       author = {{Lustig}, Peter and {Strazzullo}, Veronica and {Remus}, Rhea-Silvia and {D'Eugenio}, Chiara and {Daddi}, Emanuele and {Burkert}, Andreas and {De Lucia}, Gabriella and {Delvecchio}, Ivan and {Dolag}, Klaus and {Fontanot}, Fabio and {Gobat}, Raphael and {Mohr}, Joseph J. and {Onodera}, Masato and {Pannella}, Maurilio and {Pillepich}, Annalisa},
        title = "{Massive quiescent galaxies at z   3: A comparison of selection, stellar population, and structural properties with simulation predictions}",
      journal = {\mnras},
         year = 2023,
        month = feb,
       volume = {518},
       number = {4},
        pages = {5953-5975},
          doi = {10.1093/mnras/stac3450},
archivePrefix = {arXiv},
       eprint = {2201.09068},
 primaryClass = {astro-ph.GA},
       adsurl = {https://ui.adsabs.harvard.edu/abs/2023MNRAS.518.5953L}
}

@ARTICLE{xie24,
       author = {{Xie}, Lizhi and {De Lucia}, Gabriella and {Fontanot}, Fabio and {Hirschmann}, Michaela and {Bah{\'e}}, Yannick M. and {Balogh}, Michael L. and {Muzzin}, Adam and {Vulcani}, Benedetta and {Baxter}, Devontae C. and {Forrest}, Ben and {Wilson}, Gillian and {Rudnick}, Gregory H. and {Cooper}, M.~C. and {Rescigno}, Umberto},
        title = "{The First Quenched Galaxies: When and How?}",
      journal = {\apjl},
         year = 2024,
        month = may,
       volume = {966},
       number = {1},
          eid = {L2},
        pages = {L2},
          doi = {10.3847/2041-8213/ad380a},
archivePrefix = {arXiv},
       eprint = {2402.01314},
 primaryClass = {astro-ph.GA},
       adsurl = {https://ui.adsabs.harvard.edu/abs/2024ApJ...966L...2X}
}

@ARTICLE{chaikin26,
       author = {{Chaikin}, Evgenii and {Schaye}, Joop and {Hu{\v{s}}ko}, Filip and {Lacey}, Cedric G. and {Ploeckinger}, Sylvia and {Schaller}, Matthieu},
        title = "{The importance of super-Eddington black hole accretion for the emergence of massive quiescent galaxies at high redshift}",
      journal = {\mnras},
         year = 2026,
        month = aug,
       volume = {550},
       number = {2},
          eid = {stag1148},
        pages = {stag1148},
          doi = {10.1093/mnras/stag1148},
archivePrefix = {arXiv},
       eprint = {2601.15207},
 primaryClass = {astro-ph.GA},
       adsurl = {https://ui.adsabs.harvard.edu/abs/2026MNRAS.550g1148C}
}

@ARTICLE{chandro-gomez_inprep,
       author = {{Chandro-G{\'o}mez}, {\'A}ngel and {Lagos}, Claudia del P. and {Power}, Chris and {Baker}, Willian M. and {Ben{\'\i}tez-Llambay}, Alejandro and {Chaikin}, Evgenii and {Chittenden}, Harry G. and {Correa}, Camila and {Frenk}, Carlos S. and {Hu{\v{s}}ko}, Filip and {Ito}, Kei and {McGibbon}, Robert J. and {Nanayakkara}, Themiya and {Ploeckinger}, Sylvia and {Richings}, Alexander J. and {Schaller}, Matthieu and {Schaye}, Joop and {Trayford}, James W. and {Valentino}, Francesco},
        title = "{Unveiling the population of massive quenched galaxies at $z\ge2$ in the COLIBRE simulations -- II. The role of AGN feedback and environment on their emergence}",
      journal = {arXiv e-prints},
         year = 2026,
        month = may,
          eid = {arXiv:2605.31052},
        pages = {arXiv:2605.31052},
          doi = {10.48550/arXiv.2605.31052},
archivePrefix = {arXiv},
       eprint = {2605.31052},
 primaryClass = {astro-ph.GA},
       adsurl = {https://ui.adsabs.harvard.edu/abs/2026arXiv260531052C}
}

@ARTICLE{correa26,
       author = {{Correa}, Camila A and {Schaye}, Joop and {Schaller}, Matthieu and {Trayford}, James W and {Chaikin}, Evgenii and {Benitez-Llambay}, Alejandro and {Frenk}, Carlos S and {Ploeckinger}, Sylvia and {Richings}, Alexander J},
        title = "{A subgrid model for chemical enrichment in cosmological simulations}",
      journal = {\mnras},
         year = 2026,
        month = apr,
          doi = {10.1093/mnras/stag645},
archivePrefix = {arXiv},
       eprint = {2604.00980},
 primaryClass = {astro-ph.GA},
       adsurl = {https://ui.adsabs.harvard.edu/abs/2026MNRAS.tmp..607C}
}

@ARTICLE{ludlow26,
       author = {{Ludlow}, Aaron D. and {Proctor}, Katy L. and {Schaye}, Joop and {Hu{\v{s}}ko}, Filip and {Moreno}, Victor J. Forouhar and {Obreschkow}, Danail and {Chaikin}, Evgenii and {Schaller}, Matthieu and {Ploeckinger}, Sylvia and {Ben{\'\i}tez-Llambay}, Alejandro and {Oman}, Kyle A. and {McGibbon}, Robert J. and {Trayford}, James W. and {Frenk}, Carlos S. and {Richings}, Alexander J.},
        title = "{The evolution of the sizes and angular momentum content of galaxies in the COLIBRE simulations}",
      journal = {\mnras},
         year = 2026,
        month = jul,
          doi = {10.1093/mnras/stag1319},
archivePrefix = {arXiv},
       eprint = {2603.26200},
 primaryClass = {astro-ph.GA},
       adsurl = {https://ui.adsabs.harvard.edu/abs/2026MNRAS.tmp.1247L}
}

@ARTICLE{leung26,
       author = {{Leung}, Ho-Hin and {Carnall}, Adam C. and {Taylor}, Elizabeth and {Stevenson}, Struan D. and {Beverage}, Aliza G. and {Cullen}, Fergus and {Dunlop}, James S. and {McLeod}, Derek J. and {McLure}, Ross J. and {Begley}, Ryan and {Almaini}, Omar and {Antonogiannaki}, Stella and {Arellano-C{\'o}rdova}, Karla Z. and {Barrufet}, Laia and {Bondestam}, Cecilia and {Donnan}, Callum T. and {Holst}, Isaac J.~B. and {Liu}, Feng-Yuan F. and {Rowlands}, Kate and {Sanders}, Ryan L. and {Scholte}, Dirk and {Skarbinski}, Maya and {Stanton}, Thomas M. and {Wild}, Vivienne},
        title = "{The JWST EXCELS survey: the ages and abundances of 3 < z < 5 massive quiescent galaxies show that downsizing was already in place by z ≃ 4}",
      journal = {\mnras},
         year = 2026,
        month = jun,
       volume = {549},
       number = {1},
          eid = {stag827},
        pages = {stag827},
          doi = {10.1093/mnras/stag827},
archivePrefix = {arXiv},
       eprint = {2602.05934},
 primaryClass = {astro-ph.GA},
       adsurl = {https://ui.adsabs.harvard.edu/abs/2026MNRAS.549ag827L}
}

@ARTICLE{siegel25,
       author = {{Siegel}, Jared C. and {Setton}, David J. and {Greene}, Jenny E. and {Suess}, Katherine A. and {Whitaker}, Katherine E. and {Bezanson}, Rachel and {Leja}, Joel and {Furtak}, Lukas J. and {Cutler}, Sam E. and {de Graaff}, Anna and {Feldmann}, Robert and {Khullar}, Gourav and {Labbe}, Ivo and {Marchesini}, Danilo and {Miller}, Tim B. and {Nanayakkara}, Themiya and {Pan}, Richard and {Price}, Sedona H. and {Treiber}, Helena P. and {van Dokkum}, Pieter and {Wang}, Bingjie and {Weaver}, John R.},
        title = "{UNCOVER: Significant Reddening in Cosmic Noon Quiescent Galaxies}",
      journal = {\apj},
         year = 2025,
        month = may,
       volume = {985},
       number = {1},
          eid = {125},
        pages = {125},
          doi = {10.3847/1538-4357/adc7b7},
archivePrefix = {arXiv},
       eprint = {2409.11457},
 primaryClass = {astro-ph.GA},
       adsurl = {https://ui.adsabs.harvard.edu/abs/2025ApJ...985..125S}
}

@ARTICLE{chang26,
       author = {{Chang}, Wenjun and {Wilson}, Gillian and {Forrest}, Ben and {McConachie}, Ian and {Noble}, Allison and {Muzzin}, Adam and {Marchesini}, Danilo and {Cooper}, M.~C. and {Webb}, Tracy and {Canalizo}, Gabriela and {Gomez}, Percy and {Zhu}, Yongda and {Lei}, Han and {Edward}, Adit Harin and {Henry}, Aur{\'e}lien and {Urbano Stawinski}, Stephanie M. and {Wisz}, M.~E.},
        title = "{MAGAZ3NE: Confirming Dust Deficiency and Quiescent Nature of Ultramassive Galaxies at 3 < z < 4 with ALMA Observations}",
      journal = {\apj},
         year = 2026,
        month = jul,
       volume = {1005},
       number = {1},
          eid = {54},
        pages = {54},
          doi = {10.3847/1538-4357/ae70fa},
       adsurl = {https://ui.adsabs.harvard.edu/abs/2026ApJ..1005...54C}
}

@ARTICLE{wang26,
       author = {{Wang}, Weichen and {Cantalupo}, Sebastiano and {Galbiati}, Marta and {Travascio}, Andrea and {Pensabene}, Antonio and {Steidel}, Charles C. and {Pezzulli}, Gabriele and {Wang}, Bingjie and {Wang}, Xiaohan and {Dutta}, Rajeshwari and {Lazeyras}, Titouan and {Ledos}, Nicolas and {Mao}, Huiyang and {Quadri}, Giada},
        title = "{A quiescent galaxy in a gas-rich cosmic web node at z {\ensuremath{\sim}} 3}",
      journal = {\aap},
         year = 2026,
        month = jul,
       volume = {711},
          eid = {A84},
        pages = {A84},
          doi = {10.1051/0004-6361/202659351},
archivePrefix = {arXiv},
       eprint = {2601.20473},
 primaryClass = {astro-ph.GA},
       adsurl = {https://ui.adsabs.harvard.edu/abs/2026A&A...711A..84W}
}

@ARTICLE{harborne20,
       author = {{Harborne}, K.~E. and {van de Sande}, J. and {Cortese}, L. and {Power}, C. and {Robotham}, A.~S.~G. and {Lagos}, C.~D.~P. and {Croom}, S.},
        title = "{Recovering {\ensuremath{\lambda}}$_{R}$ and V/{\ensuremath{\sigma}} from seeing-dominated IFS data}",
      journal = {\mnras},
         year = 2020,
        month = sep,
       volume = {497},
       number = {2},
        pages = {2018-2038},
          doi = {10.1093/mnras/staa1847},
archivePrefix = {arXiv},
       eprint = {2006.12730},
 primaryClass = {astro-ph.GA},
       adsurl = {https://ui.adsabs.harvard.edu/abs/2020MNRAS.497.2018H}
}

@ARTICLE{umehata25,
       author = {{Umehata}, Hideki and {Kubo}, Mariko and {Nakanishi}, Kouichiro},
        title = "{ADF22-WEB: Detection of a Molecular Gas Reservoir in a Massive Quiescent Galaxy Located in a z ≍ 3 Protocluster Core}",
      journal = {\apjl},
         year = 2025,
        month = may,
       volume = {985},
       number = {1},
          eid = {L8},
        pages = {L8},
          doi = {10.3847/2041-8213/add1d4},
archivePrefix = {arXiv},
       eprint = {2502.06538},
 primaryClass = {astro-ph.GA},
       adsurl = {https://ui.adsabs.harvard.edu/abs/2025ApJ...985L...8U}
}




\appendix

\section{Selection criteria of MQGs}
\label{appendix:mqg-selection}

\begin{figure*}
\centering
\includegraphics[width=\textwidth]{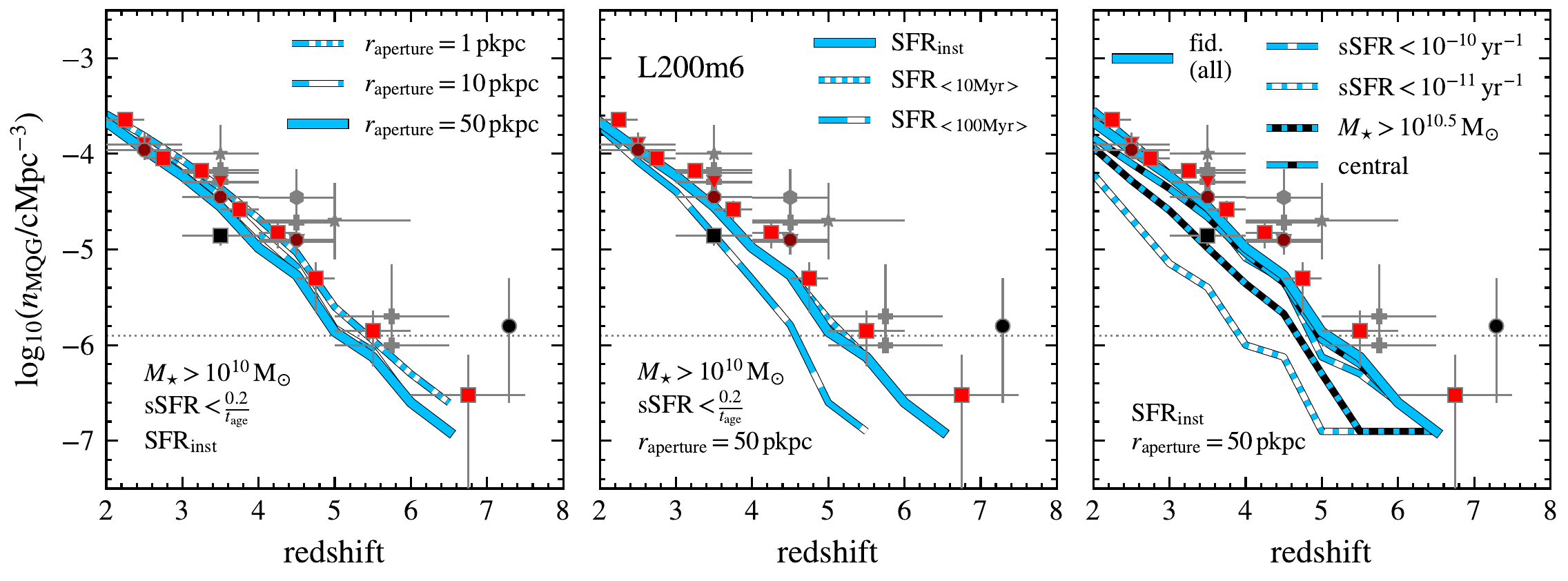}
\caption{Comoving number density of MQGs as a function of redshift for different selection criteria. The panels show variations in aperture size for galaxy properties (\textit{left}), instantaneous versus time-averaged SFR definitions (\textit{middle}), and alternative “massive” and “quenched” thresholds (\textit{right}), applied to the fiducial L200m6 simulation. Blue solid lines indicate the fiducial MQG selection described in \S~\ref{ssec:prop-mq-def}: $r_{\rm aperture}=50\,\mathrm{pkpc}$, $\mathrm{SFR_{inst}}$\, $M_{\star}>10^{10}\,\mathrm{M}_{\odot}$, and $\mathrm{sSFR}<0.2/t_{\rm age}$. Observational data are the same as in Fig.~\ref{fig:ndens-boxes}. The horizontal dotted line denotes the threshold of 10 galaxies, below which the statistics become unreliable.} 
\label{fig:ndens-selection} 
\end{figure*}

We analyse how different selection criteria for MQGs may affect our results, focusing on the evolution of number densities in our fiducial L200m6 simulation (Fig.~\ref{fig:ndens-selection}).

First, we test the impact of the aperture used by {\sc SOAP} to compute $M_{\star}$ and SFR (see \S~\ref{sec:prop}). In the left panel, we compare our fiducial $\rm 50\,pkpc$ spherical aperture with \textcolor{black}{$\rm 1\,pkpc$} (similar to the observational aperture to measure \textit{JWST} colours) and \textcolor{black}{$\rm 10\,pkpc$}. The results show minor differences \textcolor{black}{for the $\rm 10\,pkpc$ aperture, while the number densities increase slightly for the smallest aperture. This is likely because some star-forming galaxies are misclassified as quiescent when outer, star-forming regions are excluded, leading to better agreement with the observational data. This behaviour is} consistent with the compact nature of MQGs (\S~\ref{ssec:res1-size}).

Second, we examine how the definition of SFR might bias the selection. Observations typically measure SFRs averaged over different timescales depending on wavelength coverage and band availability \citep{kennicutt98}. While our fiducial choice is the instantaneous SFR, the middle panel shows results using $\rm 10\,Myr$ and $\rm 100\,Myr$ averages, representative of observational tracers. 
The number densities remain essentially unchanged \textcolor{black}{for the $\rm 10\,Myr$ average}, indicating that these galaxies have been quiescent on average over such timescales. \textcolor{black}{This is consistent with spectroscopic observations tracing star formation over similar timescales \citep{zhang25}.} \textcolor{black}{In contrast, the $\rm100\,Myr$ average produces significantly lower number densities, particularly at $z>3$. This suggests that the simulated SFHs remain bursty over $\sim100\,\mathrm{Myr}$ timescales, causing some systems to move out of the quiescent category when averaged. These galaxies therefore appear to have quenched only recently, consistent with the discussion in \S~\ref{ssec:res1-ages-results-tq} and with observational results from \citet{merlin25}. On the other hand, some observations still recover high number densities even when using $\rm 100\,Myr$-averaged SFRs \citep{baker25}, reflecting a smoother decline in star formation and a more gradual stellar mass build-up, as also seen in Appendix~\ref{appendix:sfh}. \textcolor{black}{The impact of the adopted SFR timescale on observed MQG number densities remains open.}}

Finally, in the right panel, we explore alternative definitions of “massive” and “quenched”. Excluding satellites (blue–black bicoloured solid line), identified by {\sc HBT-HERONS}, does not have a great impact, particularly at $z\gtrsim3$. \textcolor{black}{These satellite galaxies are quenched by AGN feedback, as evidenced by their more massive BHs (fig.~5 of \citetalias{chandro-gomez_inprep}).} In contrast, quenching due to environmental processes is not yet dominant at such high-$z$. The central–satellite hierarchy is determined with {\sc HBT-HERONS}: in FoF groups with multiple substructures, the central subhalo is identified by retained mass and orbital kinetic energy in the group’s centre-of-mass frame. At the high-$z$ we are interested in, the number fraction of MQG satellites relative to centrals is $\approx0.31$ at $z=2$, $\approx0.26$ at $z=3$, $\approx0.13$ at $z=4$, $\approx0.18$ at $z=5$, and $0.00$ at $z=6$ for L200m6.

For quenching, adopting a constant threshold of $\rm sSFR < 10^{-10}\,yr^{-1}$, as often in the literature \citep[e.g.][]{weibel25, zhang25}, yields results similar to our redshift-dependent definition, with differences emerging mainly at $z\gtrsim5$ where both definitions diverge more. A stricter cut of $\rm sSFR < 10^{-11}\,yr^{-1}$ produces much lower number densities, showing that most of the MQGs in {\sc COLIBRE} retain some residual star formation. This threshold has frequently been used to identify quiescent galaxies at high redshift in pre-\textit{JWST} studies \citep[e.g.][]{stefanon13}, although some \textit{JWST} sources also satisfy this criterion \citep[e.g. some galaxies in][]{nanayakkara25}. Increasing the mass threshold to $M_{\star}>10^{10.5}\,\rm M_{\odot}$ reduces number densities slightly, indicating that MQGs are massive but not extreme outliers.

\section{The effect of error convolution on the number densities of MQGs}
\label{appendix:error}

\begin{figure}
\centering
\includegraphics[width=0.48\textwidth]{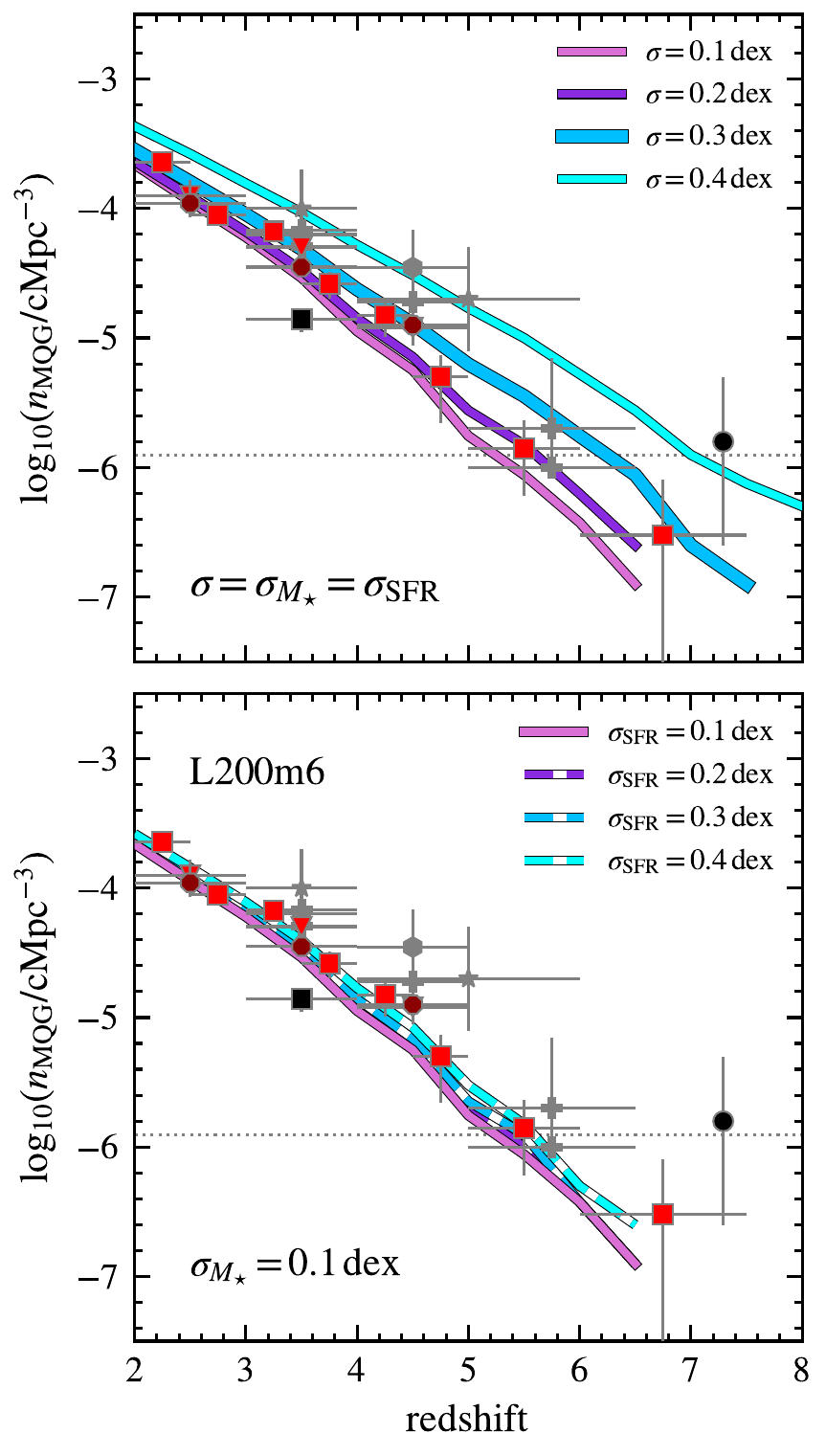}
\caption{Comoving number density of MQGs as a function of redshift for different values of the Gaussian-distributed width applied independently to $M_{\star}$ and SFR when convolving these quantities with potential observational uncertainties. The panels show variations in the convolution width applied to both $M_{\star}$ and SFR when the two are assigned the same value (\textit{top}), and variations in the SFR convolution width while fixing the $M_{\star}$ scatter to $0.1$~dex (\textit{bottom}); applied to the fiducial L200m6 simulation. The blue solid line indicates the fiducial error budget used in \S~\ref{ssec:res1-ndens}~and~\ref{ssec:res1-smf}: $0.3$~dex in both $M_{\star}$ and SFR. Observational data are the same as in Fig.~\ref{fig:ndens-boxes}. The horizontal dotted line denotes the threshold of 10 galaxies, below which the statistics become unreliable.}
\label{fig:ndens-selection-error} 
\end{figure}

In Fig.~\ref{fig:ndens-selection-error}, we illustrate the impact of adopting different values for the width of the Gaussian-distributed convolution applied to the simulation properties independently ($M_{\star}$ and SFR), intended to represent a reasonable error budget for these quantities. The top panel shows the effect of increasing the common standard deviation applied to both $M_{\star}$ and SFR. The bottom panel presents a similar analysis but fixing the width of the $M_{\star}$ convolution to $0.1$~dex, while varying the width applied to the SFR in the range $0.1-0.4$~dex. This choice is motivated by the fact that stellar mass estimates are generally better constrained than SFRs. In both panels, we find that increasing the convolution scatter leads to higher inferred number densities, as expected. Even when adopting a relatively small error budget for stellar masses (bottom panel), the resulting number densities increase sufficiently due to the SFR error to more consistently reproduce the latest observational data (in red and dark red).

\citet{chaikin25b} report a similar effect, showing that adding \textcolor{black}{only} scatter to simulated $M_{\star}$ values (comparable to the uncertainties introduced in \S~\ref{sssec:res1-ndens-results}) has a significant impact on MQG number densities. Specifically, they find that increasing the scatter raises the number densities at higher redshifts and in higher stellar-mass bins. The dependence on scatter amplitude is explored in their Fig.~D3, albeit using a slightly different sample selection and including Poisson uncertainties. In general, a larger scatter produces a stronger increase in the number densities at fixed redshift and stellar mass.

\textcolor{black}{Figure~\ref{fig:ndens-selection-error} further shows that uncertainties in stellar mass have the largest impact for these systems, as the effect is much weaker when perturbing SFRs alone (lower panel). Although not shown explicitly, we verified this by varying $\sigma_{M_{\star}}$ while keeping $\sigma_{\mathrm{SFR}}=0.1\,\mathrm{dex}$ fixed, finding a substantially stronger impact. This arises because (i) $M_{\star}$ enters both the “massive” and “quenched” selection criteria, and (ii) galaxies extends significantly into the quenched region in the SFR–$M_{\star}$ plane at intermediate masses ($10^{9}\,\mathrm{M_{\odot}} < M_{\star} < 10^{10}\,\mathrm{M_{\odot}}$), allowing them to scatter into the MQG selection, particularly at high redshift ($z \approx 6$). In contrast, the L400m7 simulation is largely unaffected by the error convolution in the lower panel of Fig.~\ref{fig:ndens-boxes}. This is partly because more efficient AGN feedback in L400m7 drives a larger quenched population in the SFR-$M_{\star}$ plane, so that the applied scatter moves galaxies into and out of the MQG selection more symmetrically, without introducing a strong net bias. By comparison, in L200m6 quenching is less efficient and galaxies retain higher residual SFRs. As a result, the quenched population is rarer, making the MQG selection more sensitive to perturbations introduced by the observational error convolution.}

The exact impact of the different sources of errors in the derived stellar masses and SFRs has been subject to extensive discussion in the literature (e.g. \citealt{robotham20,pacifici23,bellstedt25}). For the population of high-z MQGs, \citet{nanayakkara25} show that random errors in stellar masses have a magnitude $\approx 0.09$~dex, but that systematic errors can shift the derived values by as much as $\approx 0.4$~dex. Our purpose is not to isolate the magnitude of different errors, but to demonstrate the effect an error budget of $\approx 0.1-0.4$~dex can have on the derived number densities of MQGs. A more realistic calibration of these observational systematics will require forward modelling of simulated {\sc COLIBRE} MQGs to generate mock photometric and spectroscopic observations, followed by analysis with modern SED-fitting tools (Nanayakkara et al., in prep.).


\section{Error convolution on the stellar mass functions of MQGs}
\label{appendix:error-smf}

\begin{figure}
\centering
\includegraphics[width=0.45\textwidth]{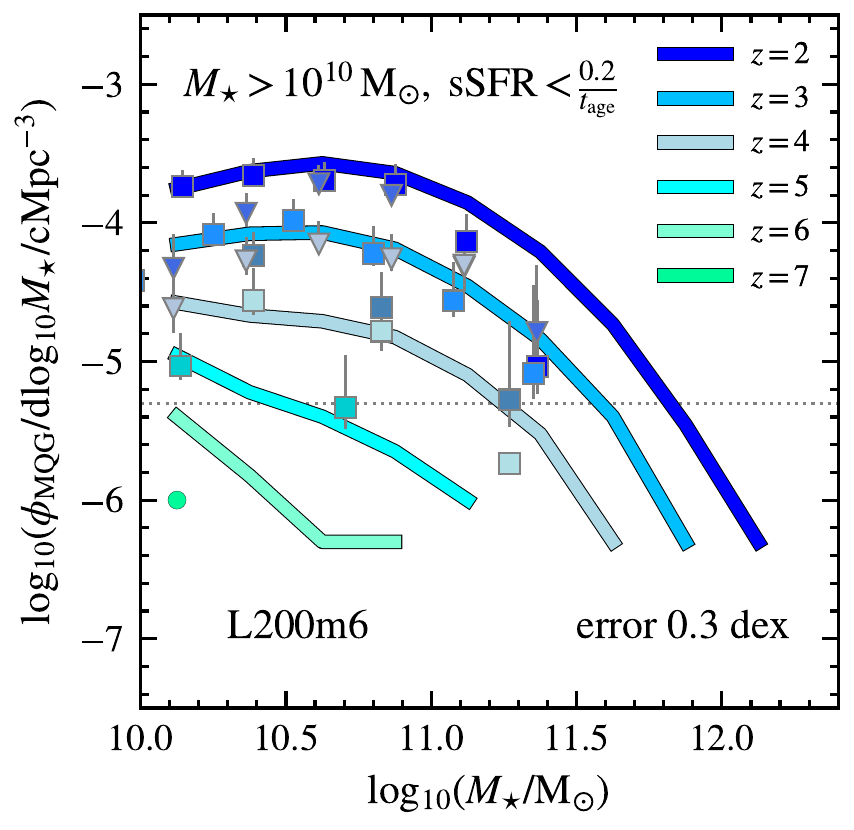}
\caption{SMF of MQGs at different redshifts $2 \le z \le 7$ for the fiducial L200m6 simulation, where $M_{\star}$ and SFR values in the simulations are convolved independently with a Gaussian distribution (mean 0, standard deviation $0.3$~dex) representing a reasonable error budget for these quantities. Observational data are the same as in Fig.~\ref{fig:smf}. The horizontal dotted line denotes the threshold of 10 galaxies, below which the statistics become unreliable.} 
\label{fig:smf-selection} 
\end{figure}

In this appendix, we examine the impact of applying a Gaussian-distributed convolution, centred at $0$ with a width of $\sigma=0.3$~dex, to the simulated stellar masses and SFRs independently when computing the SMF of MQGs in the fiducial L200m6 simulation. The error-convolved SMF at $z=3$ is shown in Fig.~\ref{fig:smf} and compared with predictions from other galaxy formation and evolution models; here, we extend this analysis to the full redshift range $2 \le z \le 7$ \textcolor{black}{in Fig.~\ref{fig:smf-selection}}. In Fig.~\ref{fig:smf}, we found that, relative to the latest \textit{JWST} observations, the highest-mass end of the SMF and the predictions at $z \gtrsim 5$ lie towards the lower end of the observationally inferred range. When the same error convolution is applied across all redshifts, we find that it increases the number of systems in both regimes, thereby alleviating these discrepancies once potential uncertainties in the derivation of stellar masses and SFRs are taken into account.

\section{Star formation histories of MQGs}
\label{appendix:sfh}

We present in this appendix a comparison between the SFH shapes of MQGs selected from our fiducial L200m6 simulation and those derived observationally via SED fitting. These SFHs are used to obtain the timescales described in \S~\ref{ssec:prop-sfh} and analysed in \S~\ref{ssec:res1-ages}.

\begin{figure}
    \centering
    \includegraphics[width=\linewidth]{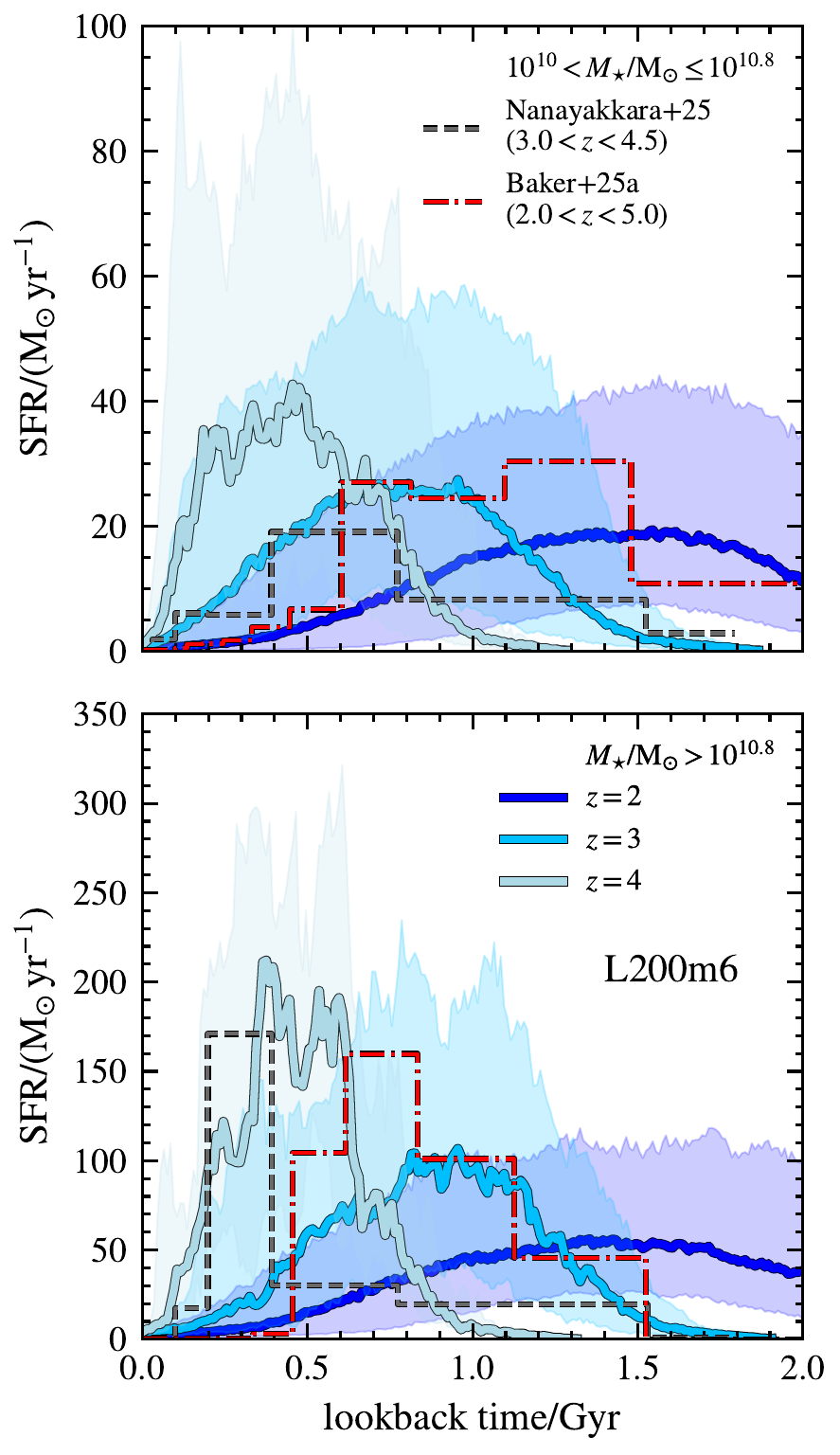}
\caption{SFHs of MQGs. Solid lines show the median predictions from the L200m6 {\sc COLIBRE} simulation at redshifts $2 \le z \le 4$, with different colours indicating selection redshift and shaded regions showing the 16th and 84th percentile range. These are compared to \textit{JWST} spectroscopic measurements: dashed lines indicate the medians from \citet{nanayakkara25} and \citet{baker25b}. \textit{Top}: MQGs with $10^{10} < M_{\star}/\mathrm{M_{\odot}} \le 10^{10.8}$. \textit{Bottom}: MQGs with $M_{\star} > 10^{10.8}\,\mathrm{M_{\odot}}$.} 
\label{fig:sfh} 
\end{figure}

Fig.~\ref{fig:sfh} shows the median SFHs from {\sc COLIBRE} (blue-palette solid lines) for different redshifts, $2 \le z \le 4$, compared with observational SFHs from \citet{nanayakkara25} in dark grey and \citet{baker25b} in red. Each panel corresponds to a stellar mass bin: the top panel shows low-mass MQGs and the bottom panel high-mass MQGs. In the simulations, higher selection redshifts correspond to more bursty SFHs. With less cosmic time available, galaxies must assemble through more intense starbursts, leading to higher peaks and distributions skewed towards later times, consistent with the discussion in \S~\ref{ssec:prop-sfh}. As expected, the less massive MQGs (top panel) require less extreme bursts, reflected in their lower SFR values. The observations show similar behaviour: less bursty and more extended SFHs in the low-mass bin. \textcolor{black}{However, the SFH peaks are lower in the observational data. This is because observational reconstructions use broader lookback-time bins. When we increase the bin size to $100\,\mathrm{Myr}$ in the simulations, the peaks from the SFHs derived from {\sc COLIBRE} become less pronounced, bringing the results into better agreement with observations.}


Interestingly, the \citet{baker25b} SFHs tend to assemble mass earlier than those of \citet{nanayakkara25}. This is partly because the \citet{baker25b} sample is more skewed towards lower redshifts, but methodological differences in the SED fitting also contribute. Both studies use {\sc Prospector}, but their assumptions differ in several ways. They do not cover the same spectral range: \citet{nanayakkara25} include the observed-frame NIR ($\lambda=1-5\ \mu$m), while \citet{baker25b} use rest-frame $\lambda>0.35\ \mu$m for the spectrum to avoid the less well-understood rest-frame UV (while the entire SED for the photometry). Both adopt non-parametric SFHs, but one uses a flat continuity prior \citep{leja19b} and the other an exponentially increasing prior in lookback time. They also use different lookback time intervals, and different SSP models: C3K \citep{conroy09} for \citet{nanayakkara25} versus MILES \citep{falcon-barroso11} for \citet{baker25b}. On the other hand, both employ MIST isochrones \citep{paxton15} (which assume fixed solar abundances), a \citet{chabrier03} IMF, and the \citet{calzetti00} dust attenuation law and uniform metallicity priors. 

These methodological differences explain the behaviours seen in Fig.~\ref{fig:ages}. Overall, despite observational biases, the simulation results reproduce the qualitative features of the observed SFHs.

\section{Sizes and kinematics of MQGs for the L400m7 simulation}
\label{appendix:l400m7-size-kin}

\begin{figure}
\centering
\includegraphics[width=0.48\textwidth]{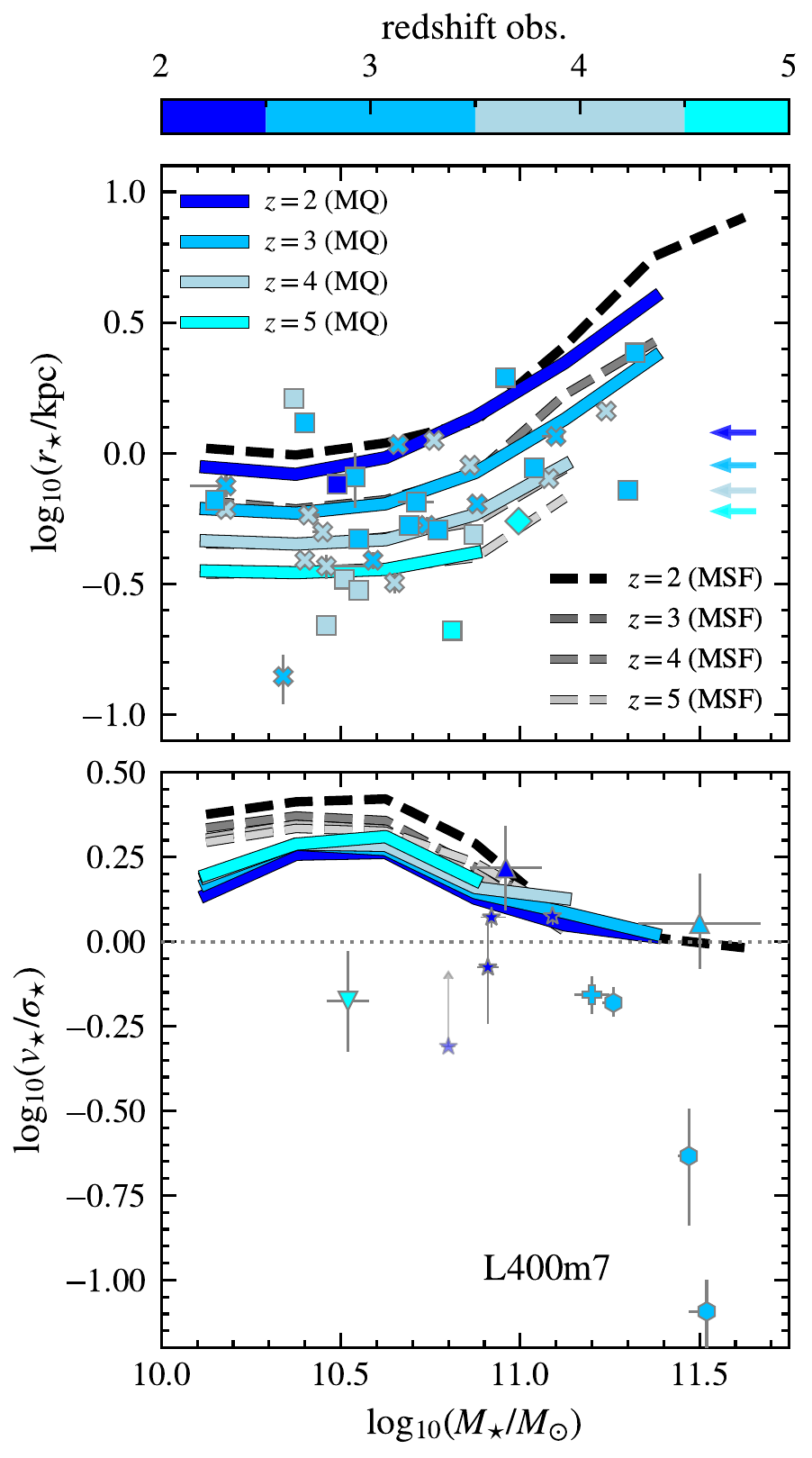}
\caption{Stellar size-mass relation, with $r_{\star}$ defined as the half-mass radius (\textit{top panel}); and $v_{\star}/\sigma_{\star}$ versus stellar mass relation (\textit{bottom panel}) for L400m7. Solid blue-palette lines show the median predictions for MQGs, and dashed grey-palette lines for MSFGs at different selection redshifts, $2 \le z \le 5$. Observational data of MQGs are the same as in Fig.~\ref{fig:sizes}. Arrows in the error bars indicate lower limits for galaxies where rotational velocity could not be constrained (also shown with a more faded colour). Blue arrows in the size–mass relation mark the gravitational softening length, $\epsilon_{\rm prop}$, at each redshift.} 
\label{fig:L400m7-size-kin} 
\end{figure}

We show results analogous to those in \S~\ref{ssec:res1-size} and \S~\ref{ssec:res1-kin} for the fiducial L200m6 simulation, but now for L400m7 in Fig.~\ref{fig:L400m7-size-kin}. We also comment on the L200m7h run where it differs. These results should be interpreted with caution because the larger gravitational softening lengths in these simulations (Table~\ref{tab:runs}) shift the non-reliable regime to less compact systems.

In the top panel, we show the galaxy size–mass relation, which extends up to $z=5$ owing to the larger volume of the L400m7 box. The blue arrows on the left-hand side indicate the corresponding softening length. Relative to L200m6, L400m7 displays a break in the power-law relation at slightly lower stellar masses ($M_{\star} \sim 10^{10.6}\,\mathrm{M_{\odot}}$). L200m7h, in contrast, shows a shallower decrease in size with stellar mass beyond this break. The flattening at the low-mass end persists, suggesting it is not a direct consequence of the softening length, \textcolor{black}{consistent with the behaviour seen across all resolutions at similarly high redshifts \citep[see fig.~10 in][]{ludlow26}}.

For the bottom panel, which presents the stellar kinematics via $v_{\star}/\sigma_{\star}$ versus stellar mass, \textcolor{black}{the contrast between MQGs and MSFGs in L400m7 is even more pronounced than in L200m6. MSFGs display substantially higher rotational support at the low-mass end, whereas MQGs are more dispersion-dominated, effectively reversing the trend found in L200m6 and suggesting that morphological transformation may precede quenching, despite the two populations exhibiting similar sizes. This behaviour may result from the lower resolution and/or the burstier stellar feedback in L400m7, which could more easily perturb the shallower gravitational potentials of low-mass galaxies, affecting their stellar kinematics without significantly altering their sizes. However, a detailed investigation of the origin of this behaviour is beyond the scope of this paper. Although not shown, in L200m7h, the difference between MQGs and MSFGs extends across the entire stellar mass range, suggesting that jet feedback may further enhance the dispersion support of MQGs.} However, these differences should be interpreted cautiously, as the larger softening lengths in these runs likely affect the reliability of the kinematic measurements.


\section{Correlations for MQGs}
\label{appendix:corr}

\begin{figure*}
\centering
\begin{subfigure}[b]{0.49\textwidth}
   \includegraphics[trim={0 0 0 0},clip, width=\textwidth]{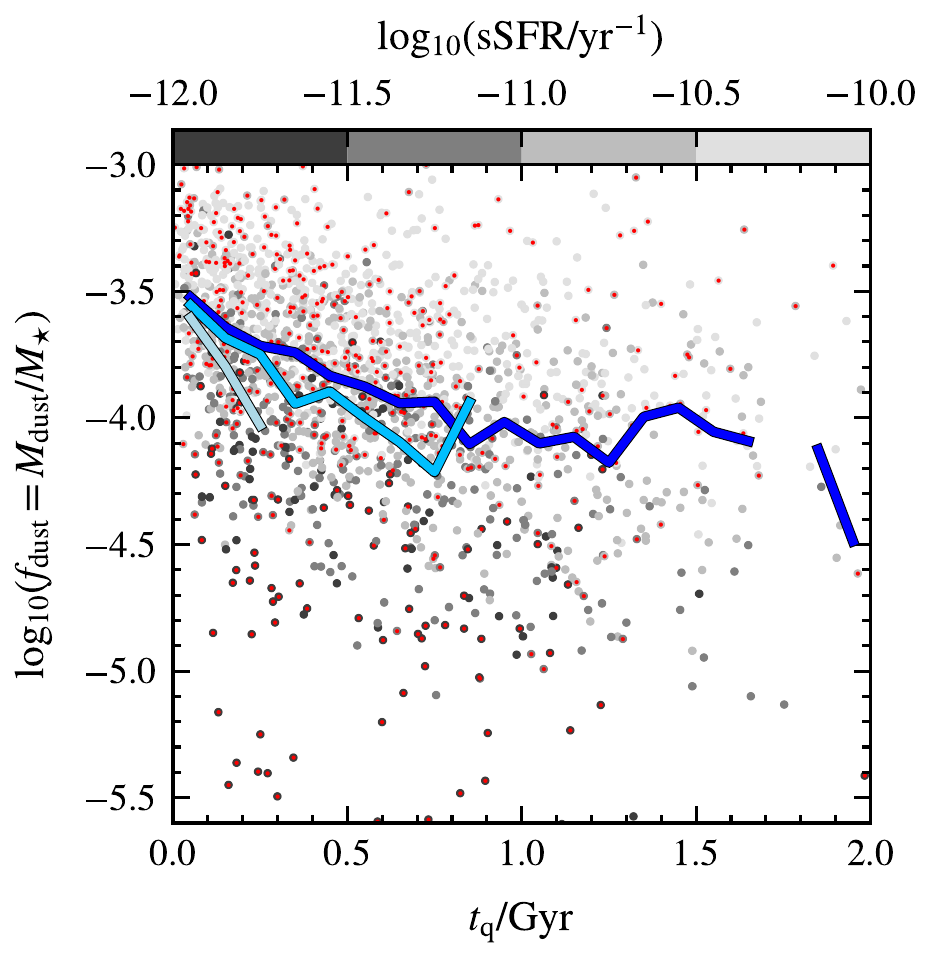}
\end{subfigure}
\begin{subfigure}[b]{0.49\textwidth}
   \includegraphics[trim={0 0 0 0},clip, width=\textwidth]{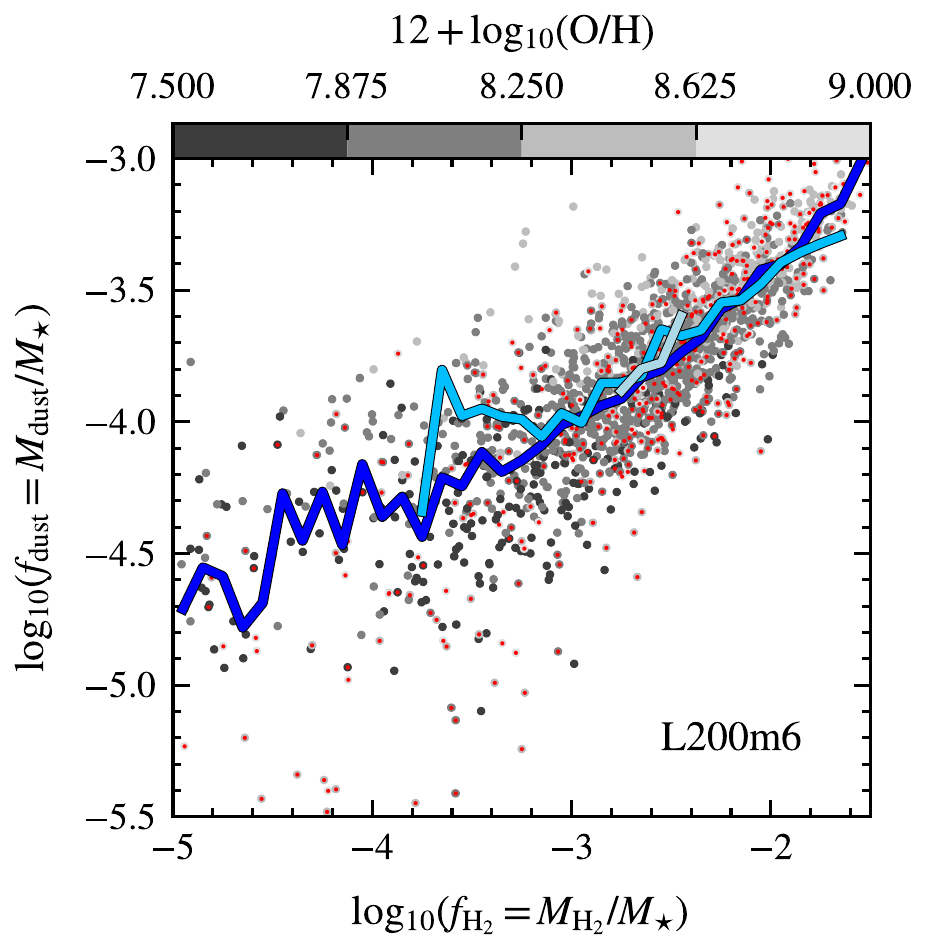}
\end{subfigure}
\\[0.4cm]
\begin{subfigure}[b]{0.49\textwidth}
   \includegraphics[trim={0 0 0 0},clip, width=\textwidth]{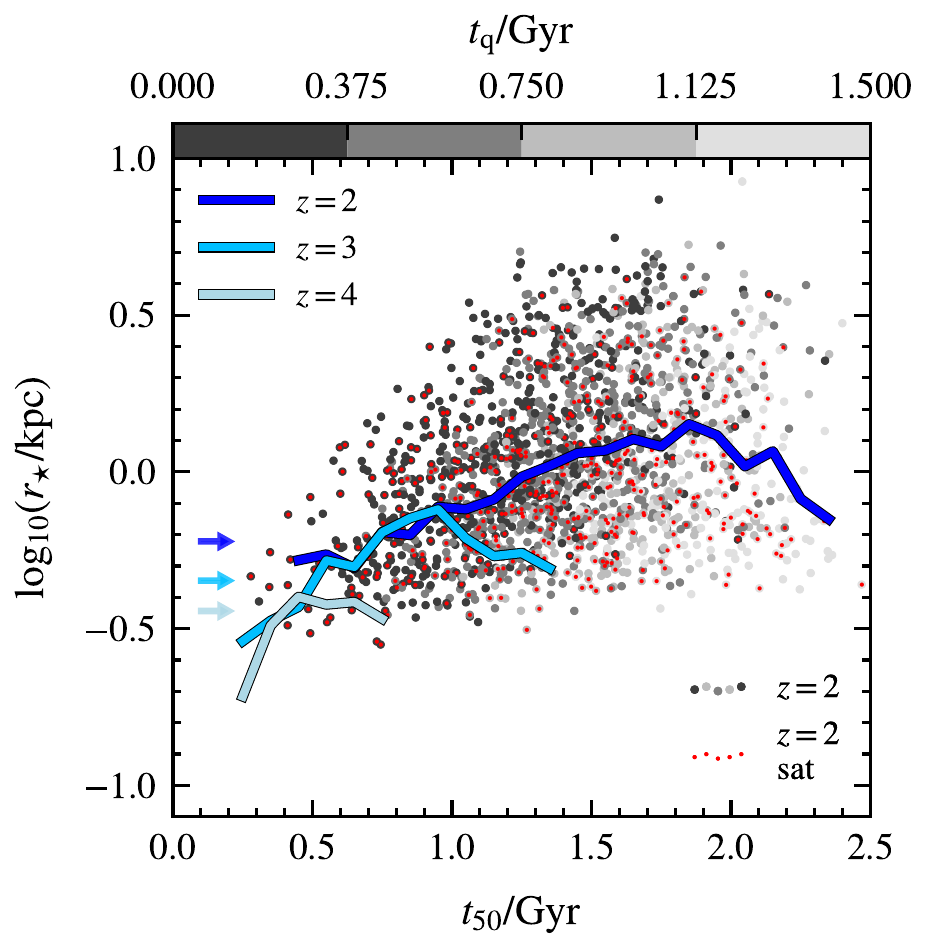}
\end{subfigure}
\begin{subfigure}[b]{0.487\textwidth}
   \includegraphics[trim={0 0 0 0},clip, width=\textwidth]{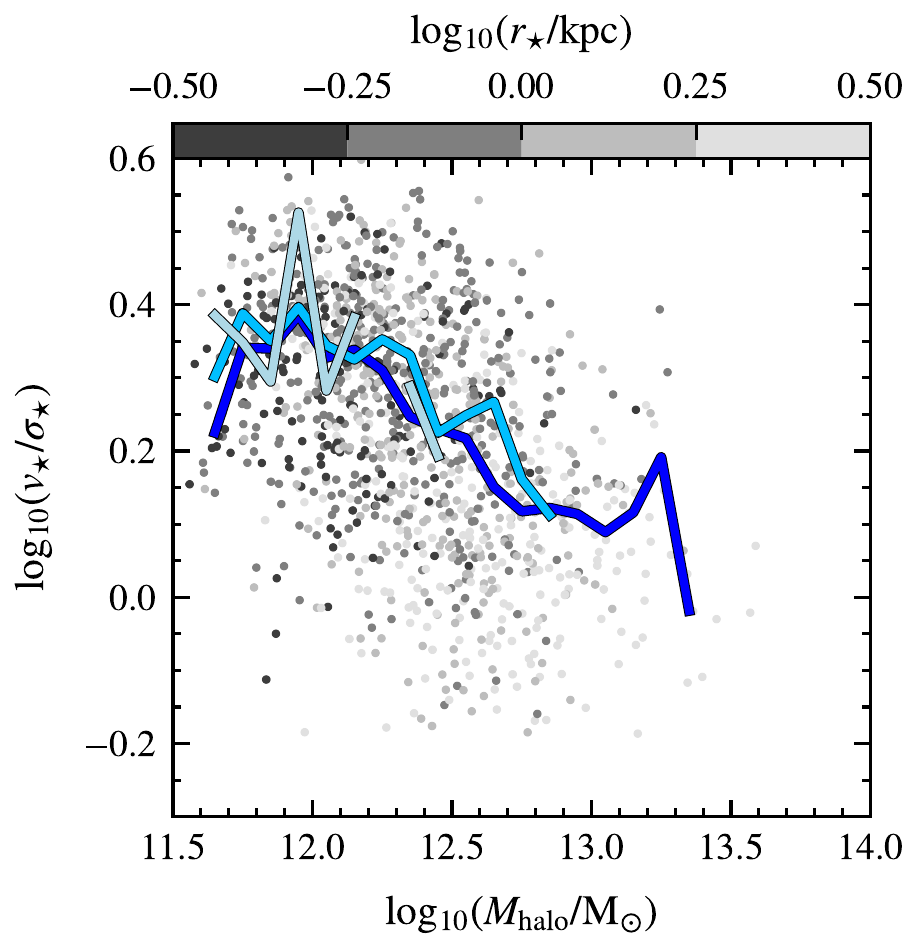}
\end{subfigure}
\caption{Dust fraction ($f_{\rm dust}=M_{\rm dust}/M_{\star}$) versus quenching time, colour-coded by sSFR (\textit{\textcolor{black}{upper left panel}}); dust fraction ($f_{\rm dust}=M_{\rm dust}/M_{\star}$) versus molecular hyfrogen fraction ($f_{\rm H_2}=M_{\rm H_2}/M_{\star}$), colour-coded by gas metallicity ($12+\rm log_{10}(O/H)$) (\textit{\textcolor{black}{upper right panel}}); half-mass radius versus formation timescale, colour-coded by quenching time (\textit{\textcolor{black}{lower left panel}}); and $v_{\star}/\sigma_{\star}$ versus halo mass \textcolor{black}{only for central galaxies}, colour-coded by half-mass radius (\textit{\textcolor{black}{lower right panel}}) for MQGs \textcolor{black}{in L200m6}. 
Solid lines show the median values at different redshifts ($z=2,3,4$), while (red) scatter points represent individual (satellite) $z=2$ MQGs, colour-coded by the properties indicated above. Blue arrows in the galaxy sizes mark the gravitational softening length, $\epsilon_{\rm prop}$, at each redshift.}
\label{fig:corr-t} 
\end{figure*}

We present correlations for MQGs predicted by the {\sc COLIBRE} model \textcolor{black}{in} Fig.~\ref{fig:corr-t}, with median values of Y versus X properties indicated in different colours for different selection redshifts ($2 \leq z \leq 4$) in the L200m6 simulation. Scatter points in the background represent individual galaxies selected at $z=2$, colour-coded by a third Z property. Satellite MQGs are highlighted with red scatter points.



In the \textcolor{black}{upper} left panel, we plot dust fraction versus quenching timescale, showing that galaxies that quenched earlier have lower dust content relative to their stellar mass. This relation becomes steeper at higher redshifts, indicating that dust is depleted after quenching. \textcolor{black}{However, the large scatter is potentially driven by dust regrowth (together with renewed $\rm H_2$ formation), as suggested in the evolutionary histories and the corresponding gas/dust images for these systems in appendix~B of \citetalias{chandro-gomez_inprep}, where some individual systems clearly exhibit this behaviour, with dust even present in the galactic outskirts}. The dust removal seems to be quicker the higher the redshift, as recently reported observationally with AGN as the main responsible mechanism \citep{lesniewska25}. The scatter points further show that dust fraction and quenching times correlate with sSFR: galaxies with lower sSFR have lower dust content, suggesting that dust directly traces star formation. 

The \textcolor{black}{upper right} panel further finds a correlation between dust fraction and $\rm H_{2}$ fraction for MQGs, indicating that molecular hydrogen and dust are removed consistently in these systems. The points are colour-coded in this case by gas metallicity (measured within the $50\,\rm pkpc$ fiducial aperture as the linear sum of the diffuse oxygen over hydrogen ratio of gas, multiplied with the gas mass), showing that, as expected, more metal-rich systems have higher dust fractions. A tail of systems with relatively high metallicity but low dust fraction ($f_{\rm dust}\lesssim10^{-4}$) corresponds to satellite galaxies (in red).

The \textcolor{black}{lower left} panel presents galaxy size versus formation timescale, indicating that earlier-formed galaxies are more extended, having had more time to grow. This trend reflects hierarchical structure formation, where objects start smaller and grow over time. Colour-coding by quenching timescales further demonstrates that formation and quenching times are correlated.
In the \textcolor{black}{lower right} panel, more rotation-supported systems tend to reside in lower-mass haloes \textcolor{black}{(where $M_{\rm halo}$ is the total mass of the 3D FoF group to which the galaxy belongs)}. 
This is in agreement with the observational results in \citep{kawinwanichakij25}, where more dispersion-supported systems live in denser environments, highlighting the role of environment in shaping morphology at such early times. \textcolor{black}{Further details on the role of the environment for these systems in \citetalias{chandro-gomez_inprep}.} Colour-coding by galaxy size in the lower right panel indicates that rotation-supported systems are more compact than dispersion-supported systems, a direct consequence of their later formation times (lower left panel of Fig.~\ref{fig:corr-t}). \textcolor{black}{Although not shown}, we also identify a shallower trend in which earlier-formed galaxies inhabit more massive haloes, a direct link between DM haloes and galaxies \citep{white91}.




\bsp	
\label{lastpage}
\end{document}